\pdfoutput=1
\documentclass[reprint,amsmath,amssymb, aps]{revtex4-2}
\usepackage{graphicx}
\usepackage{float}
\usepackage{dcolumn}
\usepackage{color}
\usepackage{latexsym,bm}
\usepackage[normalem]{ulem}
\usepackage{multirow}
\usepackage{appendix}
\usepackage{amsmath}
\usepackage{amsfonts}
\usepackage{booktabs}
\usepackage{array}
\usepackage{soul}
\usepackage{CJKutf8}

\newcommand{\MC}[1]{\textcolor{black}{{#1}}}
\newcommand{\MCC}[1]{\textcolor{black}{{#1}}}

\newcommand{\tabincell}[2]{
\begin{tabular}{@{}#1@{}}#2\end{tabular}
}
\begin{document}

\hyphenpenalty=5000
\tolerance=1000
\begin{CJK}{UTF8}{gkai} 
\title{Plane-Wave-Based Stochastic-Deterministic Density Functional Theory for Extended Systems} 

\author{Qianrui Liu (刘千锐)}
\affiliation{HEDPS, CAPT, College of Engineering and School of Physics, Peking University, Beijing 100871, P. R. China}
\author{Mohan Chen (陈默涵)}
\email{mohanchen@pku.edu.cn (Corresponding author)}
\affiliation{HEDPS, CAPT, College of Engineering and School of Physics, Peking University, Beijing 100871, P. R. China}
\date{\today}
%%%%%%%%%%%%%%%%%%%%%%%%%%%%%%%%%%%%%%%%%%%%%%%%%%%%%%%%%%%%%%%%%%%%%%%%%%%%%%%%%%%%%%%%%%%%%%%%%%%%%%%%%%%%%%%%%%%%%%%%%%%%%%%%%%%
%%%%%     Title
%%%%%%%%%%%%%%%%%%%%%%%%%%%%%%%%%%%%%%%%%%%%%%%%%%%%%%%%%%%%%%%%%%%%%%%%%%%%%%%%%%%%%%%%%%%%%%%%%%%%%%%%%%%%%%%%%%%%%%%%%%%%%%%%%%%

\begin{abstract}
{\MCC{Traditional finite-temperature Kohn-Sham density functional theory (KSDFT) that based on the diagonalization (DG) method has an unfavorable scaling with respect to the electron number or at high temperatures. 
The evaluation of the ground-state density in KSDFT can be replaced by the Chebyshev trace (CT) method.
In addition, the use of stochastic orbitals within the CT method leads to the stochastic density functional theory [Phys. Rev. Lett. 111, 106402 (2013)] (SDFT) and its improved theory, mixed stochastic-deterministic density functional theory [Phys. Rev. Lett. 125, 055002 (2020)] (MDFT).
We have implemented the above four methods based on the plane-wave basis set within the first-principles package ABACUS.
In addition, the four methods are adapted to the norm-conserving pseudopotentials and periodic boundary conditions with the use of $k$-point sampling in the Brillouin zone.
By testing the Si and C systems \MCC{from the DG method} as benchmarks, we systematically evaluate the accuracy and efficiency of the CT, SDFT, and MDFT methods by examining a series of physical properties, which include the electron density, the free energy, the atomic forces, stress, and density of states.
We conclude that the CT, SDFT, and MDFT methods not only reproduce the DG results with a sufficient accuracy, but also exhibit several advantages over the DG method. We expect these methods can be of great help in studying high-temperature and large-size extended systems such as warm dense matter and dense plasmas.
}

}

\end{abstract}
%\pacs{}
\maketitle
\end{CJK}

\section{Introduction}

The past few decades have established Kohn-Sham (KS) density functional theory~\cite{64PR-Hohenberg,65PR-Kohn} (DFT) as one of the most popular quantum-mechanics-based methods in modeling materials due to an excellent balance between accuracy and efficiency.
Traditional KSDFT solves the KS equation \MCC{by the diagonalization (DG) method} to yield the ground-state electron density of a given system consisting of electrons and nuclei.
However, the solution procedure is typically considered to scale as $O(N_e^3)$, where $N_e$ is the number of electrons simulated in the system.
It becomes expensive for traditional KSDFT to simulate large systems \MCC{over hundreds of atoms}.
Linear-scaling techniques~\cite{91L-Yang,93B-Li,96L-Kohn,99RMP-Goedecker,97MSMSE-Bowler,96B-Ordejon,07CPC-Gillan,08B-Shimojo} have been developed to avoid solving KS orbitals by taking advantage of the “nearsightedness” principle.~\cite{05PNAS-Prodan}
However, these methods typically can only be applied to non-metallic systems.~\cite{97MSMSE-Bowler,96B-Ordejon,07CPC-Gillan} 
Although some metallic systems~\cite{08B-Shimojo} can be studied, additional approximations are usually involved.
As a result, linear-scaling KSDFT cannot fully take the place of traditional KSDFT, yet they can treat large systems \MCC{up to tens of thousands of atoms}.

The Mermin finite-temperature density-functional approach~\cite{65PR-Mermin} enables KSDFT be readily applied to study extremely high-temperature materials such as a dense plasmas and warm dense matter.~\cite{82RMP-Ichimaru,14-Graziani,05PPCF-Koenig}
For instance, recent advances of applying KSDFT to study equations of state,~\cite{11PP-Wang,14E-Sheppard,15E-Hu,18E-Driver} opacity,~\cite{09HEDP-Vinko,14E-Hu,16B-Zhang,21E-Karasiev} X-ray Thomson scattering,~\cite{09RMP-Glenzer,15E-Vorberger,18L-Mo} and transport properties~\cite{02E-Desjarlais,18PP-Witte} have provided abundant microscopic insight into high energy density physics and laboratory astrophysics.
\MCC{Besides, the consideration of temperature effects in the approximated form of the exchange-correlation (XC) functional has been expected to yield better results at finite temperatures.~\cite{14L-Karasiev,16E-Karasiev,19B-Karasiev}}
However, these new applications typically suffer from a low efficiency of traditional KSDFT due to two main reasons.~\cite{01L-Surh, 10L-Hu, 14E-Sheppard} On the one hand, as the temperature $T$ rises, more electrons are ionized and a higher kinetic energy cutoff becomes a necessity to accurately describe the interactions between electrons and nuclei. 
On the other hand, the needed number of KS electron wave functions is proportional to $T^{3/2}$, resulting in an unfavorable $O(T^3)$ scaling to obtain these electron wave functions.~\cite{18B-Cytter}
Several methods have been proposed to overcome the difficulties in simulating high-temperature materials. For example, the path integral Monte Carlo (PIMC) method adopts stochastic techniques to integrate the many-body density matrix in calculations.~\cite{84B-Pollock,84JPSJ-Takahashi,86L-Ceperley,95RMP-Ceperley,15L-Militzer,17E-Zhang}
However, limited by the computational efficiency, PIMC faces great challenges in simulating high-Z materials beyond the second-row elements.~\cite{15L-Militzer,17E-Zhang,21E-Militzer} 
Orbital-free density functional theory~\cite{02-Wang,12B-Karasiev,14L-Sjostrom,20B-Luo} is fast for temperatures over tens of eV, but its accuracy is affected by the use of kinetic-energy density functional (KEDF). 
In fact, 
%the KEDF yields imprecise results~\cite{13B-Sjostrom,20PP-Blanchet,21B-Liu} 
the local KEDF such as the finite-temperature Thomas–Fermi model~\cite{27MPCPS-Thomas} is only accurate for free-electron-like systems while the nonlocal KEDF~\cite{92B-Wang,10B-Huang} can improve the results to some extent but for limited systems.~\cite{13B-Sjostrom,20JPCM-Liu}
The extended first-principles molecular dynamics method~\cite{16PP-Zhang,20PP-Blanchet} uses a plane-wave basis to approximately replace high energy orbitals when simulating extremely high-temperature systems.~\cite{21B-Liu,16B-Gao,20B-Mo}

Recently, several works~\cite{13L-Baer,18B-Cytter,14JCP-Daniel,17JCP-Aronon,19B-Cytter,19JCP-Ming,19JCP-Ming2,19JCP-Li,19WCMS-Fabian} \MCC{based on the framework of KSDFT} have been developed to focus on the stochastic density functional theory (SDFT).
SDFT owns a few advantages as compared to traditional KSDFT. First,
the stochastic orbitals instead of the orthogonal KS wave functions are used in SDFT to avoid the computationally expensive orthogonal operations acting on the KS wave functions, and the computational cost of SDFT in principle \MCC{scales} linearly with the system size.
Second, SDFT scales inversely to the temperature~\cite{18B-Cytter} because a small order of Chebyshev expansion is needed to expand the Fermi-Dirac distribution function at high temperatures.
Therefore, the SDFT method is well suitable to study high-temperature materials.
Third, the stochastic orbitals used in SDFT can be trivially parallelized, and the memory costs in SDFT remain almost unchanged with varying temperatures.
However, the SDFT unavoidably suffers from statistical errors. For instance, volume-averaged quantities such as the total energy per electron decreases as $1/\sqrt{N_e}$ and decays towards zero as the system size becomes larger.~\cite{13L-Baer}
In order to mitigate the statistical noise in SDFT, some fragmentation approaches have recently been proposed for molecular systems~\cite{14JCP-Daniel} and bonded materials.~\cite{17JCP-Aronon,19JCP-Ming} In particular, the stochastic embedding DFT~\cite{19JCP-Li} \MCC{was} applied to {\it p}-nitroaniline in water. In addition, the energy window SDFT~\cite{19JCP-Ming2} method becomes a more robust scheme to reduce the noise by breaking the electron density into different components according to orbital energies.
Notably, White {\it et al.} implemented mixed stochastic-deterministic density functional theory (MDFT)~\cite{20L-White}, which combines the advantages of traditional KSDFT and SDFT and is faster and more accurate than just using the stochastic orbitals.
However, to the best of our knowledge, the above SDFT and MDFT methods have not been
applied to sample the $k$-points in the Brillouin zone yet. \MCC{Furthermore, the formulas of implementing the methods with the plane-wave basis have not been well documented. In this regard, our work could provide useful information for adopting the SDFT and MDFT methods to study extended systems in condensed matter physics.}~\cite{13L-Baer,18B-Cytter,20L-White}

\MC{
The plane-wave pseudopotential~\cite{04-Martin,79L-Hamann} method is suitable for studying materials under periodic boundary conditions. Notably, a variety of works, which adopted KSDFT to study warm dense matter or dense plasmas, have used plane-wave packages.~\cite{11PP-Wang,14E-Sheppard,15E-Hu,18E-Driver,09HEDP-Vinko,14E-Hu,16B-Zhang,21E-Karasiev,15E-Vorberger,18L-Mo,02E-Desjarlais,18PP-Witte}
In fact, there are several advantages of using plane waves with pseudopotentials in KSDFT calculations.~\cite{92RMP-Payne}
First, periodic boundary conditions 
with the use of $k$-point sampling enable the study of various physics properties of extended systems such as condensed matter systems.
Second, a \MCC{smaller} plane-wave basis is needed for pseudopotentials when compared to all-electron methods, so the computational costs and memory costs are \MCC{reduced}.
Third, the number of plane-wave basis \MCC{functions} can be systematically increased by the kinetic energy cutoff. The formula of plane wave is simple and it can be used with Fast Fourier Transform (FFT) which improves the computing efficiency.
Fourth, the plane-wave basis is independent of atomic coordinates. Thus, it is easy to compute atomic forces directly with the Hellmann-Feynman theorem~\cite{35AP-Hellmann,39-Feynman}
%compared with the atomic basis which is dependent of the position of atoms and needs to calculate Pulay forces~\cite{16CMS-Li,69MP-Pulay} in addition.
}

\MCC{In this work, we first introduce deterministic KSDFT with the DG method and the Chebyshev trace (CT) method by using a plane-wave basis set. We then describe the detailed formulas of both SDFT and MDFT methods based on \MCC{a} plane-wave basis set, aiming to simulate high-temperature systems with periodic boundary conditions. All of the above four methods are compatible with sampling methods of the Brillouin zone such as the Monkhorst-Pack method.~\cite{73B-Chadi,76B-Monkhorst} 
The above four methods have been implemented in the ABACUS package,~\cite{16CMS-Li} we systematically compare several aspects of the DG, CT, SDFT and MDFT methods and choose Si and C as target systems. The tested physical properties include the electron density, the free energy, the pressure, the atomic forces and the density of states. We also \MCC{emphasize} justifying the statistical errors that exist in the SDFT and MDFT methods. Furthermore, we evaluate the efficiency of the SDFT and MDFT methods in terms of parallel scheme for large systems up to 512 atoms. We demonstrate that the use of stochastic orbitals is sufficiently accurate and more efficient than the KSDFT method at high temperatures.
%Several systems consisting of Si and C atoms are chosen.
%We have systematically tested the accuracy and efficiency of the MDFT method. 
%On the one hand, properties of the electron density, the free energy, the pressure, the forces, and the density of states are calculated.
%On the other hand, we compare the parallel efficiency of KSDFT and MDFT and the efficiency's varying with temperatures of KSDFT and SDFT.
The paper is organized as follows. In Sec.~II we briefly introduce the \MCC{DG, CT, SDFT and MDFT} methods, as well as the error analysis of SDFT and MDFT. In Sec.~III, \MCC{we present the results and discussions in terms of the accuracy and efficiency.} We conclude our work in Sec.~IV.
}

\section{Methods}
\MCC{
In the KSDFT~\cite{64PR-Hohenberg,65PR-Kohn,65PR-Mermin} framework, the ground-state total energy functional of a given electron system is defined as
 \begin{equation}\label{etot}
 \begin{split}
    E_{\mathrm{tot}}[\rho(\mathbf{r})]=T_s+\int v_{\mathrm{ext}}(\mathbf{r})\rho(\mathbf{r})d\mathbf{r}\\
    +\frac{1}{2}\int\int\frac{\rho(\mathbf{r})\rho(\mathbf{r^{\prime}})}{|\mathbf{r}-\mathbf{r^{\prime}}|}d\mathbf{r}d\mathbf{r^{\prime}}
    +E_{\mathrm{xc}}[\rho(\mathbf{r})],
 \end{split}
 \end{equation}
where the first term $T_s$ is the kinetic energy of a non-interacting electron system.
The second one is the energy caused by the external potential $v_{\mathrm{ext}}(\mathbf{r})$ and $\rho(\mathbf{r})$ is the electron density in real space.
The third one is the Hartree repulsion energy and the last one depicts the exchange-correlation energy.
Since the ground state energy is the global minimum of the functional in Eq.~(\ref{etot}), the one-particle Kohn-Sham equation is derived from the variational method as
  \begin{equation}\label{kseq}
  \Big[-\frac{1}{2}\nabla^2+v_{\mathrm{eff}}(\mathbf{r})-\epsilon_i\Big]\psi_i(\mathbf{r})=0,
  \end{equation}
  where $\epsilon_{i}$ and $\psi_{i}$ are the eigenvalues and the electron wave functions, respectively. 
   The effective potential $v_{\mathrm{eff}}(\mathbf{r})$ takes the form of
  \begin{equation}\label{veff}
	v_{\mathrm{eff}}(\mathbf{r})=v_{\mathrm{ext}}(\mathbf{r})+\int{\frac{\rho(\mathbf{r^{\prime}})}{|\mathbf{r}-\mathbf{r^{\prime}}|}\mbox{d}\mathbf{r^{\prime}}}
    +v_{\mathrm{xc}}(\mathbf{r}).
  \end{equation}
  Here
  \begin{equation}
      v_\mathrm{xc}(\mathbf{r})=\frac{\delta E_\mathrm{xc}[\rho(\mathbf{r})]}{\delta\rho(\mathbf{r})}.
 \end{equation}
  }

\MCC{In this work, KSDFT is implemented with the DG, CT, SDFT, and MDFT methods. In principle, the DG and CT methods are equivalent methods, which use deterministic orbitals while the other two methods adopt stochastic orbitals.
The four methods have been implemented in the ABACUS package~\cite{10JPCM-Mohan,16CMS-Li} with the usage of norm-conserving pseudopotentials and a plane-wave basis set with periodic boundary conditions. Furthermore, the four methods support the usage of multiple $k$-points in the Brillouin zone.}

We choose bulk Si and C systems in order to compare the accuracy and efficiency of the above mentioned methods. The pseudopotential of Si \MCC{describes} 4 valence electrons with a cutoff radius of 1.3 Bohr and the cutoff kinetic energy is set to 50 Ry. The LDA~\cite{65PR-Kohn,81B-Perdew} exchange-correlation functional is used for Si. 
In addition, we prepare two pseudopotentials for C, which share the same cutoff radius of 0.8 Bohr. One \MCC{describes} 6 valence electrons and the cutoff kinetic energy is set to 240 Ry. The other one has 4 valence electrons and the energy cutoff is 70 Ry. In practice, we use the 4-valence-electron pseudopotential for temperatures no more than 20 eV and the 6-valence-electron pseudopotential for higher temperatures. The XC functional is chosen to be PBE~\cite{96L-PBE} for the C systems.

  \subsection{\MCC{Diagonalization Method}}

  Traditional KSDFT is generally implemented with the diagonalization (DG) method.
  \MCC{Solving Eq.~(\ref{kseq}) can be accomplished by diagonalizing the KS Hamiltonian $\hat{H}=-\frac{1}{2}\nabla^2+v_{\mathrm{eff}}(\mathbf{r})$}
   and the electron density can be evaluated from the Kohn-Sham orbitals as
 \begin{equation}\label{eq:rho}
    \rho(\mathbf{r})=2\sum_{i=1}^{N_{oc}}f(\epsilon_i;\mu)|\psi_{i}(\mathbf{r})|^2,
  \end{equation}
  where $f(\epsilon_i;\mu)$ is the Fermi-Dirac function and $\mu$ is the chemical potential. The value of $f(\epsilon_i;\mu)$ represents the occupation of the $i$-th orbital. Orbitals with occupation larger than a specific threshold are considered as occupied and $N_{oc}$ is the number of occupied orbitals.
  
  \begin{figure*}
	\centering
	\includegraphics[width=16cm]{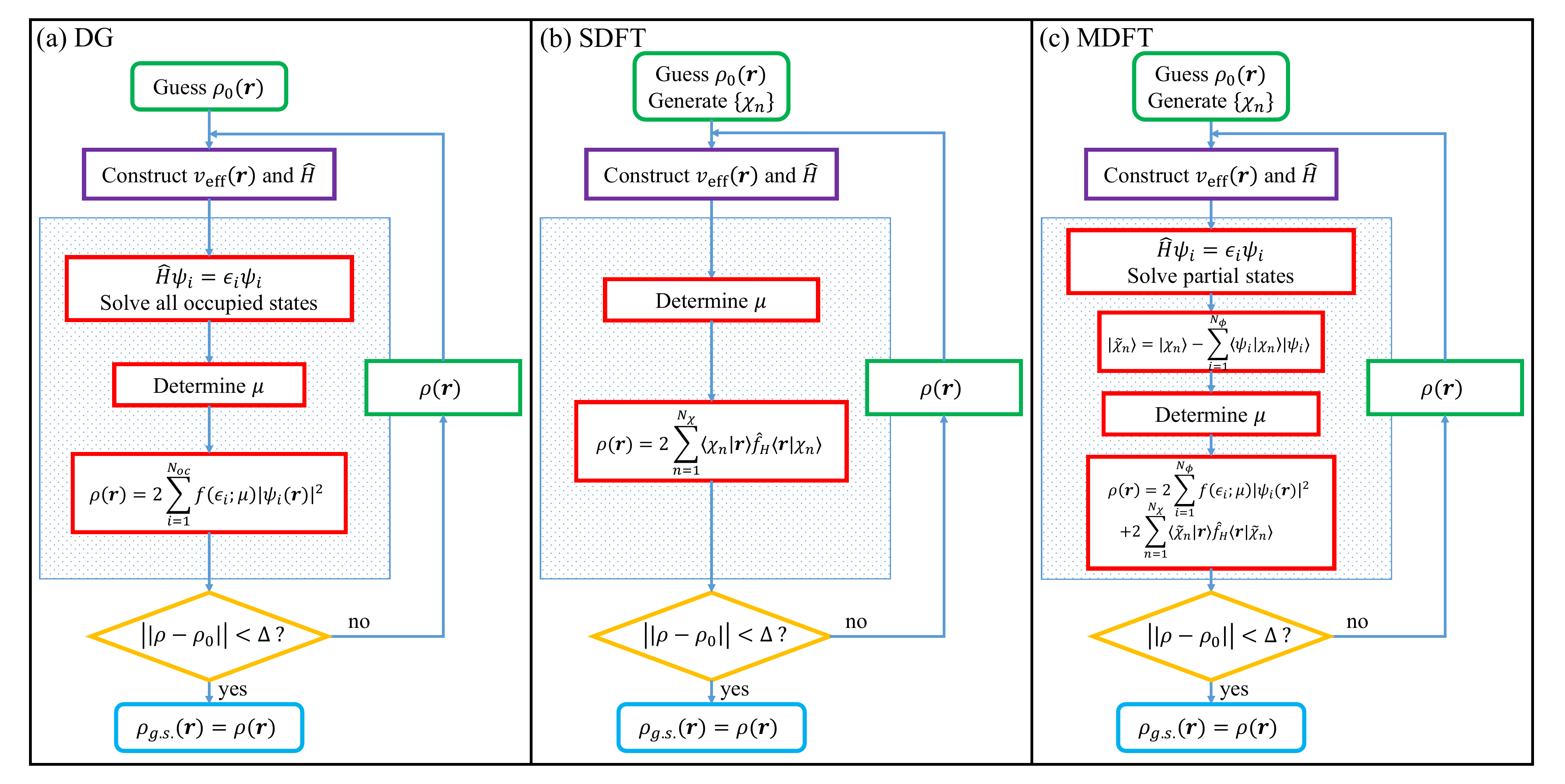}\\
	\caption{\MC{(Color online) Workflows of (a) the \MCC{diagonalization (DG)} method, (b) the stochastic DFT (SDFT) method, and (c) the mixed stochastic-deterministic DFT (MDFT) method as implemented in ABACUS.
	$\rho_0$, $\rho$, $\rho_{g.s.}$ are the initial guess of electron density, the computed electron density, and the ground-state electron density, respectively.
	$v_{eff}(\mathbf{r})$ is the effective potential in real space and the Hamiltonian of the system is $\hat{H}=-\frac{1}{2}\nabla^2+v_{eff}(\mathbf{r})$.
	$\epsilon_i$ and $\psi_i(\mathbf{r})$ are the eigenvalue and the electronic wave function of the $i$th state, while $\mu$ depicts the chemical potential.
	$\Delta$ defines the threshold of convergence.  
    $N_{\chi}$ is the number of stochastic orbitals $\{\chi_i\}$.
    The Fermi-Dirac function is $\hat{f}_H$ with the hamiltonian operator being $\hat{H}$.
     The stochastic orbitals $\tilde{\chi}_n$ in MDFT are orthogonal to the deterministic orbitals $\{\psi_i\}$ with the number of $\psi_i$ being $N_{\phi}$.
    The major difference exists in the procedure to compute $\rho(\mathbf{r})$ from the Hamiltonian $\hat{H}$, which are labelled with the red boxes in the above three workflows.}
	}\label{fig:workflow}
  \end{figure*}
  
  \MCC{
  In practice, the electron density is computed via a self-consistent loop. In order to clarify the procedure,\MCC{ Fig.~\ref{fig:workflow} illustrates three different self-consistent schemes of the KSDFT method. Fig.~\ref{fig:workflow}(a) shows the DG method with the following seven steps:} 
  (i) Guess an initial electron density $\rho_0(\mathbf{r})$. 
  (ii) Construct an effective potential $v_{\mathrm{eff}}(\mathbf{r})$ with Eq.~(\ref{veff}) and obtain  \MCC{the Hamiltonian $\hat{H}$}.
  (iii) Solve the Kohn-Sham equation shown in Eq.~(\ref{kseq}) \MCC{by the DG method} and obtain occupied electronic wave functions \{$\psi_i$\} and eigenvalues \{$\epsilon_i$\} \MCC{with the Fermi weights smaller than a specific threshold.}
  (iv) Determine the chemical potential $\mu$ according to $N_e=2\sum_{i=1}^{N_{oc}}{f(\epsilon_i;\mu)}$. 
  (v) Evaluate a new electron density $\rho(\mathbf{r})$ via Eq.~(\ref{eq:rho}).
  (vi) \MCC{Obtain a new guessing electron density by mixing the new electron density $\rho(\mathbf{r})$ with previous ones.~\cite{80CPL-Pulay,88B-Johnson}}
  (vii) Perform iterations for steps (ii)-(vi) until the difference between $\rho(\mathbf{r})$ and $\rho_0(\mathbf{r})$ is under a specific threshold. The ground-state electron density is $\rho(\mathbf{r})$.}
  
  \MC{
  Next, the total energy of the electron system takes the form of
  \begin{equation}
  \begin{aligned}
    &E_{\mathrm{tot}}=\sum_{i=1}^{N_{oc}}{f(\epsilon_i;\mu)\epsilon_i}-\frac{1}{2}\int\int{\frac{\rho(\mathbf{r})\rho(\mathbf{r'})}{|\mathbf{r}-\mathbf{r'}|}d\mathbf{r}d\mathbf{r'}}\\
    &-\int{v_\mathrm{xc}(\mathbf{r})\rho(\mathbf{r})d\mathbf{r}}+E_{\mathrm{xc}}[\rho(\mathbf{r})].
  \end{aligned}
  \end{equation}
  Thus, the total energy of electrons and ions in the system is
  \begin{equation}
    E=E_{\mathrm{tot}}+E_{\mathrm{II}},
  \end{equation}
  where $E_{II}$ is the ion-ion repulsion energy.
  By considering the temperature effect, the free energy of the system $A$ has the form of
  \begin{equation}
    A=E-TS,
  \end{equation}
  where the electron entropy term is
  \begin{equation}
  \begin{aligned}
    S=-&2\sum_{i=1}^{N_{oc}}\big\{f(\epsilon_i;\mu)\mathrm{ln}f(\epsilon_i;\mu)\\
    &+[1-f(\epsilon_i;\mu)]\mathrm{ln}[1-f(\epsilon_i;\mu)]\big\}.
  \end{aligned}
  \end{equation}
  }
  \MCC{The formulas of expanding the Kohn-Sham equation in a plane wave basis set are provided in Appendix A, while the formulas of forces and stress as obtained from the DG method can be found in Appendix B.}

% Method B: 
  \subsection{Chebyshev Trace Method}
  The Chebyshev trace (CT) ~\cite{95CPL-Huang,97JCP-Baer,97L-Baer} method is equivalent to \MCC{the DG method in obtaining the electron density}. However, the CT method skips the \MCC{diagonalization} procedure to solve the Kohn-Sham equation in Eq.~(\ref{kseq}). Instead, the CT method directly obtains the electron density $\rho(\mathbf{r})$, which takes the form of
  \begin{equation}\label{eq:rhoct}
      \rho(\mathbf{r})=2{\rm Tr} [f(\hat{H};\mu)\delta(\hat{\mathbf{r}}-\mathbf{r})]=
      2{\rm Tr}[f(\hat{H};\mu)|\mathbf{r}\rangle \langle \mathbf{r}|].
  \end{equation}
  Here, the Hamiltonian operator of a give system is $\hat{H}$ while the chemical potential is $\mu$. In this regard, 
  $f(\hat{H};\mu)$ refers to the density matrix with $f$ being the Fermi-Dirac function as defined in Eq.~(\ref{eq:rho}).
 For short, we define
 \begin{equation}
      \hat{f}_H\equiv f(\hat{H};\mu).
 \end{equation}
 It is worth mentioning that Eq.~(\ref{eq:rhoct}) is equivalent to Eq.~(\ref{eq:rho}) in principle, and hence the two methods should yield the same results.
 We \MCC{divide the $\hat{f}_H$ operator into $\hat{f}_H^{1/2}\cdot\hat{f}_H^{1/2}$} and choose the plane-wave basis set to compute the trace in Eq.~(\ref{eq:rhoct}):
  \begin{equation}\label{eq:rhoctpw}
  \begin{aligned}
      \rho(\mathbf{r})&=2\sum_{\mathbf{k}}W(\mathbf{k})\sum_{\mathbf{G}}{\langle\mathbf{k+G}|\hat{f}_{H}^{1/2}|\mathbf{r}\rangle \langle \mathbf{r}|\hat{f}_{H}^{1/2}|\mathbf{k+G}\rangle}\\
      &=2\sum_{\mathbf{k}}W(\mathbf{k})\sum_{\mathbf{G}}{\left|\langle\mathbf{r}|\hat{f}_{H}^{1/2}|\mathbf{k+G}\rangle\right|^2}.
  \end{aligned}
  \end{equation}
  \MCC{Here the prefactor 2 accounts for the electron spin while} $\mathbf{G}$ and $\mathbf{k}$ denote the wave vectors of a plane wave and a $k$-point, respectively. $W(\mathbf{k})$ is the weight of the $\mathbf{k}$ vector.
  %and $\langle r|\mathbf{k+G}\rangle$ is the plane-wave basis which is defined in Eq.~(\ref{eq:pw}).
  
  \MC{For a given operator $\hat{a}$ whose eigenvalues have absolute values less than one, an analytic function $g$ can be evaluated by the Chebyshev expansion method\cite{86CPL-Kosloff,94B-Wang,95CPL-Huang,97JCP-Baer,97L-Baer} with the formula
  \begin{equation}
  g(\hat{a})=\sum_{n=0}^{N_c} C_n[g]T_n(\hat{a}),
  \end{equation}
  where $N_c$ is the expansion order.
  The coefficients are $\{C_n\}$ and 
  $T_n$ is the $n$th-order Chebyshev polynomial.}
  \MC{
  The coefficients can be evaluated by
  \begin{equation}\label{eq:cn}
  \begin{aligned}
      C_n[g]&=\frac{2-\delta_{n0}}{\pi}\int_{-1}^1{\frac{g(x)}{\sqrt{1-x^2}}dx}\\
            &=\frac{2-\delta_{n0}}{\pi}\int_0^\pi{g\Big({\rm cos}(\theta)\Big){\rm cos}(n\theta)d\theta}.
  \end{aligned}
  \end{equation}
  Furthermore, the integration in Eq.~(\ref{eq:cn}) can be calculated according to the Simpson's 1/3 rule~\cite{89-Atkinson} by splitting the integral interval into $N_d$ fragments
  \begin{equation}
  \begin{aligned}
      C_n[g]=\frac{2-\delta_{0n}}{6N_d}\Re\Big\{\tilde{g}_n\Big[{{\rm cos}\Big(\frac{\pi k}{N_d}\Big)}\Big] \\
      + 2e^{\frac{in\pi}{2N_d}}\tilde{g}_n\Big[{{\rm cos}\Big(\frac{\pi (k+\frac{1}{2})}{N_d}\Big)}\Big]\Big\},
  \end{aligned}
  \end{equation}
  where $\Re$ depicts the real part and 
  \begin{equation}
      \tilde{g}_n\Big[p(k)\Big]\equiv\sum_{k=0}^{2N_d-1}{g\Big[p(k)\Big]e^{i\frac{2\pi k}{2N_d}n}}
  \end{equation}
  can be evaluated via fast Fourier transform. We choose $N_d=16N_c$ to avoid accuracy loss in high orders of expansion because the number of fragments $N_d$ should be no less than the frequency of ${\rm cos}(n\theta)$ in Eq.~(\ref{eq:cn}).
  For the $n$th-order Chebyshev polynomial $T_n$, one sets $T_0(\hat{a})=1$, $T_1(\hat{a})=\hat{a}$ and for $n > 1$, $T_n(\hat{a})=2\hat{a}T_{n-1}(\hat{a})-T_{n-2}(\hat{a})$.}
  
  \MC{
  One key operation in applying the CT method to solve the KS equation is to calculate $\hat{f}_{H}^{1/2}|\mathbf{k+G}\rangle$ in Eq.~(\ref{eq:rhoctpw}).
  First, we normalize $\hat{H}$ as
  \begin{equation}
      \hat{h} = \frac{\hat{H}-\frac{1}{2}(E_\mathrm{max}+E_\mathrm{min})}{\frac{1}{2}(E_\mathrm{max}-E_\mathrm{min})}\equiv\frac{\hat{H}-\Bar{E}}{\Delta E},
  \end{equation}
  where $E_\mathrm{max}$ ($E_\mathrm{min}$) is the upper (lower) bound of eigenvalues of $\hat{H}$. If we set $|\mathbf{k+G}\rangle$ as $|\xi_0\rangle$, then
  \begin{equation}\label{cheby}
  \begin{aligned}
      \sqrt{f(\hat{h}\cdot\Delta E+\Bar{E};\mu)}|\mathbf{k+G}\rangle&\equiv g(\hat{h})|\xi_0\rangle\\
      &=\sum_{n=0}^{N_c} C_n[g]|\xi_n\rangle.
  \end{aligned}
  \end{equation}
  Here $|\xi_n\rangle$ equals to $T_n(\hat{h})|\xi_0\rangle$, which satisfies $|\xi_n\rangle=2\hat{h}|\xi_{n-1}\rangle-|\xi_{n-2}\rangle$.
  The chemical potential $\mu$ is determined by tuning the electron number to be $N_e$.
%  In order to determine the chemical potential $\mu$, different numbers of electrons $N_e$ will be calculated when setting different $\mu$. 
  If we consider obtaining $N_e$ by integrating $\rho(\mathbf{r})$, the memory cost is too high because all of the $|\xi_n\rangle$ expanded in plane waves must be stored. Instead, we compute $N_e$ through the formula
  \begin{equation}\label{eq:ne}
  \begin{aligned}
      N_e&=2\sum_{\mathbf{k}}W(\mathbf{k})\sum_{\mathbf{G}}\langle\mathbf{k+G}|f_{\hat{H}}|\mathbf{k+G}\rangle\\
         &=\sum_{n=0}^{N_c}{C_n[f]\left(2\sum_{\mathbf{k}}W(\mathbf{k})\sum_{\mathbf{G}}{\langle\mathbf{k+G}|T_n(\hat{h})|\mathbf{k+G}\rangle}\right)}\\
         &\equiv \sum_{n=0}^{N_c}{C_n[f]P_n}.
  \end{aligned}
  \end{equation}
  We emphasize that since $\{P_n\}$ is independent of $f$ and $\mu$, so $\{P_n\}$ can be computed once and stored. In this regard, the coefficients $\{C_n[f]\}$ with respect to different $\mu$ can be efficiently computed. It is worth mentioning that any quantity $O$ taking the form of
  \begin{equation}
      O={\rm Tr}[g(\hat{H})]=\sum_{n=0}^{N_c}{C_n[g]P_n}\label{trace}
  \end{equation}
  can be calculated with $\{P_n\}$ directly.}
  
  \MC{
  The total free energy of the system is given by
  \begin{equation}\label{eq:ctfe} 
  \begin{aligned}
    &A=2{\rm Tr}[\hat{f}_{H}\hat{H}]-TS+E_{xc}[\rho(\mathbf{r})]\\
    &-\int{v_{xc}(\mathbf{r})\rho(\mathbf{r})d\mathbf{r}}-\frac{1}{2}\int\int{\frac{\rho(\mathbf{r})\rho(\mathbf{r'})}{|\mathbf{r}-\mathbf{r'}|}d\mathbf{r}d\mathbf{r'}}+E_{II},
  \end{aligned}
  \end{equation}
  where
  \begin{equation}\label{eq:ctentrophy}
  \begin{aligned}
    S=-2{\rm Tr}\Big[\hat{f}_{H}\mathrm{ln}\hat{f}_{H}+(1-\hat{f}_{H})\mathrm{ln}(1-\hat{f}_{H})\Big].
  \end{aligned}
  \end{equation}
  }
  
  \MCC{
  Forces and stress can be evaluated by the CT method. On the one hand, the forces acting on atoms can be divided into three main parts (See Eq.~(\ref{eq:force})). Among them, only the formula to compute the non-local pseudopotential force differs from that in the DG method.} In the CT method, the non-local force of atom $I$ of species $\tau$ ($\mathbf{F}_{I,\tau}^{\mathrm{NL}}$) takes the form of
  \begin{equation}
  \begin{aligned}
    \mathbf{F}_{I,\tau}^\mathrm{NL}=&2{\rm Tr}\Big[-i\hat{f}_{H}\sum_{\mathbf{G,G'}}(\mathbf{G'-G})
                   e^{i(\mathbf{G'-G})\cdot\mathbf{R}_I}\\
                   &\times v_\tau^\mathrm{NL}(\mathbf{k+G,k+G'})|\mathbf{k+G}\rangle\langle\mathbf{k+G'}|\Big],
  \end{aligned}
  \label{eq:fnl}
  \end{equation}
  where $\mathbf{R}_I$ and $v_\tau^\mathrm{NL}$ are the atomic coordinate of $I$-th atom and non-local pseudopotential of atom species $\tau$, respectively.
  On the other hand, the stress contains five terms (See Eq.~(\ref{eq:stress})) and three of them can be evaluated by the same way as \MCC{in the DG method}. The other two terms are the kinetic energy term $\sigma_{\alpha\beta}^{T}$ and the non-local potential term $\sigma_{\alpha\beta}^{NL}$.
  The stress associated with the kinetic energy is
  \begin{equation}
  \begin{aligned}
    \sigma_{\alpha\beta}^{T}=&\frac{2}{V}{\rm Tr}\Big[\hat{f}_{H}\sum_{\mathbf{G,G'}}(\mathbf{k+G})_\alpha
                         (\mathbf{k+G'})_\beta\\
                         &\times |\mathbf{k+G}\rangle\delta(\mathbf{G,G'})\langle\mathbf{k+G'}|\Big]
  \end{aligned}
  \label{eq:stress_t}
  \end{equation}
  while the stress of the non-local pseudopotential term is
  \begin{equation}
  \begin{aligned}
    \sigma_{\alpha\beta}^\mathrm{NL}=&-\frac{2}{V}{\rm Tr}\Big[\hat{f}_{H}\sum_{\mathbf{G,G',\tau}}
                         S_\tau(\mathbf{G'-G})\\
                         &\times\frac{\partial v_\tau^\mathrm{NL}(\mathbf{G+k,G'+k})}{\partial \epsilon_{\alpha\beta}}|\mathbf{k+G}\rangle\langle\mathbf{k+G'}|\Big],
  \end{aligned}
   \label{eq:stress_nl}
  \end{equation}
  where $\epsilon_{\alpha\beta}$ is the strain tensor, $V$ is the volume of the system and $S_\tau(\mathbf{G})$ is the structure factor of atom species $\tau$.
  
 \MC{
 We test four systems to verify the equivalence of the CT and \MCC{DG} methods, the detailed setup and the results including free energy, stress, and forces are listed in Table~\ref{tab:free-energy} and \ref{tab:force}.
 \MCC{The temperature is set to be 5 or 10 eV. At such a high temperature, we choose a large number of KS orbitals to ensure the occupation number of the last orbital to be smaller than $10^{-8}$.}
 We can see both methods yield quantitatively the same results. Notably, the CT method can be accurately applied to extended systems with multiple $k$-points. For example, for the disordered Si with 8 atoms and a $k$-point sampling of $4\times4\times4$, both CT and \MCC{DG} methods yield the similar free energy of -1316.13273028 and -1316.13273909 eV, the difference is smaller than 1e-5 eV.}
  
  \begin{table*}[htbp]
  	\centering
  	\caption{\MC{Free energy $A$ and stress $\sigma$ of four systems as computed from the DG and CT methods. 
  	  The four systems are listed with parameters including the $k$-point sampling ($\Gamma$ or $4\times4\times4$), the temperature $T$ (in eV), and the density $\rho$ (in $\rm{g/cm}^3$).
  	  $N$, $N_\phi$ and $N_{pw}$ depict 
  	  the number of atoms, the number of KS orbitals used in the DG method and the number of plane waves employed in the CT method, respectively.}
  	}
  	\label{tab:free-energy}
  	\begin{tabular}{cccccccccccccccc}
  	\toprule
  	\hline
  	\specialrule{0.05em}{1pt}{3pt}
  	&$N$ & \tabincell{c}{$A^{DG}$\\(eV)} &\ && \tabincell{c}{$\sigma^{DG}$\\(kBar)} &&\ &$N_\phi$
  	     & \tabincell{c}{$A^{CT}$\\(eV)} &\ && \tabincell{c}{$\sigma^{CT}$\\(kBar)} &&\ &$N_{pw}$\\
  	\midrule
  	\specialrule{0em}{3pt}{0pt}
  	\hline
  	\specialrule{0em}{0pt}{3pt}
  	\vspace{3mm}
  	\tabincell{c}{Diamond C\\$\Gamma$ point\\$T$=10, $\rho$=3.51 }&8
  	&-1490.92197492&&\tabincell{c}{5700.925\\0\\0}&\tabincell{c}{0\\5700.925\\0}&\tabincell{c}{0\\0\\5700.925}&&500
  	&-1490.92197491&&\tabincell{c}{5700.926\\0\\0}&\tabincell{c}{0\\5700.926\\0}&\tabincell{c}{0\\0\\5700.926}&&3743\\
  	\vspace{3mm}
  	\tabincell{c}{Disordered C\\  $\Gamma$ point\\$T$=5, $\rho$=5.00 }&16
  	&-2527.69075155&&\tabincell{c}{5218.598\\-77.079\\45.122}&\tabincell{c}{-77.079\\5404.248\\97.585}&\tabincell{c}{45.122\\97.585\\5473.140}&&400
  	&-2527.69075156&&\tabincell{c}{5218.594\\-77.080\\45.120}&\tabincell{c}{-77.080\\5404.242\\97.585}&\tabincell{c}{45.120\\97.585\\5473.142}&&5185\\
  	\vspace{3mm}
  	\tabincell{c}{Diamond Si\\  $\Gamma$ point\\$T$=10, $\rho$=2.33 }&8
  	&-1339.74099262&&\tabincell{c}{1863.768\\0\\0}&\tabincell{c}{0\\1863.768\\0}&\tabincell{c}{0\\0\\1863.768}&&1000
  	&-1339.74100200&&\tabincell{c}{1863.768\\0\\0}&\tabincell{c}{0\\1863.768\\0}&\tabincell{c}{0\\0\\1863.768}&&6451\\
  	\tabincell{c}{Disordered Si\\ $4\times4\times4$ $k$-points\\$T$=10, $\rho$=2.33}&8
  	&-1316.13273028&&\tabincell{c}{1882.256\\202.845\\36.518}&\tabincell{c}{202.845\\1959.906\\-74.171}&\tabincell{c}{36.518\\-74.171\\2426.388}&&1000
  	&-1316.13273909&&\tabincell{c}{1882.256\\202.845\\36.518}&\tabincell{c}{202.845\\1959.908\\-74.170}&\tabincell{c}{36.518\\-74.170\\2426.390}&&6488\\
  	\bottomrule
  	\specialrule{0.05em}{3pt}{1pt}
  	\hline
  \end{tabular}
  \end{table*}
%   -1490.921974915110   -1490.921974912696
%   -2527.690751554755  -2527.690751560491
%   -1339.740992622172  -1339.741001995337
%   -1316.132730281594  -1316.132739094026

  \begin{table}[htbp]
  	\centering
  	\caption{Comparison of atomic forces (in eV/\AA) in two systems as obtained from the DG ($F^{DG}_x$, $F^{DG}_y$ and $F^{DG}_z$) and CT ($F^{CT}_x$, $F^{CT}_y$ and $F^{CT}_z$ \MCC{as compared to the DG data}) methods. The two systems are a disordered C system (16 atoms) and a disordered Si system (8 atoms), the conditions of which are listed in Table~\ref{tab:free-energy}. %One is 5.00 $\rm g/cm^3$, 5 eV disordered C calculated with the $\Gamma$ $k$-point. The other is 2.33 $\rm g/cm^3$, 10 eV disordered Si calculated with $4\times4\times4$ $k$-points.
  	  In DG calculations, 400 and 1000 KS orbitals are used for C and Si systems, respectively. In the CT method, 5185 and 6488 plane waves for C and Si are utilized, respectively.
  	}
  	\label{tab:force}
  	\begin{tabular}{cccccccc}
  		\toprule
  		\hline
  		\specialrule{0.05em}{1pt}{3pt}
        && $F^{DG}_x$& $F^{DG}_y$&$F^{DG}_z$&\MCC{$F^{CT}_x$}&\MCC{$F^{CT}_y$}&\MCC{$F^{CT}_z$}\\
  		\midrule
  		\specialrule{0em}{3pt}{0pt}
  	    \hline
  	    \specialrule{0em}{0pt}{3pt}
  	    ~ & 1 & 3.4434 & -0.9882 & 0.454 & -0.0005 & -0.0002 & -0.0006 \\
        ~ & 2 & -1.9595 & 5.7627 & 3.1996 & -0.0006 & 0.0003 & 0.0002  \\
        ~ & 3 & 7.0587 & -1.2141 & -3.5839 & -0.0003 & 0.0004 & 0.0001 \\
        ~ & 4 & -0.2327 & 0.0505 & 3.0127 & 0.0004 & 0.0003 & -0.0006 \\
        ~ & 5 & 4.142 & 5.4697 & -0.4322 & 0.0002 & -0.0003 & 0.0001 \\
        ~ & 6 & 10.1551 & -0.6472 & -0.5713 & -0.0001 & 0.0000 & -0.0005 \\
        ~ & 7 & 3.6263 & 2.5401 & -3.3339 & 0.0004 & -0.0003 & -0.0004 \\
        C & 8 & -6.5981 & -4.8153 & -2.3112 & -0.0002 & 0.0002 & 0.0000 \\
        ~ & 9 & -6.4496 & 5.6967 & 1.2765 & 0.0003 & -0.0003 & 0.0005 \\
        ~ & 10 & -3.6473 & -2.8595 & 4.5359 & 0.0004 & 0.0001 & -0.0002 \\
        ~ & 11 & 3.3913 & 2.0238 & -5.4687 & -0.0004 & 0.0000 & 0.0005 \\
        ~ & 12 & -1.0327 & -1.5951 & 2.5317 & 0.0003 & 0.0001 & 0.0004 \\
        ~ & 13 & -5.7166 & 1.3521 & -0.2795 & 0.0001 & -0.0003 & -0.0005 \\
        ~ & 14 & -6.0667 & 0.1237 & -2.1423 & 0.0000 & 0.0002 & -0.0003 \\
        ~ & 15 & 1.1041 & -8.5614 & -0.455 & 0.0004 & 0.0002 & 0.0006 \\
        ~ & 16 & -1.2176 & -2.3386 & 3.5676 & -0.0005 & -0.0004 & 0.0005 \\
        ~ & 1 & 5.9738 & 11.2804 & -18.4636 & 0.0001 & 0.0002 & -0.0001 \\
        ~ & 2 & 27.1567 & 23.4301 & -16.0622 & 0.0000 & 0.0001 & 0.0000 \\
        ~ & 3 & -2.5108 & -4.0214 & -0.5398 & 0.0002 & -0.0001 & 0.0000 \\
        ~ & 4 & -18.6459 & -8.4193 & -14.3648 & 0.0001 & -0.0001 & 0.0000 \\
        Si & 5 & -8.492 & -11.0528 & 13.9717 & 0.0000 & -0.0001 & 0.0001 \\
        ~ & 6 & -1.4202 & 0.3509 & 5.5613 & -0.0001 & 0.0000 & 0.0001 \\
        ~ & 7 & 2.3269 & 2.5932 & -3.6515 & -0.0001 & 0.0000 & 0.0000 \\
        ~ & 8 & -4.3885 & -14.1611 & 33.5488 & -0.0001 & 0.0000 & 0.0000 \\
  		\bottomrule
  		\specialrule{0.05em}{3pt}{1pt}
  		\hline
  	\end{tabular}
  \end{table}

  % Method C:
\subsection{Stochastic Density Functional Theory}
Although the CT method with a complete plane-wave basis set is as accurate as the DG method, the number of plane waves used in CT is much larger than the KS orbitals used in DG (See Tables ~\ref{tab:free-energy}) and results in an inefficient CT method for practical usage.
However, SDFT~\cite{13L-Baer,18B-Cytter,19JCP-Ming} adopts quasi-complete stochastic orbitals,~\cite{89CSSC-Hutchinson,94B-Wang,22ARPC-Baer} the number of which is much smaller than that of a complete basis set. In this regard, the SDFT has a potential to substantially speedup the CT method.
Recently, the stochastic orbitals have been successfully used in the stochastic GW method,~\cite{14L-Neuhauser,17JCTC-Vlcek,18B-Vlcek} the stochastic Kubo-Greenwood method,~\cite{19B-Cytter} calculation of the multiexciton generation rates,~\cite{12NL-Baer} the second order M{\o}ller-Plesset perturbation theory~\cite{14JPCL-Ge} and the stochastic optimally tuned range-separated hybrid density functional theory.~\cite{16JPCA-Neuhauser}
\MC{
For any orthogonal and complete basis set $\{\varphi_j\}$ (see Fabian {\it et al.}~\cite{19WCMS-Fabian} for non-orthogonal basis), the stochastic orbitals $\{\chi_n\}$ in the $\varphi$ representation can be defined as~\cite{13L-Baer,18B-Cytter,20L-White,19B-Cytter}
\begin{equation}
    \langle\varphi_j|\chi_n\rangle=\frac{1}{\sqrt{N_\chi}}\exp{(i2\pi\theta_j^n)},
\end{equation}
where $\theta_j^n$ is randomly generated by the uniform distribution between 0 and 1. $N_\chi$ is the number of stochastic orbitals. 
Other works~\cite{89CSSC-Hutchinson,19JCP-Ming,19JCP-Ming2} also use the definition of
$\langle\varphi_j|\chi_n\rangle=\pm\frac{1}{\sqrt{N_\chi}}$ with the probability of plus and minus signs being 1/2.}
Notably, Baer {\it et al.}~\cite{22ARPC-Baer} have proved that both definitions lead to the same expectation and similar variance values of a given Hermitian matrix with comparable real and imaginary parts.
\MC{
Ideally,
infinite stochastic orbitals form a complete basis set, which satisfies~\cite{89CSSC-Hutchinson, 13L-Baer}
\begin{equation}
  \lim_{N_\chi\to+\infty}{\sum_{n=1}^{N_\chi}|\chi_n\rangle\langle\chi_n|}=\hat{I}
\end{equation}
and
\begin{equation}
\begin{aligned}
  {\rm Tr}(\hat{O})&=\lim_{N_\chi\to+\infty}{{\rm Tr}\left(\sum_{n=1}^{N_\chi}|\chi_n\rangle\langle\chi_n|\hat{O}\right)}\\
  &=\lim_{N_\chi\to+\infty}\sum_{n=1}^{N_\chi}\langle\chi_n|\hat{O}|\chi_n\rangle
\end{aligned}
\label{eq:traceo_sdft}
\end{equation}
for any linear operator $\hat{O}$.
In this regard, sufficient stochastic orbitals are regarded as an approximated identity operator, so traces such as Eqs.~(\ref{eq:rhoctpw}), (\ref{eq:ne}), (\ref{eq:ctfe}), and (\ref{eq:ctentrophy}) can be approximately evaluated with stochastic orbitals and the CT method.
In this work, we choose $\{\varphi_j\}$ as plane-wave basis so the stochastic orbitals can be defined as
\begin{equation}
    \langle \mathbf{k+G}|\chi_n\rangle=\frac{1}{\sqrt{N_\chi}}\exp{(i2\pi\theta_\mathbf{k,G}^n)},
\end{equation}
where $\theta_\mathbf{k,G}^n$ is an independent random number between 0 and 1.
The self-consistent loop of SDFT (See Fig.~\ref{fig:workflow}(b)) is summarized as the following six steps:
(i) Guess an initial electron density $\rho_0(\mathbf{r})$ and generate stochastic orbitals $\{\chi_n\}$.
(ii) Construct the effective potential $v_\mathrm{eff}(\mathbf{r})$ and the Hamiltonian $\hat{H}$ with Eq.~(\ref{veff}).
(iii) Determine the chemical potential $\mu$ according to evaluating the correct electron number  $N_e=2\sum_{n=1}^{N_\chi}{\langle\chi_n|\hat{f}_{H}|\chi_n\rangle}$. 
(iv) Evaluate a new electron density
$\rho(\mathbf{r})=2\sum_{n=1}^{N_\chi}{\left|\langle\mathbf{r}|\hat{f}_{H}^{1/2}|\chi_n\rangle\right|^2}$.
(v) Mix the new electron density $\rho(\mathbf{r})$ with previous ones.
(vi) Reiterate (ii-v) until the difference between the $\rho(\mathbf{r})$ and $\rho_0(\mathbf{r})$ is under a given threshold and then the ground-state electron density is $\rho(\mathbf{r})$.}

\MCC{
The forces and stress formulas from the SDFT method share similar forms with those used in the CT method. However, one should pay attention to the trace operations that appear in Eqs.~\ref{eq:fnl},~\ref{eq:stress_t} and~\ref{eq:stress_nl}, which should be done with Eq.~\ref{eq:traceo_sdft} using stochastic orbitals.
}

\subsection{Mixed Stochastic-Deterministic DFT}
\MC{
The MDFT method adopts both deterministic orbitals $\{\phi_i\}$ and stochastic orbitals $\{\tilde{\chi_n}\}$,~\cite{20L-White} where $\{\phi_i\}$ is \MCC{an} orthogonal but incomplete basis. $\tilde{\chi_n}$ is obtained by orthogonalizing the initial stochastic orbital $\chi_{n}$ to the set of $\{\phi_i\}$ as
\begin{equation}\label{eq:ortho}
    |\tilde{\chi_n}\rangle=|\chi_n\rangle-\sum_{i=1}^{N_\phi}\langle\phi_i|\chi_n\rangle|\phi_i\rangle.
\end{equation}
The two sets of basis \MCC{functions} satisfy
\begin{equation}
  \lim_{N_\chi\to+\infty}{\sum_{n=1}^{N_\chi}|\tilde{\chi}_n\rangle\langle\tilde{\chi}_n|}+\sum_{i=1}^{N_\phi}|\phi_i\rangle\langle\phi_i|=\hat{I},
\end{equation}
where $N_\chi$ and $N_\phi$ are the number of stochastic and deterministic orbitals, respectively.
Based on the mixed orbitals, the trace of a given operator $\hat{O}$ can be evaluated by
\begin{equation}
    {\rm Tr}(\hat{O})=\lim_{N_\chi\to+\infty}{\sum_{n=1}^{N_\chi}\langle\tilde{\chi}_n|\hat{O}|\tilde{\chi}_n\rangle}+\sum_{i=1}^{N_\phi}\langle\phi_i|\hat{O}|\phi_i\rangle.
    \label{eq:traceo_mdft}
\end{equation}
In the MDFT method,
\MCC{if the operator $\hat{O}$ is a function of Hamiltonian $\hat{O}=g(\hat{H})$ and part of the eigenstates $\{\psi_i\}$ of $\hat{H}$ are chosen as deterministic orbitals, the above equation can be simplified as}
\begin{equation}\label{eq:mdft}
    {\rm Tr}(\hat{O})=\lim_{N_\chi\to+\infty}{\sum_{n=1}^{N_\chi}\langle\tilde{\chi}_n|\hat{O}|\tilde{\chi}_n\rangle}+\sum_{i=1}^{N_\phi}g(\epsilon_i).
\end{equation}
%Eq.~(\ref{eq:mdft}) demonstrates both SDFT and KSDFT are special cases of MDFT. KSDFT or SDFT is the case when $N_\chi$ or $N_\phi$ equals 0. 
%In this work, even though they are special cases and can be uniformly implemented in one MDFT program, we distinguish MDFT, SDFT and KSDFT in the statement.
The self-consistent loop of MDFT (See Fig.~\ref{fig:workflow}(c)) includes the following eight steps:
(i) Guess an initial electron density $\rho_0(\mathbf{r})$ and generate a set of stochastic orbitals $\{\chi_n\}$.
(ii) Construct the effective potential $v_{eff}(\mathbf{r})$ and the Hamiltonian $\hat{H}$ with Eq.~(\ref{veff}).
(iii) Solve a selected set of eigenstats $\{\psi_i\}$ of $\hat{H}$.
(iv) Orthogonalize $\{\chi_n\}$ to $\{\psi_i\}$ via Eq.~(\ref{eq:ortho}) and obtain a new set of stochastic orbitals $\{\tilde{\chi}_n\}$.
(v) Determine the chemical potential $\mu$ according to evaluating the correct electron number with the formula of $N_e=2\sum_{n=1}^{N_\chi}{\langle\tilde{\chi}_n|\hat{f}_{H}|\tilde{\chi}_n\rangle}+2\sum_{i=1}^{N_\phi}f(\epsilon_i;\mu)$. 
(vi) Evaluate a new electron density
$\rho(\mathbf{r})=2\sum_{n=1}^{N_\chi}{\left|\langle\mathbf{r}|\hat{f}_{H}^{1/2}|\tilde{\chi}_n\rangle\right|^2}+2\sum_{i=1}^{N_\phi}{f(\epsilon_i;\mu)|\psi_i(\mathbf{r})|^2}$.
(vii) Mix the new electron density $\rho(\mathbf{r})$ with previous ones.
(viii) Reiterate (ii-vii) until the difference between the $\rho(\mathbf{r})$ and $\rho_0(\mathbf{r})$ is under a specific threshold and then the ground-state density $\rho(\mathbf{r})$ is obtained.}

\MCC{
In order to compute the forces and stress formulas from the MDFT method, one should use the formulas introduced in the CT method. Furthermore, Eq.~\ref{eq:traceo_mdft} with both stochastic orbitals and deterministic orbitals should be adopted to evaluate the trace operator.
}

\section{Results and Discussion}
 We systematically compare several aspects of the DG, CT, SDFT and MDFT methods and the results are shown below. Specifically, we test physical properties \MCC{including} the electron density, the free energy, the pressure, the atomic forces and the density of states. In particular, we analyze the statistical errors introduced in the SDFT and MDFT methods, \MCC{and the related formulas are described in Appendix C.} We also evaluate the efficiency of the SDFT and MDFT methods by parallel calculations of systems containing 512 atoms. We conclude that the use of stochastic orbitals can be more efficient than the DG method at extremely high temperatures (above a few tens of eV).
 Several systems consisting of Si and C atoms are chosen. %Specifically, we simulate Si at densities of 2.33, 5.0 and 10.0 ${\rm g/cm^3}$ and the temperature of 10 eV. 
 %We use 64, 96, 128, 256, 512, 1024 %stochastic orbitals in SDFT %calculations.
 Both Diamond and disordered structures are selected with 8 or 64 atoms.
 \MCC{Our results are demonstrated with different $k$-point sampling schemes.}
 We adopt the $\Gamma$ $k$-point in all cases except $2\times2\times2$ $k$-points when simulating Si of 8 atoms and $4\times4\times4$ $k$-points when calculating the density of states.
 %We simulate an 8-atom C system with the diamond structure at temperatures of 10, 50, 100 eV with the density of 3.51 ${\rm g/cm^3}$.

 \subsection{Electron Density}
 
   \begin{figure}
	\centering
	\includegraphics[width=8.6cm]{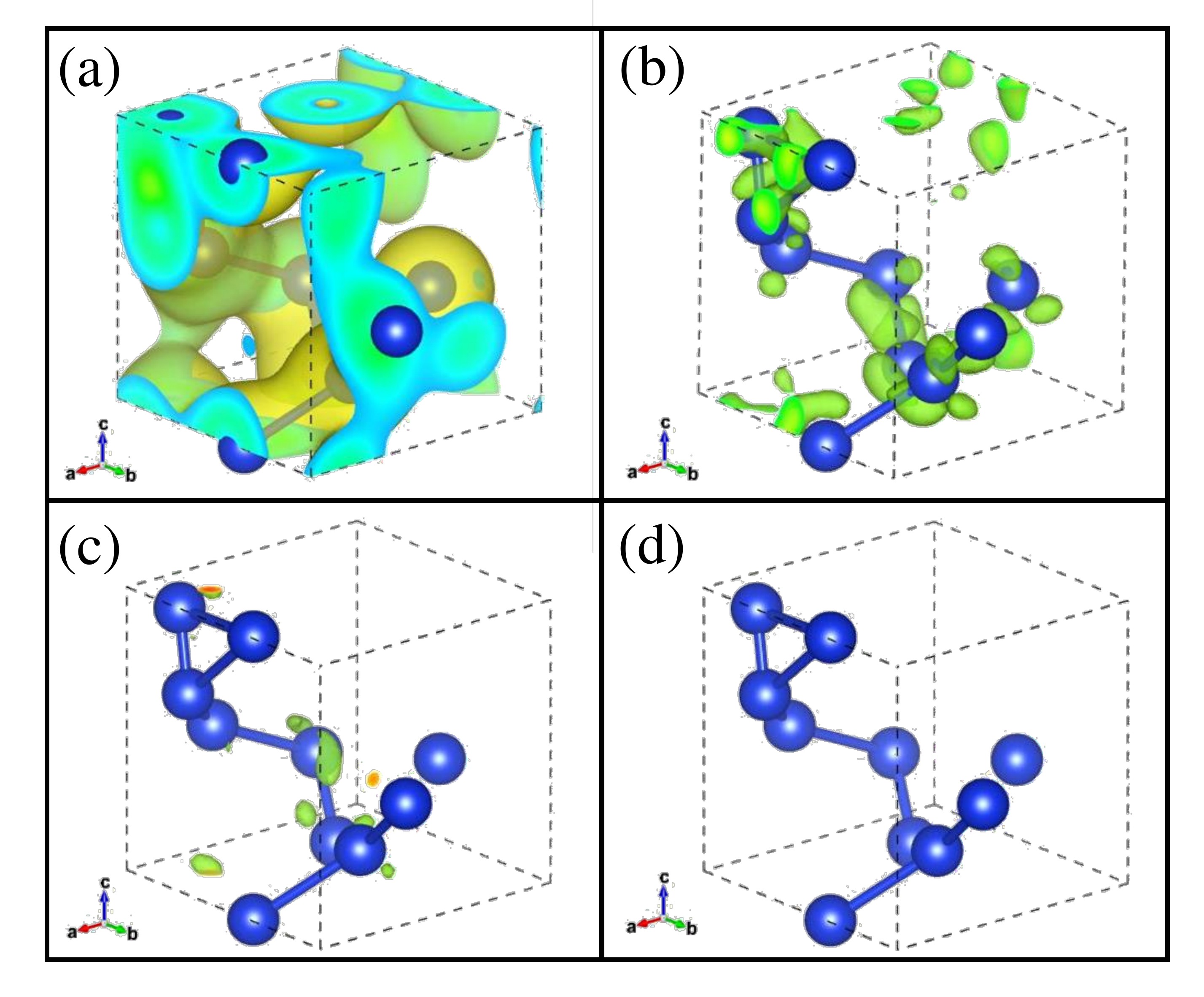}\\
	\caption{\MCC{(Color online) Electron density profile of disordered Si (8 atoms, depicted in blue) as obtained from different methods, including (a) the DG method with the number of deterministic orbitals being $N_\phi=1000$, (b) the SDFT method with $N_\chi=64$ (the number of stochastic orbitals), (c) the SDFT method with $N_\chi=256$, (d) the MDFT method with $N_\chi=256$ and $N_\phi=32$ (the number of deterministic orbitals). The electron density in (a) serves as reference, while (b-d) illustrate the electron density differences between the one from the stochastic methods and the one from the DG method. The density of Si is 2.33 ${\rm g/cm^3}$ and the temperature is 10 eV.}
	}\label{fig:rho}
\end{figure}

 The ground-state electron density is a key variable that can be obtained from DG, SDFT, and MDFT.
 We test a disordered Si systems consisting of 8 atoms. The density is 2.33 ${\rm g/cm^3}$ and the temperature is 10 eV. The energy cutoff is 50 Ry.
 %directly predicted by DFT. It decides the 
 %Hamiltonian $\hat{H}$ and then decides all %properties of the electron system.
 Fig.~\ref{fig:rho} compares the results from DG, SDFT, and MDFT. 
 \MCC{The benchmark electron density obtained from DG is shown in Fig.~\ref{fig:rho}(a), which is obtained via the DG method with 1000 deterministic orbitals.
 We observe valence electrons that form chemical bonds among the Si atoms.
 Figs.~\ref{fig:rho}(b) and (c) represent the electron density differences between the one from the SDFT method and the one from the DG method; the number of stochastic orbitals used in Figs.~\ref{fig:rho}(b) and (c) is $N_{\chi}$=64 and 256, respectively.
 Based on the $N_{\chi}$=256 adopted in Fig.~\ref{fig:rho}(c), additional 32 deterministic orbitals are used in Fig.~\ref{fig:rho}(d). 
 Interestingly, the usage of 64 stochastic orbitals in Fig.~\ref{fig:rho}(b) roughly captures the electron density, although differences are randomly scattered due to the statistical errors caused by the stochastic orbitals.
 We notice that the electron density differences are largely eliminated when using 256 stochastic orbitals in Fig.~\ref{fig:rho}(c) and there are no obvious differences when additional 32 deterministic orbitals are used in Fig.~\ref{fig:rho}(d). In fact, Figs.~\ref{fig:rho}(a) , (c) and (d) yield close electron densities, demonstrating that both SDFT and MDFT with a sufficiently large number of stochastic orbitals can yield the same electron density as the DG method, \MCC{which is expected.}}

 \subsection{Free energy, Pressure and Forces}
\MC{
We analyze the statistical errors of the SDFT and MDFT methods, \MCC{the formulas of which can be found in the Appendix C.}
We prepare $N_s$=20 replicas with different random seeds and define the errors of free energy $\Delta A$, pressure $\Delta P$, and force $\Delta F$ as
 \begin{equation}
     \Delta A = \sqrt{\frac{\sum_{i=1}^{N_s}{|A^i-A^\mathrm{ref}|^2}}{N_s}},
 \end{equation}
 \begin{equation}
     \Delta P = \sqrt{\frac{\sum_{i=1}^{N_s}{|P^i-P^\mathrm{ref}|^2}}{N_s}},
 \end{equation}
 and
 \begin{equation}
     \Delta F = \sqrt{\frac{\sum_{i=1}^{N_s}\sum_{j=1}^N{|\mathbf{F}_{j}^i-\mathbf{F}_{j}^\mathrm{ref}|^2}}{N_sN}},
 \end{equation}
 respectively.
 Here, $N$ is the number of atoms, $i$ is the index of replicas, the ``ref" index represents the reference results obtained from \MCC{the DG} and CT methods,
 which are listed in Table.~\ref{tab:ref} \MCC{in Appendix D}.
 $\mathbf{F}_j^{i}$ is the force acting on atom $j$ in the replica $i$.}
 
\begin{figure}
	\centering
	\includegraphics[width=8.6cm]{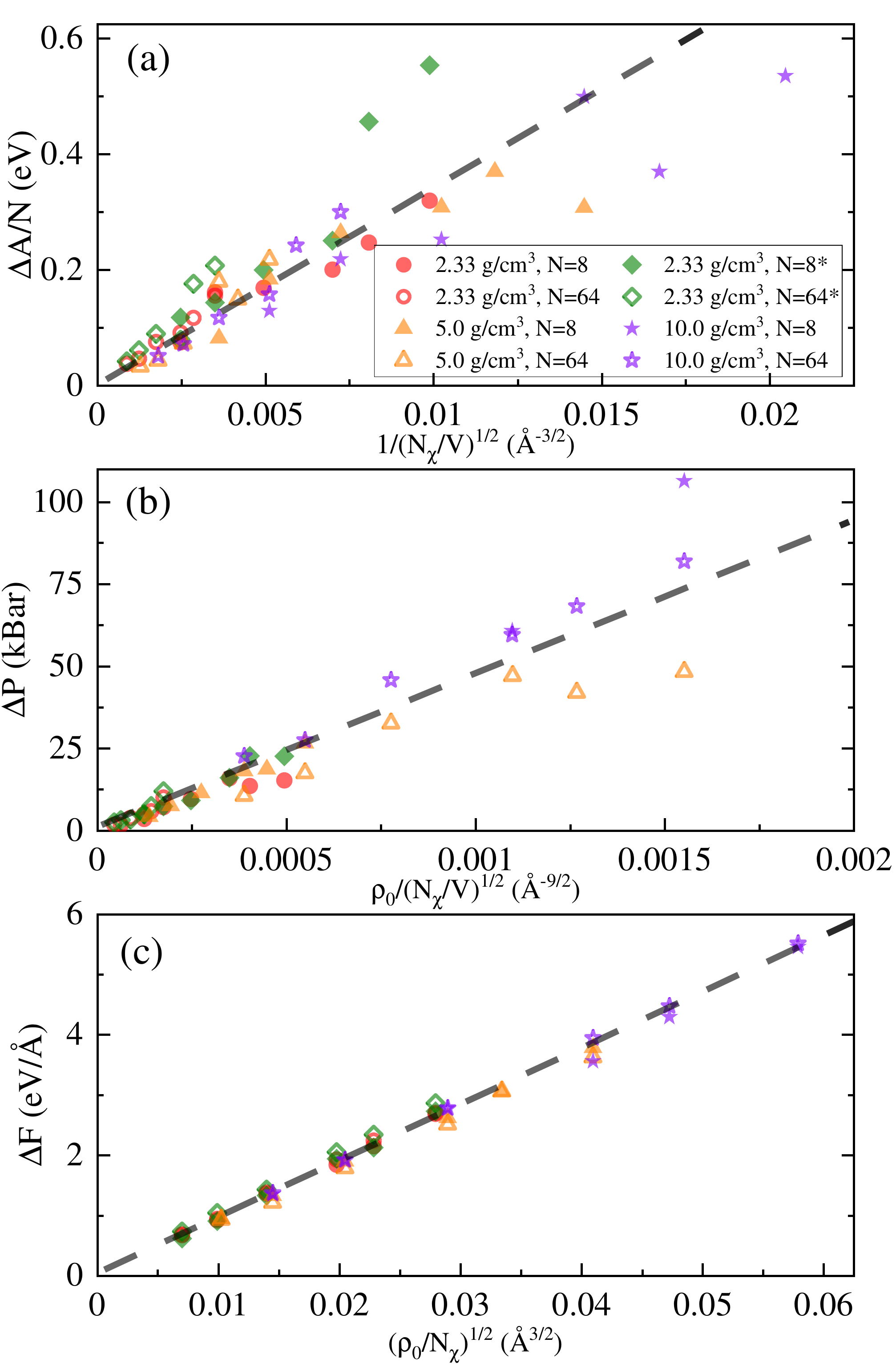}\\
	\caption{\MC{(Color online) Statistical errors of the (a) average atomic free energy $\Delta A$, (b) pressure $\Delta P$ and (c) forces $\Delta F$ acting on atoms with respect to the number of stochastic orbitals $N_\chi$. We use $N_\chi$ of 64, 96, 128, 256, 512, 1024. Eight different Si systems are tested with $N$ and $V$ being the number of atoms and the volume of the system, respectively. All of the systems are simulated at the temperature of 10 eV.}
	}\label{fig:std}
\end{figure}

% \subsubsection{Statistical Errors of SDFT}
 \MCC{
 To analyze the statistical errors from the SDFT method, we perform several tests for Si at a temperature of 10 eV. Three densities (2.33, 5.0, 10.0 g/cm$^3$) are chosen with the diamond structure, while a disordered structure is adopted with a density of 2.33 g/cm$^3$.
 In each case, we use two sizes of system (8 and 64 atoms). The energy cutoff is set to 50 Ry for the plane-wave basis.
 Fig.~\ref{fig:std} shows the statistical errors of SDFT and we observe the following features.
 First, although all of the data points roughly stay in a straight line, $\Delta F$ exhibits a better linear behavior than $\Delta A$ and $\Delta P$ because more samples are included.
 In fact, each point of $\Delta F$ is averaged from $N_s\times N$ points but only $N_s$ points are used to evaluate $\Delta A$ and $\Delta P$.
 Second, the statistical errors are reduced when more stochastic orbitals are included, as we find the three quantities rise inversely with $N_\chi^{1/2}$. This feature is consistent with the above conclusion that more stochastic orbitals lead to a more complete basis set.
 Third, for a given density $\rho_0$ and a fixed $N_\chi$, the $\Delta A$ per atom and $\Delta P$ are proportional to $V^{-1/2}$, suggesting larger systems exhibit smaller errors.
 Impressively, to keep the same precision for large systems, SDFT do not need to include more orbitals.
 Therefore, SDFT is well suitable for studying large systems.}
 %
 %Fourth, the atomic forces remain unaltered under a specific $N_\chi$. 
 %Interestingly, one distinct property of SDFT is that we do not need to include more orbitals for larger systems. Therefore, SDFT is of more advantages when treating large systems.

 %\subsubsection{Statistical Errors of MDFT}

\begin{figure}
	\centering
	\includegraphics[width=8cm]{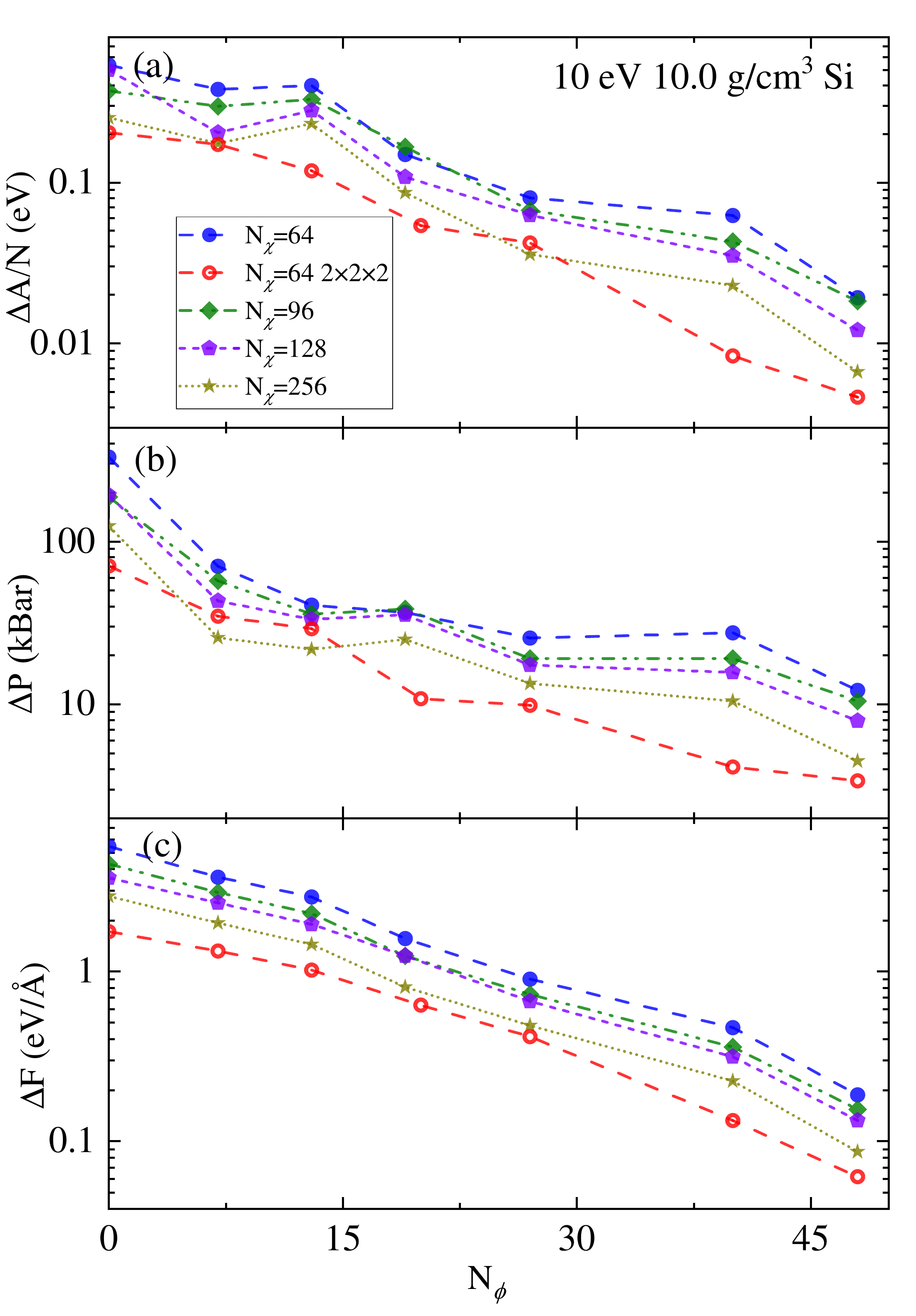}\\
	\caption{\MC{(Color online) Statistical errors of the average atomic free energy $\Delta A$, pressure $\Delta P$ and forces acting on atoms $\Delta F$ of 8-atom diamond Si at the temperature of 10 eV and the density of 10.0 ${\rm g/cm^3}$. We use 7, 13, 19, 27, 40, and 48 deterministic orbitals for each set of stochastic orbitals $N_{\chi}$ (64, 96, 128, and 256).}
	}\label{fig:mix}
\end{figure}

\begin{figure}
	\centering
	\includegraphics[width=8.8cm]{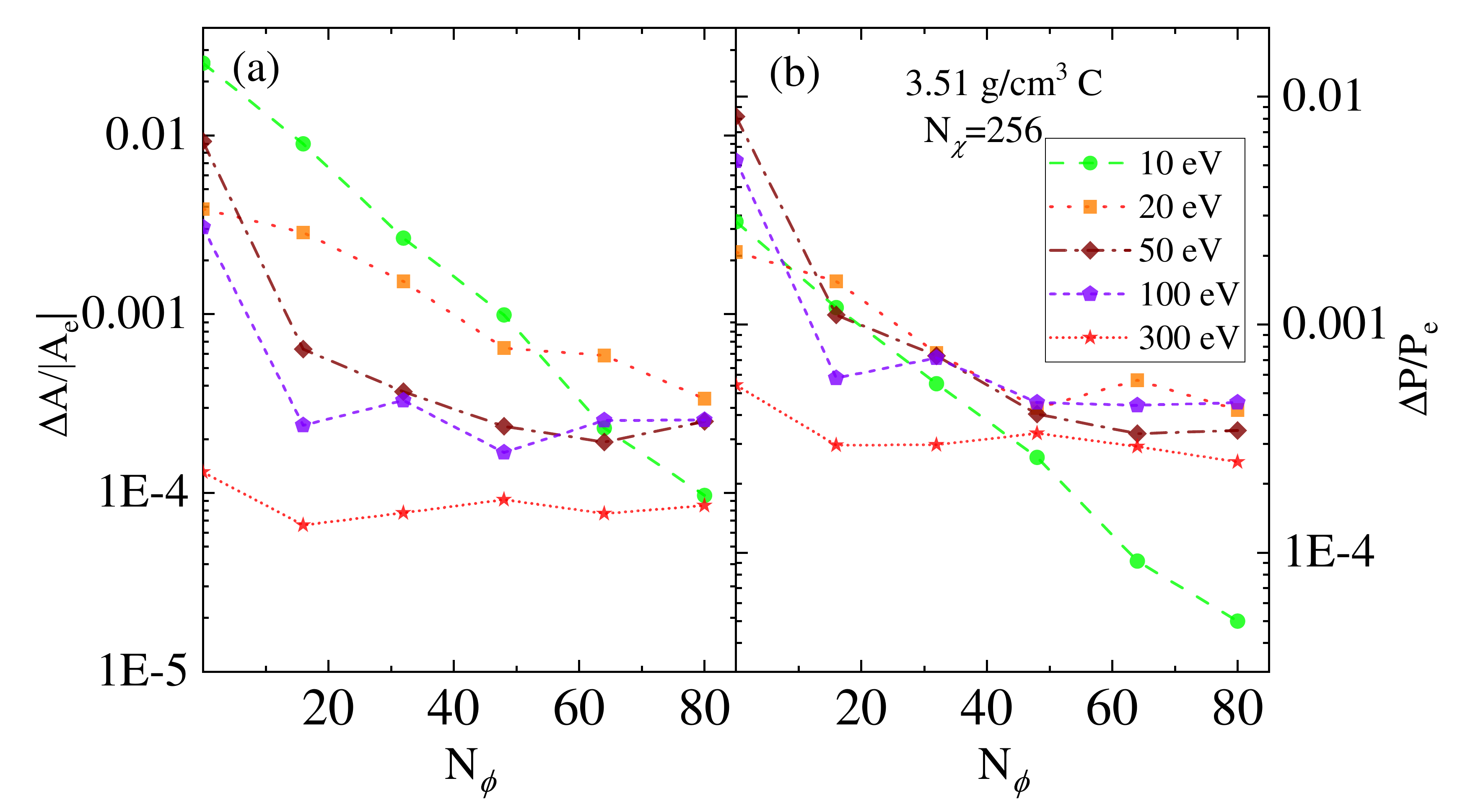}\\
	\caption{\MC{(Color online) Relative statistical errors of the average atomic free energy and the pressure for C system consists of 8 atoms in a diamond structure. The temperature is set to 10, 50, 100 and 300 eV while the density is 3.51 ${\rm g/cm^3}$. 
	We use 256 stochastic orbitals and additional 0-80 deterministic orbitals. 
	$\Delta A$ and $\Delta P$ are the statistical errors of the free energy and the pressure.}
	Since the statistical errors only result from the electronic part, we choose the electronic parts of the free energy ($A_e$) and the pressure $P_e$ as denominators, which exclude ionic contributions from the reference free energy $A$ and pressure $P$ listed in Table~\ref{tab:ref}. The
	$A_e$ values are -101.1, -729.2, -8372.6, -15787.9 , -64128.9 eV for systems at temperatures of 10, 20, 50, 100, 300 eV, respectively.
	The	$P_e$ values are 21988, 30312, 80577, 148158, 496196 kBar for systems at temperatures of 10, 20, 50, 100, 300 eV, respectively.
	}\label{fig:mixT}
\end{figure}

%MDFT results are plotted in Fig.~\ref{fig:mix} and Fig.~\ref{fig:mixT}. 

\MCC{Besides the stochastic orbitals,
the MDFT method utilizes a set of deterministic orbitals (the Kohn-Sham orbitals). We analyze the statistical errors of MDFT with respect to the used number of deterministic orbitals and the results are illustrated in Fig.~\ref{fig:mix}.
A diamond structure of Si with 8 atoms is tested at 10 eV with a density of 10.0 ${\rm g/cm^3}$.}
We choose several sets of stochastic orbitals, i.e., 64, 96, 128, 256.
We then utilize 7, 13, 18, 27, 40, and 48 deterministic orbitals for each set of stochastic orbitals and evaluate the errors of the free energy per atom ($\Delta A/N$), the pressure $\Delta P$, and the force $\Delta F$.
The above tests demonstrate that increasing the deterministic orbitals in MDFT can systematically decrease the statistical errors of the free energy ($\Delta A$), the pressures ($\Delta P$) and the forces ($\Delta F$).
For example, if no deterministic orbitals but 256 stochastic orbitals are used, the errors are $\Delta A/N$=0.25 eV, $\Delta P$=124.4 kBar, and $\Delta F$=2.78 ${\rm eV/\AA}$, where $N$ is the number of atoms. 
Notably, when 48 deterministic orbitals are added onto the 256 stochastic orbitals, the errors are substantially reduced to $\Delta A/N$=0.007 eV,  $\Delta P$=4.5 kBar and $\Delta F$=0.09 ${\rm eV/\AA}$. 
We conclude that MDFT can achieve almost the same accuracy as the DG method but need much less deterministic orbitals \MCC{(1000 orbitals from DG)}.

\MC{Fig.~\ref{fig:mixT} shows the relative statistical errors of C at different temperatures. We find the errors of the free energy $\Delta A$ and the pressure $\Delta P$ at a temperature of 10 eV decrease quickly with the increase of the number of deterministic orbitals $N_\phi$. However, the relative statistical errors at higher temperatures such as 300 eV are small even without deterministic orbitals, suggesting the SDFT method is sufficiently accurate at high temperatures.} In other words, it is worth mentioning that the error of MDFT depends on $(\Phi-\Psi)$ in Eq.~(\ref{eq:merror}) \MCC{in Appendix C}. As the temperature rises, the number of partially occupied orbitals becomes larger, resulting in a big difference between $\Phi$ and $\Psi$. Hence, a small portion of deterministic orbitals used in MDFT cannot substantially increase the accuracy.

\MC{Increasing the number of $k$-points can substantially converge the results for systems with a small number of atoms.
Table~\ref{tab:ref} \MCC{in Appendix D} shows that both setups, i.e., an 8-atom Si with $2\times2\times2$ $k$-points and a 64-atom Si with the $\Gamma$ $k$-point, yield almost the same free energy and pressure; the difference of free energy and pressure is on the order of 0.001 eV and 0.1 kBar, respectively.
For the two sizes of Si systems, we also try SDFT using 64 stochastic orbitals. We utilize  $2\times2\times2$ $k$-points and the single $\Gamma$-point for the 8- and 64-atom systems, respectively.}
As a result, the statistical errors are close as we find the pressure error (the average free energy error) are $\Delta P$=71 kBar ($\Delta A/N$=0.2 eV) and $\Delta P$=82 kBar ($\Delta A/N$=0.3 eV) for the 8-atom Si system [the first point of the red dotted line in Fig.~\ref{fig:mix}(a,b)] and the 64-atom Si system [the last hollow purple star in Fig.~\ref{fig:std}(a,b)], respectively.

\subsection{Density of States}

\begin{figure}
  		\centering
  		\includegraphics[width=8.6cm]{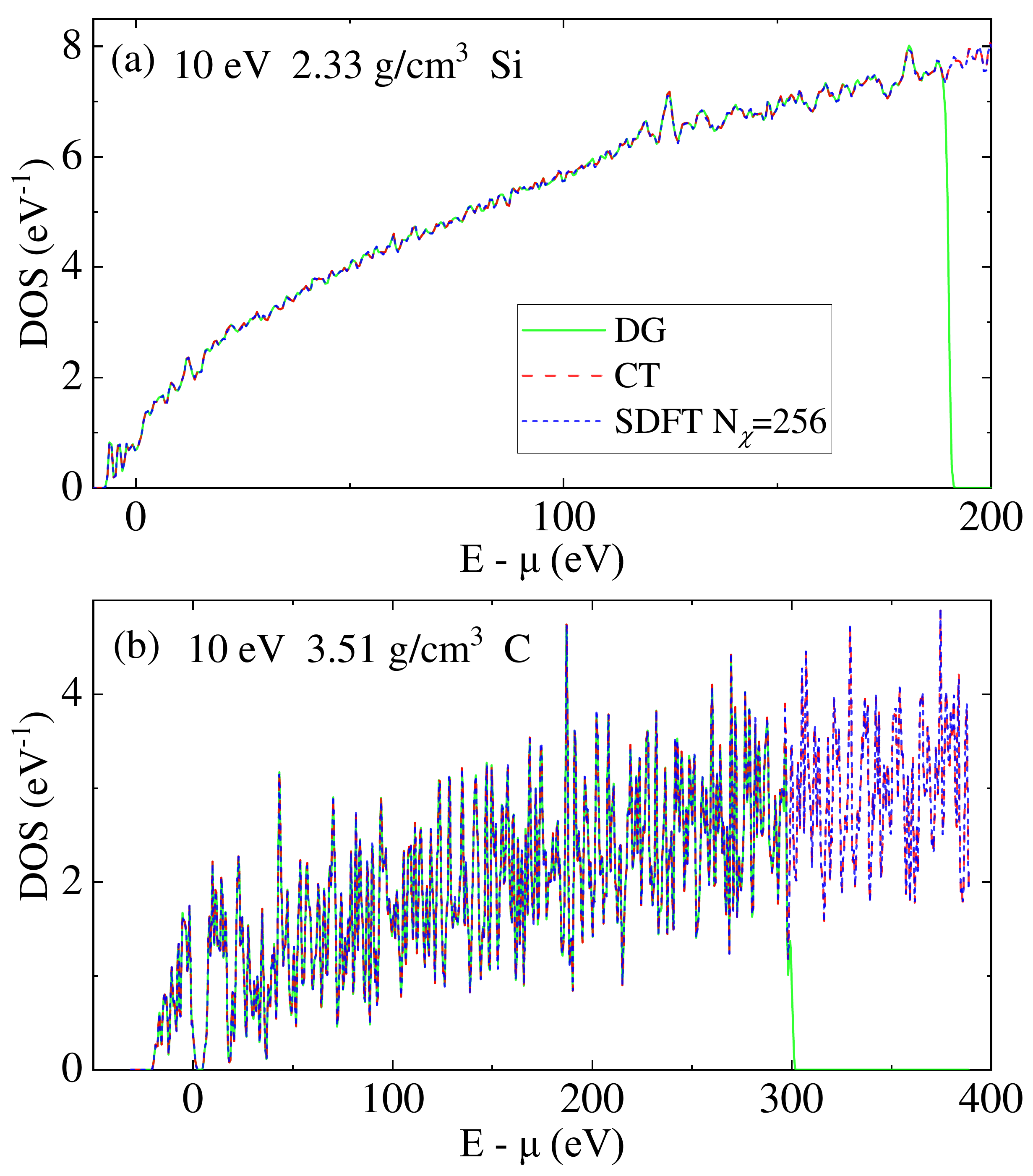}\\
  		\caption{\MC{(Color online) Density of states (DOS) of Si (a) and C (b) at the temperature of 10 eV. The Fermi level is set to 0. The densities of C and Si are 3.51 ${\rm g/cm^3}$ and 2.33 ${\rm g/cm^3}$, respectively. We compare the results from the diagonalization (DG) method, the Chebyshev trace (CT) method, and the SDFT method. We use 256 stochastic orbitals in SDFT. The $k$-point sampling is chosen to be $4\times4\times4$.}
  		}\label{fig:dos}
\end{figure}

The density of states (DOS) of a given system can be evaluated through either SDFT or MDFT methods, even though the eigenvalues of $\hat{H}$ cannot be directly obtained.
The formula of DOS has the form of
\begin{equation}
    g(E) = 2{\rm Tr}\left[\frac{1}{\sqrt{2\pi}\sigma}\exp\left(-\frac{(E-\hat{H})^2}{2\sigma^2}\right)\right],
    \label{eq:dos}
\end{equation}
where $\sigma$ \MCC{controls} the width of smearing. The DOS $g(E)$ can be computed via Eq.~\ref{trace}.
Figs.~\ref{fig:dos}(a) and (b) respectively plot the DOS of disordered Si and diamond C at 10 eV. The density of Si and C is 2.33 ${\rm g/cm^3}$ and
3.51 ${\rm g/cm^3}$, respectively. 
We choose the energy cutoffs to be 50 and 70 Ry for the Si and C systems, respectively.
A Monkhorst-Pack $4\times4\times4$ $k$-point mesh is adopted. The number of stochastic orbitals ($N_{\chi}$) used in SDFT is 256, while the number of bands in DG is chosen to be 1000 and 600 for the Si and C systems, respectively.

\MC{
As illustrated in Figs.~\ref{fig:dos}, three methods are used to compute the DOS, including the the DG method, the CT method with the full plane-wave basis, and SDFT. The Fermi level is set to zero.
We observe the high-energy DOS obtained from the DG method is limited by the number of bands used, while the other two methods can yield high-energy DOS directly from evaluating Eq.~\ref{eq:dos}.
In fact, for a given energy cutoff, the CT method uses all of the plane-wave basis, and the result serves as a benchmark for SDFT.
Overall, We find all of the three methods yield similar DOS, demonstrating the accuracy of utilizing the CT method and the stochastic orbitals compares well with the DG method.}
%
%Second, the diamond C system with more symmetries own more peaks than disordered Si. Even so, SDFT can still acquire the same results, which shows the validity better. Last, KSDFT has a truncation since only some occupied bands are calculated while SDFT can obtain DOS of all bands with energies under the kinetic cutoff energy.

\subsection{Efficiency}
\begin{figure}
  		\centering
  		\includegraphics[width=8.6cm]{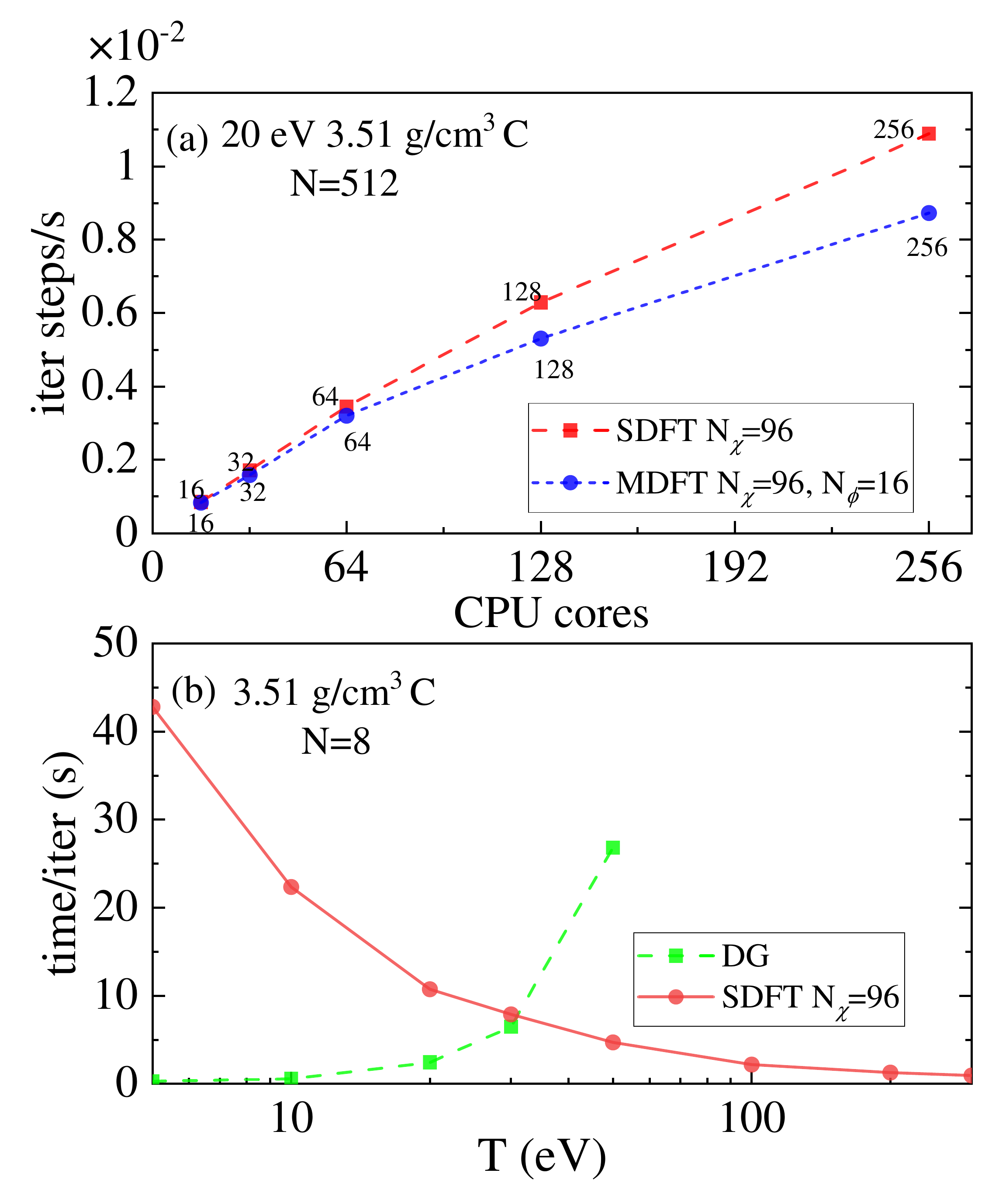}\\
  		\caption{\MCC{(Color online) Efficiency tests of the SDFT, MDFT and DG methods. (a) Parallel efficiency (iteration steps per second) of SDFT and MDFT for a C system with the number of atoms $N$ being 512.
  		The number of stochastic orbitals $N_{\chi}$ is chosen to be 96 for both SDFT and MDFT, while additional 16 deterministic orbitals ($N_{\phi}$) are adopted in MDFT.
  		  The temperature is 20 eV and the density is 3.51 g/cm$^3$. (b) Efficiency of the DG and SDFT methods when calculating a C system ($N$=8) at temperatures ranging from 5 to 300 eV. The recorded time is the averaged wall time for an electronic iteration step.
     		  All of the tests are performed with Intel(R) Xeon(R) Platinum 9242 CPU @ 2.30GHz nodes.}
  		}\label{fig:efficiency}
\end{figure}

\MC{
Besides the accuracy of SDFT and MDFT, we also test their parallel efficiency. 
Fig.~\ref{fig:efficiency}(a) shows a test for a 512-atom C system with the diamond structure. 
The temperature is 20 eV and the density is 3.51 $\rm{g/cm^3}$. 
We adopt a norm-conserving pseudopotential for C with 4 electrons and choose the energy cutoff to be 70 Ry.
The number of stochastic orbitals ($N_{\chi}$) is 96 for both SDFT and MDFT methods, and 16 deterministic orbitals ($N_{\phi}$) are utilized in the MDFT.
We adopt 16, 32, 64, 128 and 256 CPU cores and record the average number of electronic iteration steps \MCC{per second}.  
This test demonstrates the advantages of adopting either SDFT or MDFT to efficiently compute large extended systems at high temperatures, which pose challenges for the DG method.
}
On the one hand, we can see that SDFT owns a well-behaved scaling because all of the stochastic orbitals are independent and can be trivially parallelized.
\MC{
On the other hand, MDFT needs to compute KS orbitals with extra data communications among the CPU cores,
which lowers the efficiency as compared to SDFT. For example, SDFT is 2.4\% and 19.8\% faster than MDFT with 16 and 256 CPU cores, respectively. Even so, MDFT is still more efficient than the DG method because MDFT
needs much less deterministic orbitals than the DG method.
}

\MC{
Fig.~\ref{fig:efficiency}(b) compares the efficiency of running the SDFT and DG methods at different temperatures by recording 
the average time of one electronic iteration step of DG and SDFT. 
We set up a diamond structure with 8 C atoms. The density is chosen to be 3.51 $\rm{g/cm^3}$ and the temperature ranges from 5 to 300 eV.
A norm conserving pseudopotential with 6 valence electrons is used for C atoms. The energy cutoff is set to be 240 Ry.
The number of bands in the DG calculations is chosen to ensure the occupation number of the highest band less than 1e-5.
In this regard, 100, 200, 460, 750, 1400 bands are used for calculations of temperatures at 5, 10, 20, 30, 50 eV, respectively.
On the other hand, the number of stochastic orbitals is 96 in the SDFT method.
The number of Chebyshev expansion order in SDFT calculations is chosen to ensure the error of the electron number smaller than 1e-9. 
Specifically, the expansion orders are set to be 1200, 620, 300, 220, 130, 60, 35, 25 for 5, 10, 20, 30, 50, 100, 200, 300 eV, respectively.
All of the calculations utilize 64 CPU cores.
We notice that the computational cost for the DG method calculations increases \MCC{rapidly} with elevated temperatures.
On the contrary, SDFT becomes more efficient when the temperature rises because less expansion orders are needed. 
%In particular, the SDFT method is more %efficient than KSDFT when the temperature is %larger than 30 eV.
%
%Therefore, we conclude that SDFT is more suitable to study high-temperature materials than the DG method in terms of the efficiency.
}

\section{Conclusions}

\MCC{
The traditional KSDFT method uses the diagonalization (DG) method to yield the Kohn-Sham orbitals and the ground-state electron density of a given system.
We demonstrate that the DG method and the Chebyshev trace (CT) method share similar accuracy.
Furthermore, the combination of the stochastic orbitals and the CT method leads to the SDFT method. 
Additionally, an improved theory of SDFT is the MDFT method, which mixes a set of KS orbitals with stochastic orbitals and generally yields better accuracy.
%Note that 
\MCC{On the one hand,} the SDFT is a special case of MDFT when no deterministic orbitals are used in MDFT.
%
%In addition, 
\MCC{On the other hand,} when the number of stochastic orbitals is set to zero and all of the Kohn-Sham orbitals are used to evaluate the electron density, the MDFT method becomes the traditional KSDFT method.
\MCC{Both SDFT and MDFT methods are more efficient than the DG method when the electron temperature is high and a large number of occupied states need to be solved.}
}

\MCC{Based on the above theories, we have implemented all of the four methods, i.e., DG, CT, SDFT, and MDFT methods in the ABACUS package~\cite{10JPCM-Mohan,16CMS-Li}, \MCC{which can be freely downloaded (https://github.com/deepmodeling/abacus-develop). The four methods are implemented with the use of periodic boundary conditions, plane-wave basis set, $k$-point samplings, and the norm-conserving pseudopotentials.}
To the best of our knowledge, this is the first implementation of the four methods within a single package for extended systems.
%\MCC{The results show that both SDFT and MDFT implemented by the plane-wave basis set achieve sufficient accuracy and efficiency as compared to the traditional KSDFT method, and can be readily applied to study extended systems at high temperatures.}
}

\MCC{We systematically tested the accuracy and efficiency of SDFT and MDFT by choosing Si and C systems with different \MCC{sizes, densities, and temperatures. We first compared the electron densities obtained from the SDFT, MDFT and DG methods.} Both SDFT and MDFT yield\MCC{ed} similar electron density \MCC{as compared to the one from} DG, and the use of deterministic orbitals in MDFT led to smaller deviations of electron density than SDFT.
We \MCC{analyzed} the statistical errors of \MCC{the free energies, the pressures, and the atomic forces from SDFT and MDFT.} Our results show\MCC{ed} that both SDFT and MDFT yielded
%similar results as KSDFT.
\MCC{sufficiently small errors when compared to the DG results.}} 
%There are some ways to reduce the statistical errors. For SDFT, the statistical errors are inversely proportional to the square of the number stochastic orbitals $N_{\chi}^{1/2}$, which means more stochastic orbitals have smaller errors. Simulating larger systems can also reduces the errors of properties like the free energy per atom and the pressure but that of forces remain unchanged. Besides, using more $k$-points is able to reduce the errors including the forces. For MDFT, not only the methods mentioned above but adopting more determined orbitals can also reduce the errors. However, it is worth mentioning that the errors do not nearly change at too high temperatures and it is economical to use SDFT in such cases. Last, we plotted the density of states with $4\times4\times4$ $k$-points. Because the large number of $k$-points, SDFT only with 256 stochastic orbitals agrees well with KSDFT. 
%
\MCC{To validate the parallel efficiency of SDFT and MDFT methods in dealing with large-size systems, we used a 512-atom C system \MCC{and} found both \MCC{methods} exhibit excellent parallel efficiency up to 256 CPU cores. Furthermore, we tested the efficiency of an 8-atom C system with temperatures ranging from 5 to 300 eV. Notably, SDFT became more efficient at higher temperatures due to the less expansion order needed in the CT method, which was more efficient than the DG method at extremely high temperatures.}

\MCC{
In summary, our work shows that the implementation of SDFT and MDFT methods in plane-wave basis set can be readily applied to study extended systems such as condensed matter systems with several advantages. 
First, at extremely high temperatures, these methods
adopt stochastic wave functions and are more efficient than the DG method used in traditional KSDFT. 
Second, both SDFT and MDFT methods
can be easily parallelized for large-size and high-temperature systems. 
Third, by enabling sampling of $k$-points in the Brillouin-zone, physical properties in extended systems can be converged by sampling a sufficiently large number of $k$-points within the SDFT and MDFT methods. For example, even though hundreds of atoms are adopted in a simulation cell, properties like the electronic conductivity needs %multiple 
a fairly large number of $k$-points to be converged.~\cite{11B-Pozzo,21MRE-Liu} 
In future works, we expect these methods can be combined with molecular dynamics to obtain the equation of state (EOS) or electron transport properties of systems at high temperatures.}

\acknowledgements
\MC{This work is supported by the National Science Foundation of China under Grant No.12122401 and No.12074007. The numerical simulations were performed on the High Performance Computing Platform of CAPT.}

%$\\$
%{\bf Availability of Data}
%$\\$
%The data that support the findings of %this study are available from the %corresponding author upon reasonable %request.

\appendix

\MCC{
\section{Kohn-Sham Equation in Plane-Wave Basis}
}

The Kohn-Sham equation in Eq.~(\ref{kseq}) can be solved by a plane-wave basis set with periodic boundary conditions, the corresponding algorithms have been implemented in ABACUS.~\cite{10JPCM-Mohan,16CMS-Li}
According to the Bloch theorem, the electronic wave function takes the form of
  \begin{equation}
      \psi_{n\mathbf{k}}(\mathbf{r})=\frac{1}{\sqrt{V}}\sum_{\mathbf{G}}{c_{n\mathbf{k}}(\mathbf{G})e^{i(\mathbf{k+G})\cdot\mathbf{r}}},
  \end{equation}
  where $V$ is the volumn of the system, $\mathbf{G}$ and $\mathbf{k}$ denote the wave vector of a plane wave and a $k$-point vector, respectively. $n$ represents the index of electronic states.
  The $c_{n\mathbf{k}}(\mathbf{G})$ coefficients are the expansion coefficients of the plane-wave basis with the form of
  \begin{equation}
      c_{n\mathbf{k}}(\mathbf{G})=\langle\mathbf{k+G}|\psi_{n\mathbf{k}}\rangle,
  \end{equation}
  where
  \begin{equation}\label{eq:pw}
     \langle\mathbf{r}|\mathbf{k+G}\rangle=\frac{1}{\sqrt{V}}e^{i(\mathbf{k+G})\cdot\mathbf{r}}.
  \end{equation}
  As a result, Eq.~(\ref{kseq}) is transformed to
  \begin{equation}
  \begin{aligned}
      \sum_{\mathbf{G'}}{\Big[\frac{1}{2}(\mathbf{k+G})^2\delta_{\mathbf{GG'}}+\tilde{v}_\mathrm{eff}(\mathbf{G-G'})\Big]c_{n\mathbf{k}}(\mathbf{G'})}\\
      =\epsilon_{nk}c_{n\mathbf{k}}(\mathbf{G}).
  \end{aligned}
  \end{equation}
  Solving the above equation is equal to solving an eigenvalue problem of a Hermitian matrix with linear algebra techniques and the dimension of the matrix is equal to the number of plane waves. Typically, the diagonalization of this equation has an $O(N^3)$ scaling. With the aid of pseudopotentials,~\cite{04-Martin,79L-Hamann} the ionic coulomb potential becomes more smooth by neglecting atomic core states. Thus, less plane waves are needed which significantly reduces the computational cost. The plane-wave basis set is chosen by satisfying $\frac{1}{2}(\mathbf{k+G})^2<E_{cut}$, where $E_{cut}$ is a specific kinetic energy cut-off for plane waves.
  
\MCC{
\section{Forces and Stress}
}

Forces and stress are two key quantities that can be obtained from \MCC{the diagonalization (DG) method in traditional KSDFT}. On the one hand, the force acting on an atom $I$ has the definition of
  \begin{equation}
      \mathbf{F}_I=-\frac{\partial E}{\partial \mathbf{R}_I},
  \end{equation}
  while the stress is defined as
   \begin{equation}
      \sigma_{\alpha\beta}=-\frac{1}{V}\frac{\partial E}{\partial \epsilon_{\alpha\beta}},
  \end{equation}
  where $E$ is the total energy, $\mathbf{R}_I$ is the position of the atom and $\epsilon_{\alpha\beta}$ is the strain with the spatial coordinates $\alpha$ and $\beta$.
  As a result, the force of atom $I$ of species $\tau$ can be divided into three parts,~\cite{04-Martin} which takes the form of
  \begin{equation}\label{eq:force}
      \mathbf{F}_{I,\tau}=\mathbf{F}_{I,\tau}^\mathrm{Ewald} + \mathbf{F}_{I,\tau}^\mathrm{L} + \mathbf{F}_{I,\tau}^\mathrm{NL},
  \end{equation}
   where $\mathbf{F}_{I,\tau}^\mathrm{Ewald}$ is the Ewald force, $\mathbf{F}_{I,\tau}^\mathrm{L}$ is the local pseudopotential force and $\mathbf{F}_{I,\tau}^\mathrm{NL}$ is the non-local pseudopotential force. Specifically, the local term has a form of
   \begin{equation}
       \mathbf{F}_{I,\tau}^\mathrm{L}=-iV\sum_{\mathbf{G}}{\mathbf{G}e^{i\mathbf{G}\cdot\mathbf{R}_I}v_\tau^\mathrm{L}(\mathbf{G})\rho^*(\mathbf{G})},
   \end{equation}
   while the non-local term is written as
   \begin{equation}
   \begin{aligned}
       \mathbf{F}_{I,\tau}^\mathrm{NL} = &-2i\sum_{\mathbf{k},n}W(\mathbf{k})\sum_{\mathbf{G,G'}}f(\epsilon_n;\mu)c^*_{n\mathbf{k}}(\mathbf{G})\Big[(\mathbf{G'-G})\\
                   &\times e^{i(\mathbf{G'-G})\cdot\mathbf{R}_I}v_\tau^\mathrm{NL}(\mathbf{k+G,k+G'})\Big]c_{n\mathbf{k}}(\mathbf{G'}),
   \end{aligned}
   \end{equation}
  where $v_\tau^\mathrm{L}$ is the local pseudopotential of atom species $\tau$.
  On the other hand,
  the stress~\cite{04-Martin,85B-Nielsen} is divided into the Ewald term $\sigma_{\alpha\beta}^\mathrm{Ewald}$, the Hartree term $\sigma_{\alpha\beta}^\mathrm{H}$, the exchange-correlation term $\sigma_{\alpha\beta}^\mathrm{xc}$, the kinetic term $\sigma_{\alpha\beta}^\mathrm{T}$, the local pseudopotential term $\sigma_{\alpha\beta}^\mathrm{L}$ and the non-local pseudopotential term $\sigma_{\alpha\beta}^\mathrm{NL}$ with the form of
  \begin{equation}\label{eq:stress}
      \sigma_{\alpha\beta}=\sigma_{\alpha\beta}^\mathrm{Ewald}+\sigma_{\alpha\beta}^\mathrm{Hartree}+\sigma_{\alpha\beta}^\mathrm{xc}+\sigma_{\alpha\beta}^\mathrm{T}+\sigma_{\alpha\beta}^\mathrm{L}+\sigma_{\alpha\beta}^\mathrm{NL},
  \end{equation}
  where
  \begin{equation}
      \sigma_{\alpha\beta}^\mathrm{Hartree}=-2\pi\sum_{\mathbf{G}\ne 0}\frac{|\rho(\mathbf{G})|^2}{G^2}\left[2\frac{\mathbf{G}_\alpha\mathbf{G}_\beta}{G^2}-\delta_{\alpha\beta}\right],
  \end{equation}
  \begin{equation}
      \sigma_{\alpha\beta}^\mathrm{xc}=-\delta_{\alpha\beta}\sum_{\mathbf{G}}[\epsilon_\mathrm{xc}(\mathbf{G})-v_\mathrm{xc}(\mathbf{G})]\rho^*(\mathbf{G}),
  \end{equation}
  %$E_{xc}[\rho(\mathbf r)]=\int{\epsilon_\mathrm{xc}(\mathbf r)\rho(\mathbf r)d\mathbf r}$,
  \begin{equation}
  \begin{aligned}
      \sigma_{\alpha\beta}^\mathrm{T}=&\frac{2}{V}\sum_{\mathbf{k},n}W(\mathbf{k})\sum_{\mathbf{G}}f(\epsilon_n;\mu)c^*_{n\mathbf{k}}(\mathbf{G})(\mathbf{k+G})_\alpha\\
                         &\times\delta(\mathbf{G,G'})(\mathbf{k+G'})_\beta c_{n\mathbf{k}}(\mathbf{G'}),
  \end{aligned}
  \end{equation}
  
  \begin{equation}
  \begin{aligned}
      \sigma_{\alpha\beta}^\mathrm{L} = &\sum_{\mathbf{G},\tau}S_\tau(\mathbf{G})\Big[\frac{\partial v_\tau^\mathrm{L}(\mathbf{G})}{\partial (G^2)}2\mathbf{G}_\alpha\mathbf{G}_\beta+\\
                         &v_\tau^{L}(\mathbf{G})\delta_{\alpha\beta}\Big]\rho^*(\mathbf{G})
  \end{aligned}
  \end{equation}
  and
  \begin{equation}
  \begin{aligned}
      \sigma_{\alpha\beta}^\mathrm{NL} = &-\frac{2}{V}\sum_{\mathbf{k},n}W(\mathbf{k})\sum_{\mathbf{G,G',\tau}}f(\epsilon_n;\mu)c^*_{n\mathbf{k}}(\mathbf{G})\\
                         &\times S_\tau(\mathbf{G'-G})\frac{\partial v_\tau^\mathrm{NL}(\mathbf{G+k,G'+k})}{\partial \epsilon_{\alpha\beta}}c_{n\mathbf{k}}(\mathbf{G'}).
  \end{aligned}
  \end{equation}
  Here $\epsilon_{\alpha\beta}$ is the strain tensor.
  $S_\tau(\mathbf{G})$ is the structure factor of atom species $\tau$ and the electron density in the plane-wave basis is calculated via the formula of
  \begin{equation}
      \rho(\mathbf{G}) = \frac{1}{V}\int{\rho (\mathbf r)\exp(-i\mathbf G\cdot \mathbf r) \mathrm{d}\mathbf{r}}.
  \end{equation}
 
\section{Error Analysis}
\MCC{Since stochastic or mixed orbitals only form a complete basis when the number of stochastic orbitals approaches infinity, the use of a finite number of stochastic orbitals in practical calculations unavoidably introduces statistical errors for both SDFT and MDFT methods.}

\MCC{
By using the stochastic orbitals in SDFT, the deviation of a given Hermitian operator $\hat{O}$ takes the form of
\begin{equation}
  (\Delta O)_s\equiv {\rm Tr}[\hat{X}\hat{O}]-{\rm Tr}[\hat{O}]={\rm Tr}[(\hat{X}-\hat{I})\hat{O}],
\end{equation}
where $\hat{I}$ is the identity operator and
$\hat{X}$ forms by a set of stochastic orbitls $\{\chi_{n}\}$ as
\begin{equation}
\hat{X}\equiv\sum_{n=1}^{N_\chi}|\chi_n\rangle\langle\chi_n|.
\end{equation}
According to previous studies,~\cite{89CSSC-Hutchinson,19WCMS-Fabian,22ARPC-Baer} the expectation and variance of $(\Delta O)_s$ are
\begin{equation}\label{eq:se}
  {\rm E}[(\Delta O)_s]=0
\end{equation}
and
\begin{equation}\label{eq:svar}
  {\rm Var}[(\Delta O)_s]=\frac{2}{N_\chi}\sum_{i\ne j}{|O_{ij}|^2}\equiv\sigma_s^2,
\end{equation}
where 
\begin{equation}
O_{ij}=\langle\varphi_i|\hat{O}|\varphi_j\rangle.
\end{equation}
Thus, the statistical fluctuation $\sigma_s$ is proportional to $1/\sqrt{N_\chi}$. It is worth mentioning that although Eq.~(\ref{eq:se}) shows an unbiased trace result, the SDFT result still has an $O(1/N_\chi)$ bias due to the nonlinear dependence of $\hat{H}$ of the operator.~\cite{19WCMS-Fabian} 
However, the bias decreases much faster than the fluctuation when $N_\chi$ increases. Therefore, the error mainly comes from the statistical fluctuation $\sigma_s$.}

\MCC{
Importantly, the errors of the volume-integrated, atom-averaged, local-volume-integrated, and the volume-averaged quantities can be estimated as follows.
If we set $\hat{O}$ in Eq.~(\ref{eq:svar}) as the density operator, $\sum_{i\ne j}{|O_{ij}|^2}$ can be roughly estimated by $\rho^2(\mathbf{r})$,~\cite{19JCP-Ming} which is proportional to $\rho_0^2$ ($\rho_0=N/V$ with $N$ being the number of atoms). }
%
%Importantly, the errors from different types of quantities can now be estimated. For example,
%
\MCC{
In this regard, for volume-integrated quantities such as the free energy which includes the integration of the density, the error fluctuation is proportional to $\sqrt{V/N_\chi}\rho_0$; 
for atom-averaged quantities such as the free energy per atom, the error fluctuations are proportional to $1/\sqrt{N_\chi V}$; 
for volume-averaged quantities such as the pressure, the error fluctuations are proportional to $\rho_0/\sqrt{N_\chi V}$; %
for local-volume-integrated quantities such as forces acting on atoms, the fluctuations are proportional to $\sqrt{\rho_0/N_\chi}$.}

\MCC{
To calculate the deviation in terms of mixed orbitals, we first assume that operator $\hat{O}$ contains a factor of $f(\hat{H})$ and it satisfies
\begin{equation}
{\rm Tr}[\hat{O}]={\rm Tr}\left[\sum_{i=1}^{N_{oc}}|\psi_i\rangle\langle\psi_i|\hat{O}\right],
\end{equation}
where $\{\psi_i\}$ are the eigenstates of $\hat{H}$ and we consider $f(\epsilon_i)\approx0$ for $i>N_{oc}$.
We define
\begin{equation}
\hat{\Psi}=\sum_{i=1}^{N_{oc}}|\psi_i\rangle\langle\psi_i| {\rm \ and \ }
\hat{\Phi}=\sum_{i=1}^{N_\phi}|\psi_i\rangle\langle\psi_i|,
\end{equation}
where $N_\phi$ is the number of deterministic orbitals within mixed ones and $N_\phi \leq N_{oc}$.
Then the deviation of mixed orbitals has the form of
\begin{equation}
\begin{aligned}
  (\Delta O)_m&\equiv {\rm Tr}[(\hat{\tilde{X}}+\hat{\Phi})\hat{O}]-{\rm Tr}[\hat{O}]\\
  &= {\rm Tr}[(\hat{\tilde{X}}+\hat{\Phi})\hat{\Psi}\hat{O}]-{\rm Tr}[\hat{\Psi}\hat{O}]\\
  &={\rm Tr}[(\hat{X}-\hat{I})\hat{\tilde{O}}],
\end{aligned}
\end{equation}
where
\begin{equation}
\hat{\tilde{X}}=\sum_{n=1}^{N_\chi}|\tilde{\chi_n}\rangle\langle\tilde{\chi_n}|=(\hat{I}-\hat{\Phi})\hat{X}(\hat{I}-\hat{\Phi}),
\end{equation}
and 
\begin{equation}
\hat{\tilde{O}}=(\hat{\Psi}-\hat{\Phi})\hat{O}(\hat{\Psi}-\hat{\Phi}).\label{eq:merror}
\end{equation}
 The above formulas show that the error of mixed orbitals is smaller than that of stochastic orbitals and as $N_\phi$ approaches $N_{oc}$ or $\hat{\Phi}$ approaches $\hat{\Psi}$,  the error decreases towards zero.}

\MCC{
\section{Reference Data}
}
To analyze the statistical errors of SDFT and MDFT methods as mentioned in Sec. III B, we provide the reference data with both DG and CT methods. They are listed in Table III.

  \begin{table}[htbp]
  	\centering
  	\caption{\MCC{Reference data for analyzing the statistical errors of SDFT and MDFT methods, which include the free energy per atom $A/N$ (in eV, $N$ is the number of atoms) and pressure $P$ (in kBar) for Si and C systems at several temperatures $T$ (in eV). For the 8-atom systems, we use both DG and CT methods, which yield the same results. For the 64-atom Si and high-temperature C systems, we adopt the CT method. All of the systems are in the diamond structure except those with the * label, which depicts a disordered structure. Regarding the Brillouin zone sampling method, the "\#" index implies that a $2\times2\times2$ $k$-point mesh is used, otherwise only the $\Gamma$ point is adopted.}
  	}
  	
  	\label{tab:ref}
  	\begin{tabular}{ccccccc}
  		\toprule
  		\hline
  		\specialrule{0.05em}{1pt}{3pt}
  		 &$T$ (eV)&$\rho$ (${\rm g/cm^3}$)& $N$ & $N_c$& $A$/$N$ (eV)& $P$ (kBar)\\
  		\midrule
  		\specialrule{0em}{3pt}{0pt}
  	    \hline
  	    \specialrule{0em}{0pt}{3pt}
  		
  		Si&10 &2.33&  8  & 240 & -167.468 & 1863.8\\
  		Si&10 &2.33& 64  & 240 & -167.468 & 1863.9\\
  		Si&10 &2.33&  8* & 240 &-164.517 & 2089.4\\
  		Si&10 &2.33& 64* & 240 &-164.241 & 2070.2\\
  		Si&10 &5.0 & 8  & 240 &-146.353 & 5889.4\\
  		Si&10 &5.0 & 64  & 240 &-146.325 & 5908.3\\
  		Si&10 &10.0&  8  & 240 &-113.850 & 21904.1\\
  		$\rm{Si}^\#$&10 &10.0&  8  &240 & -113.667 & 22003.7\\
  		Si&10 &10.0& 64  & 240 &-113.667& 22003.8\\
  		C &10 &3.51& 8  & 240 &-186.365 & 5700.9\\
  		C &20 &3.51& 8  & 160 &-264.878& 14025.2\\
  		C &50 &3.51& 8  & 140 &-1437.459& 43931.6\\
  	    C &100&3.51& 8  & 80  &-2364.369& 111513.4\\
  		C &300&3.51& 8  & 35  &-8406.994& 459551.3\\
 
  		\bottomrule
  		\specialrule{0.05em}{3pt}{1pt}
  		\hline
  	\end{tabular}
  \end{table}

\clearpage
\bibliography{sDFT}

\begin{thebibliography}{94}%
\makeatletter
\providecommand \@ifxundefined [1]{%
 \@ifx{#1\undefined}
}%
\providecommand \@ifnum [1]{%
 \ifnum #1\expandafter \@firstoftwo
 \else \expandafter \@secondoftwo
 \fi
}%
\providecommand \@ifx [1]{%
 \ifx #1\expandafter \@firstoftwo
 \else \expandafter \@secondoftwo
 \fi
}%
\providecommand \natexlab [1]{#1}%
\providecommand \enquote  [1]{``#1''}%
\providecommand \bibnamefont  [1]{#1}%
\providecommand \bibfnamefont [1]{#1}%
\providecommand \citenamefont [1]{#1}%
\providecommand \href@noop [0]{\@secondoftwo}%
\providecommand \href [0]{\begingroup \@sanitize@url \@href}%
\providecommand \@href[1]{\@@startlink{#1}\@@href}%
\providecommand \@@href[1]{\endgroup#1\@@endlink}%
\providecommand \@sanitize@url [0]{\catcode `\\12\catcode `\$12\catcode
  `\&12\catcode `\#12\catcode `\^12\catcode `\_12\catcode `\%12\relax}%
\providecommand \@@startlink[1]{}%
\providecommand \@@endlink[0]{}%
\providecommand \url  [0]{\begingroup\@sanitize@url \@url }%
\providecommand \@url [1]{\endgroup\@href {#1}{\urlprefix }}%
\providecommand \urlprefix  [0]{URL }%
\providecommand \Eprint [0]{\href }%
\providecommand \doibase [0]{https://doi.org/}%
\providecommand \selectlanguage [0]{\@gobble}%
\providecommand \bibinfo  [0]{\@secondoftwo}%
\providecommand \bibfield  [0]{\@secondoftwo}%
\providecommand \translation [1]{[#1]}%
\providecommand \BibitemOpen [0]{}%
\providecommand \bibitemStop [0]{}%
\providecommand \bibitemNoStop [0]{.\EOS\space}%
\providecommand \EOS [0]{\spacefactor3000\relax}%
\providecommand \BibitemShut  [1]{\csname bibitem#1\endcsname}%
\let\auto@bib@innerbib\@empty
%</preamble>
\bibitem [{\citenamefont {Hohenberg}\ and\ \citenamefont
  {Kohn}(1964)}]{64PR-Hohenberg}%
  \BibitemOpen
  \bibfield  {author} {\bibinfo {author} {\bibfnamefont {P.}~\bibnamefont
  {Hohenberg}}\ and\ \bibinfo {author} {\bibfnamefont {W.}~\bibnamefont
  {Kohn}},\ }\bibfield  {title} {\bibinfo {title} {Inhomogeneous electron
  gas},\ }\href@noop {} {\bibfield  {journal} {\bibinfo  {journal} {Phys.
  Rev.}\ }\textbf {\bibinfo {volume} {136}},\ \bibinfo {pages} {864B} (\bibinfo
  {year} {1964})}\BibitemShut {NoStop}%
\bibitem [{\citenamefont {Kohn}\ and\ \citenamefont {Sham}(1965)}]{65PR-Kohn}%
  \BibitemOpen
  \bibfield  {author} {\bibinfo {author} {\bibfnamefont {W.}~\bibnamefont
  {Kohn}}\ and\ \bibinfo {author} {\bibfnamefont {L.~J.}\ \bibnamefont
  {Sham}},\ }\bibfield  {title} {\bibinfo {title} {Thermal properties of the
  inhomogeneous electron gas},\ }\href@noop {} {\bibfield  {journal} {\bibinfo
  {journal} {Phys. Rev.}\ }\textbf {\bibinfo {volume} {140}},\ \bibinfo {pages}
  {1133A} (\bibinfo {year} {1965})}\BibitemShut {NoStop}%
\bibitem [{\citenamefont {Yang}(1991)}]{91L-Yang}%
  \BibitemOpen
  \bibfield  {author} {\bibinfo {author} {\bibfnamefont {W.}~\bibnamefont
  {Yang}},\ }\bibfield  {title} {\bibinfo {title} {Direct calculation of
  electron density in density-functional theory},\ }\href
  {https://doi.org/10.1103/PhysRevLett.66.1438} {\bibfield  {journal} {\bibinfo
   {journal} {Phys. Rev. Lett.}\ }\textbf {\bibinfo {volume} {66}},\ \bibinfo
  {pages} {1438} (\bibinfo {year} {1991})}\BibitemShut {NoStop}%
\bibitem [{\citenamefont {Li}\ \emph {et~al.}(1993)\citenamefont {Li},
  \citenamefont {Nunes},\ and\ \citenamefont {Vanderbilt}}]{93B-Li}%
  \BibitemOpen
  \bibfield  {author} {\bibinfo {author} {\bibfnamefont {X.-P.}\ \bibnamefont
  {Li}}, \bibinfo {author} {\bibfnamefont {R.~W.}\ \bibnamefont {Nunes}},\ and\
  \bibinfo {author} {\bibfnamefont {D.}~\bibnamefont {Vanderbilt}},\ }\bibfield
   {title} {\bibinfo {title} {Density-matrix electronic-structure method with
  linear system-size scaling},\ }\href
  {https://doi.org/10.1103/PhysRevB.47.10891} {\bibfield  {journal} {\bibinfo
  {journal} {Phys. Rev. B}\ }\textbf {\bibinfo {volume} {47}},\ \bibinfo
  {pages} {10891} (\bibinfo {year} {1993})}\BibitemShut {NoStop}%
\bibitem [{\citenamefont {Kohn}(1996)}]{96L-Kohn}%
  \BibitemOpen
  \bibfield  {author} {\bibinfo {author} {\bibfnamefont {W.}~\bibnamefont
  {Kohn}},\ }\bibfield  {title} {\bibinfo {title} {Density functional and
  density matrix method scaling linearly with the number of atoms},\ }\href
  {https://doi.org/10.1103/PhysRevLett.76.3168} {\bibfield  {journal} {\bibinfo
   {journal} {Phys. Rev. Lett.}\ }\textbf {\bibinfo {volume} {76}},\ \bibinfo
  {pages} {3168} (\bibinfo {year} {1996})}\BibitemShut {NoStop}%
\bibitem [{\citenamefont {Goedecker}(1999)}]{99RMP-Goedecker}%
  \BibitemOpen
  \bibfield  {author} {\bibinfo {author} {\bibfnamefont {S.}~\bibnamefont
  {Goedecker}},\ }\bibfield  {title} {\bibinfo {title} {Linear scaling
  electronic structure methods},\ }\href
  {https://doi.org/10.1103/RevModPhys.71.1085} {\bibfield  {journal} {\bibinfo
  {journal} {Rev. Mod. Phys.}\ }\textbf {\bibinfo {volume} {71}},\ \bibinfo
  {pages} {1085} (\bibinfo {year} {1999})}\BibitemShut {NoStop}%
\bibitem [{\citenamefont {Bowler}\ \emph {et~al.}(1997)\citenamefont {Bowler},
  \citenamefont {Aoki}, \citenamefont {Goringe}, \citenamefont {Horsfield},\
  and\ \citenamefont {Pettifor}}]{97MSMSE-Bowler}%
  \BibitemOpen
  \bibfield  {author} {\bibinfo {author} {\bibfnamefont {D.~R.}\ \bibnamefont
  {Bowler}}, \bibinfo {author} {\bibfnamefont {M.}~\bibnamefont {Aoki}},
  \bibinfo {author} {\bibfnamefont {C.~M.}\ \bibnamefont {Goringe}}, \bibinfo
  {author} {\bibfnamefont {A.~P.}\ \bibnamefont {Horsfield}},\ and\ \bibinfo
  {author} {\bibfnamefont {D.~G.}\ \bibnamefont {Pettifor}},\ }\bibfield
  {title} {\bibinfo {title} {A comparison of linear scaling tight-binding
  methods},\ }\href {https://doi.org/10.1088/0965-0393/5/3/002} {\bibfield
  {journal} {\bibinfo  {journal} {Modell. Simul. Mater. Sci. Eng.}\ }\textbf
  {\bibinfo {volume} {5}},\ \bibinfo {pages} {199} (\bibinfo {year}
  {1997})}\BibitemShut {NoStop}%
\bibitem [{\citenamefont {Ordej\'on}\ \emph {et~al.}(1996)\citenamefont
  {Ordej\'on}, \citenamefont {Artacho},\ and\ \citenamefont
  {Soler}}]{96B-Ordejon}%
  \BibitemOpen
  \bibfield  {author} {\bibinfo {author} {\bibfnamefont {P.}~\bibnamefont
  {Ordej\'on}}, \bibinfo {author} {\bibfnamefont {E.}~\bibnamefont {Artacho}},\
  and\ \bibinfo {author} {\bibfnamefont {J.~M.}\ \bibnamefont {Soler}},\
  }\bibfield  {title} {\bibinfo {title} {Self-consistent order-$n$
  density-functional calculations for very large systems},\ }\href
  {https://doi.org/10.1103/PhysRevB.53.R10441} {\bibfield  {journal} {\bibinfo
  {journal} {Phys. Rev. B}\ }\textbf {\bibinfo {volume} {53}},\ \bibinfo
  {pages} {R10441} (\bibinfo {year} {1996})}\BibitemShut {NoStop}%
\bibitem [{\citenamefont {Gillan}\ \emph {et~al.}(2007)\citenamefont {Gillan},
  \citenamefont {Bowler}, \citenamefont {Torralba},\ and\ \citenamefont
  {Miyazaki}}]{07CPC-Gillan}%
  \BibitemOpen
  \bibfield  {author} {\bibinfo {author} {\bibfnamefont {M.}~\bibnamefont
  {Gillan}}, \bibinfo {author} {\bibfnamefont {D.}~\bibnamefont {Bowler}},
  \bibinfo {author} {\bibfnamefont {A.}~\bibnamefont {Torralba}},\ and\
  \bibinfo {author} {\bibfnamefont {T.}~\bibnamefont {Miyazaki}},\ }\bibfield
  {title} {\bibinfo {title} {Order-n first-principles calculations with the
  conquest code},\ }\href
  {https://doi.org/https://doi.org/10.1016/j.cpc.2007.02.075} {\bibfield
  {journal} {\bibinfo  {journal} {Comput. Phys. Commun.}\ }\textbf {\bibinfo
  {volume} {177}},\ \bibinfo {pages} {14} (\bibinfo {year} {2007})},\ \bibinfo
  {note} {proceedings of the Conference on Computational Physics
  2006}\BibitemShut {NoStop}%
\bibitem [{\citenamefont {Shimojo}\ \emph {et~al.}(2008)\citenamefont
  {Shimojo}, \citenamefont {Kalia}, \citenamefont {Nakano},\ and\ \citenamefont
  {Vashishta}}]{08B-Shimojo}%
  \BibitemOpen
  \bibfield  {author} {\bibinfo {author} {\bibfnamefont {F.}~\bibnamefont
  {Shimojo}}, \bibinfo {author} {\bibfnamefont {R.~K.}\ \bibnamefont {Kalia}},
  \bibinfo {author} {\bibfnamefont {A.}~\bibnamefont {Nakano}},\ and\ \bibinfo
  {author} {\bibfnamefont {P.}~\bibnamefont {Vashishta}},\ }\bibfield  {title}
  {\bibinfo {title} {Divide-and-conquer density functional theory on
  hierarchical real-space grids: Parallel implementation and applications},\
  }\href {https://doi.org/10.1103/PhysRevB.77.085103} {\bibfield  {journal}
  {\bibinfo  {journal} {Phys. Rev. B}\ }\textbf {\bibinfo {volume} {77}},\
  \bibinfo {pages} {085103} (\bibinfo {year} {2008})}\BibitemShut {NoStop}%
\bibitem [{\citenamefont {Prodan}\ and\ \citenamefont
  {Kohn}(2005)}]{05PNAS-Prodan}%
  \BibitemOpen
  \bibfield  {author} {\bibinfo {author} {\bibfnamefont {E.}~\bibnamefont
  {Prodan}}\ and\ \bibinfo {author} {\bibfnamefont {W.}~\bibnamefont {Kohn}},\
  }\bibfield  {title} {\bibinfo {title} {Nearsightedness of electronic
  matter},\ }\href {https://doi.org/10.1073/pnas.0505436102} {\bibfield
  {journal} {\bibinfo  {journal} {Proc. Natl. Acad. Sci.}\ }\textbf {\bibinfo
  {volume} {102}},\ \bibinfo {pages} {11635} (\bibinfo {year}
  {2005})}\BibitemShut {NoStop}%
\bibitem [{\citenamefont {Mermin}(1965)}]{65PR-Mermin}%
  \BibitemOpen
  \bibfield  {author} {\bibinfo {author} {\bibfnamefont {N.~D.}\ \bibnamefont
  {Mermin}},\ }\bibfield  {title} {\bibinfo {title} {Thermal properties of the
  inhomogeneous electron gas},\ }\href@noop {} {\bibfield  {journal} {\bibinfo
  {journal} {Phys. Rev.}\ }\textbf {\bibinfo {volume} {137}},\ \bibinfo {pages}
  {A1441} (\bibinfo {year} {1965})}\BibitemShut {NoStop}%
\bibitem [{\citenamefont {Ichimaru}(1982)}]{82RMP-Ichimaru}%
  \BibitemOpen
  \bibfield  {author} {\bibinfo {author} {\bibfnamefont {S.}~\bibnamefont
  {Ichimaru}},\ }\bibfield  {title} {\bibinfo {title} {Strongly coupled
  plasmas: high-density classical plasmas and degenerate electron liquids},\
  }\href {https://doi.org/10.1103/RevModPhys.54.1017} {\bibfield  {journal}
  {\bibinfo  {journal} {Rev. Mod. Phys.}\ }\textbf {\bibinfo {volume} {54}},\
  \bibinfo {pages} {1017} (\bibinfo {year} {1982})}\BibitemShut {NoStop}%
\bibitem [{\citenamefont {Graziani}\ \emph {et~al.}(2014)\citenamefont
  {Graziani}, \citenamefont {Desjarlais}, \citenamefont {Redmer},\ and\
  \citenamefont {Trickey}}]{14-Graziani}%
  \BibitemOpen
  \bibfield  {author} {\bibinfo {author} {\bibfnamefont {F.}~\bibnamefont
  {Graziani}}, \bibinfo {author} {\bibfnamefont {M.~P.}\ \bibnamefont
  {Desjarlais}}, \bibinfo {author} {\bibfnamefont {R.}~\bibnamefont {Redmer}},\
  and\ \bibinfo {author} {\bibfnamefont {S.~B.}\ \bibnamefont {Trickey}},\
  }\href@noop {} {\emph {\bibinfo {title} {Frontiers and challenges in warm
  dense matter}}},\ Vol.~\bibinfo {volume} {96}\ (\bibinfo  {publisher}
  {Springer Science \& Business},\ \bibinfo {year} {2014})\BibitemShut
  {NoStop}%
\bibitem [{\citenamefont {Koenig}\ \emph {et~al.}(2005)\citenamefont {Koenig},
  \citenamefont {Benuzzi-Mounaix}, \citenamefont {Ravasio}, \citenamefont
  {Vinci}, \citenamefont {Ozaki}, \citenamefont {Lepape}, \citenamefont
  {Batani}, \citenamefont {Huser}, \citenamefont {Hall}, \citenamefont {Hicks},
  \citenamefont {MacKinnon}, \citenamefont {Patel}, \citenamefont {Park},
  \citenamefont {Boehly}, \citenamefont {Borghesi}, \citenamefont {Kar},\ and\
  \citenamefont {Romagnani}}]{05PPCF-Koenig}%
  \BibitemOpen
  \bibfield  {author} {\bibinfo {author} {\bibfnamefont {M.}~\bibnamefont
  {Koenig}}, \bibinfo {author} {\bibfnamefont {A.}~\bibnamefont
  {Benuzzi-Mounaix}}, \bibinfo {author} {\bibfnamefont {A.}~\bibnamefont
  {Ravasio}}, \bibinfo {author} {\bibfnamefont {T.}~\bibnamefont {Vinci}},
  \bibinfo {author} {\bibfnamefont {N.}~\bibnamefont {Ozaki}}, \bibinfo
  {author} {\bibfnamefont {S.}~\bibnamefont {Lepape}}, \bibinfo {author}
  {\bibfnamefont {D.}~\bibnamefont {Batani}}, \bibinfo {author} {\bibfnamefont
  {G.}~\bibnamefont {Huser}}, \bibinfo {author} {\bibfnamefont
  {T.}~\bibnamefont {Hall}}, \bibinfo {author} {\bibfnamefont {D.}~\bibnamefont
  {Hicks}}, \bibinfo {author} {\bibfnamefont {A.}~\bibnamefont {MacKinnon}},
  \bibinfo {author} {\bibfnamefont {P.}~\bibnamefont {Patel}}, \bibinfo
  {author} {\bibfnamefont {H.~S.}\ \bibnamefont {Park}}, \bibinfo {author}
  {\bibfnamefont {T.}~\bibnamefont {Boehly}}, \bibinfo {author} {\bibfnamefont
  {M.}~\bibnamefont {Borghesi}}, \bibinfo {author} {\bibfnamefont
  {S.}~\bibnamefont {Kar}},\ and\ \bibinfo {author} {\bibfnamefont
  {L.}~\bibnamefont {Romagnani}},\ }\bibfield  {title} {\bibinfo {title}
  {Progress in the study of warm dense matter},\ }\href
  {https://doi.org/10.1088/0741-3335/47/12b/s31} {\bibfield  {journal}
  {\bibinfo  {journal} {Plasma Phys. Controlled Fusion}\ }\textbf {\bibinfo
  {volume} {47}},\ \bibinfo {pages} {B441} (\bibinfo {year}
  {2005})}\BibitemShut {NoStop}%
\bibitem [{\citenamefont {Wang}\ \emph {et~al.}(2011)\citenamefont {Wang},
  \citenamefont {He},\ and\ \citenamefont {Zhang}}]{11PP-Wang}%
  \BibitemOpen
  \bibfield  {author} {\bibinfo {author} {\bibfnamefont {C.}~\bibnamefont
  {Wang}}, \bibinfo {author} {\bibfnamefont {X.-T.}\ \bibnamefont {He}},\ and\
  \bibinfo {author} {\bibfnamefont {P.}~\bibnamefont {Zhang}},\ }\bibfield
  {title} {\bibinfo {title} {Thermophysical properties for shock compressed
  polystyrene},\ }\href {https://doi.org/10.1063/1.3625273} {\bibfield
  {journal} {\bibinfo  {journal} {Phys. Plasmas}\ }\textbf {\bibinfo {volume}
  {18}},\ \bibinfo {pages} {082707} (\bibinfo {year} {2011})}\BibitemShut
  {NoStop}%
\bibitem [{\citenamefont {Sheppard}\ \emph {et~al.}(2014)\citenamefont
  {Sheppard}, \citenamefont {Kress}, \citenamefont {Crockett}, \citenamefont
  {Collins},\ and\ \citenamefont {Desjarlais}}]{14E-Sheppard}%
  \BibitemOpen
  \bibfield  {author} {\bibinfo {author} {\bibfnamefont {D.}~\bibnamefont
  {Sheppard}}, \bibinfo {author} {\bibfnamefont {J.~D.}\ \bibnamefont {Kress}},
  \bibinfo {author} {\bibfnamefont {S.}~\bibnamefont {Crockett}}, \bibinfo
  {author} {\bibfnamefont {L.~A.}\ \bibnamefont {Collins}},\ and\ \bibinfo
  {author} {\bibfnamefont {M.~P.}\ \bibnamefont {Desjarlais}},\ }\bibfield
  {title} {\bibinfo {title} {Combining kohn-sham and orbital-free
  density-functional theory for hugoniot calculations to extreme pressures},\
  }\href {https://doi.org/10.1103/PhysRevE.90.063314} {\bibfield  {journal}
  {\bibinfo  {journal} {Phys. Rev. E}\ }\textbf {\bibinfo {volume} {90}},\
  \bibinfo {pages} {063314} (\bibinfo {year} {2014})}\BibitemShut {NoStop}%
\bibitem [{\citenamefont {Hu}\ \emph {et~al.}(2015)\citenamefont {Hu},
  \citenamefont {Collins}, \citenamefont {Goncharov}, \citenamefont {Kress},
  \citenamefont {McCrory},\ and\ \citenamefont {Skupsky}}]{15E-Hu}%
  \BibitemOpen
  \bibfield  {author} {\bibinfo {author} {\bibfnamefont {S.~X.}\ \bibnamefont
  {Hu}}, \bibinfo {author} {\bibfnamefont {L.~A.}\ \bibnamefont {Collins}},
  \bibinfo {author} {\bibfnamefont {V.~N.}\ \bibnamefont {Goncharov}}, \bibinfo
  {author} {\bibfnamefont {J.~D.}\ \bibnamefont {Kress}}, \bibinfo {author}
  {\bibfnamefont {R.~L.}\ \bibnamefont {McCrory}},\ and\ \bibinfo {author}
  {\bibfnamefont {S.}~\bibnamefont {Skupsky}},\ }\bibfield  {title} {\bibinfo
  {title} {First-principles equation of state of polystyrene and its effect on
  inertial confinement fusion implosions},\ }\href
  {https://doi.org/10.1103/PhysRevE.92.043104} {\bibfield  {journal} {\bibinfo
  {journal} {Phys. Rev. E}\ }\textbf {\bibinfo {volume} {92}},\ \bibinfo
  {pages} {043104} (\bibinfo {year} {2015})}\BibitemShut {NoStop}%
\bibitem [{\citenamefont {Driver}\ \emph {et~al.}(2018)\citenamefont {Driver},
  \citenamefont {Soubiran},\ and\ \citenamefont {Militzer}}]{18E-Driver}%
  \BibitemOpen
  \bibfield  {author} {\bibinfo {author} {\bibfnamefont {K.~P.}\ \bibnamefont
  {Driver}}, \bibinfo {author} {\bibfnamefont {F.}~\bibnamefont {Soubiran}},\
  and\ \bibinfo {author} {\bibfnamefont {B.}~\bibnamefont {Militzer}},\
  }\bibfield  {title} {\bibinfo {title} {Path integral monte carlo simulations
  of warm dense aluminum},\ }\href {https://doi.org/10.1103/PhysRevE.97.063207}
  {\bibfield  {journal} {\bibinfo  {journal} {Phys. Rev. E}\ }\textbf {\bibinfo
  {volume} {97}},\ \bibinfo {pages} {063207} (\bibinfo {year}
  {2018})}\BibitemShut {NoStop}%
\bibitem [{\citenamefont {Vinko}\ \emph {et~al.}(2009)\citenamefont {Vinko},
  \citenamefont {Gregori}, \citenamefont {Desjarlais}, \citenamefont {Nagler},
  \citenamefont {Whitcher}, \citenamefont {Lee}, \citenamefont {Audebert},\
  and\ \citenamefont {Wark}}]{09HEDP-Vinko}%
  \BibitemOpen
  \bibfield  {author} {\bibinfo {author} {\bibfnamefont {S.~M.}\ \bibnamefont
  {Vinko}}, \bibinfo {author} {\bibfnamefont {G.}~\bibnamefont {Gregori}},
  \bibinfo {author} {\bibfnamefont {M.~P.}\ \bibnamefont {Desjarlais}},
  \bibinfo {author} {\bibfnamefont {B.}~\bibnamefont {Nagler}}, \bibinfo
  {author} {\bibfnamefont {T.~J.}\ \bibnamefont {Whitcher}}, \bibinfo {author}
  {\bibfnamefont {R.~W.}\ \bibnamefont {Lee}}, \bibinfo {author} {\bibfnamefont
  {P.}~\bibnamefont {Audebert}},\ and\ \bibinfo {author} {\bibfnamefont
  {J.~S.}\ \bibnamefont {Wark}},\ }\bibfield  {title} {\bibinfo {title}
  {Free–free opacity in warm dense aluminum},\ }\href
  {https://doi.org/https://doi.org/10.1016/j.hedp.2009.04.004} {\bibfield
  {journal} {\bibinfo  {journal} {High Energy Density Phys.}\ }\textbf
  {\bibinfo {volume} {5}},\ \bibinfo {pages} {124} (\bibinfo {year}
  {2009})}\BibitemShut {NoStop}%
\bibitem [{\citenamefont {Hu}\ \emph {et~al.}(2014)\citenamefont {Hu},
  \citenamefont {Collins}, \citenamefont {Goncharov}, \citenamefont {Boehly},
  \citenamefont {Epstein}, \citenamefont {McCrory},\ and\ \citenamefont
  {Skupsky}}]{14E-Hu}%
  \BibitemOpen
  \bibfield  {author} {\bibinfo {author} {\bibfnamefont {S.~X.}\ \bibnamefont
  {Hu}}, \bibinfo {author} {\bibfnamefont {L.~A.}\ \bibnamefont {Collins}},
  \bibinfo {author} {\bibfnamefont {V.~N.}\ \bibnamefont {Goncharov}}, \bibinfo
  {author} {\bibfnamefont {T.~R.}\ \bibnamefont {Boehly}}, \bibinfo {author}
  {\bibfnamefont {R.}~\bibnamefont {Epstein}}, \bibinfo {author} {\bibfnamefont
  {R.~L.}\ \bibnamefont {McCrory}},\ and\ \bibinfo {author} {\bibfnamefont
  {S.}~\bibnamefont {Skupsky}},\ }\bibfield  {title} {\bibinfo {title}
  {First-principles opacity table of warm dense deuterium for
  inertial-confinement-fusion applications},\ }\href
  {https://doi.org/10.1103/PhysRevE.90.033111} {\bibfield  {journal} {\bibinfo
  {journal} {Phys. Rev. E}\ }\textbf {\bibinfo {volume} {90}},\ \bibinfo
  {pages} {033111} (\bibinfo {year} {2014})}\BibitemShut {NoStop}%
\bibitem [{\citenamefont {Zhang}\ \emph
  {et~al.}(2016{\natexlab{a}})\citenamefont {Zhang}, \citenamefont {Zhao},
  \citenamefont {Kang}, \citenamefont {Zhang},\ and\ \citenamefont
  {He}}]{16B-Zhang}%
  \BibitemOpen
  \bibfield  {author} {\bibinfo {author} {\bibfnamefont {S.}~\bibnamefont
  {Zhang}}, \bibinfo {author} {\bibfnamefont {S.}~\bibnamefont {Zhao}},
  \bibinfo {author} {\bibfnamefont {W.}~\bibnamefont {Kang}}, \bibinfo {author}
  {\bibfnamefont {P.}~\bibnamefont {Zhang}},\ and\ \bibinfo {author}
  {\bibfnamefont {X.-T.}\ \bibnamefont {He}},\ }\bibfield  {title} {\bibinfo
  {title} {Link between $k$ absorption edges and thermodynamic properties of
  warm dense plasmas established by an improved first-principles method},\
  }\href {https://doi.org/10.1103/PhysRevB.93.115114} {\bibfield  {journal}
  {\bibinfo  {journal} {Phys. Rev. B}\ }\textbf {\bibinfo {volume} {93}},\
  \bibinfo {pages} {115114} (\bibinfo {year} {2016}{\natexlab{a}})}\BibitemShut
  {NoStop}%
\bibitem [{\citenamefont {Karasiev}\ and\ \citenamefont
  {Hu}(2021)}]{21E-Karasiev}%
  \BibitemOpen
  \bibfield  {author} {\bibinfo {author} {\bibfnamefont {V.~V.}\ \bibnamefont
  {Karasiev}}\ and\ \bibinfo {author} {\bibfnamefont {S.~X.}\ \bibnamefont
  {Hu}},\ }\bibfield  {title} {\bibinfo {title} {Unraveling the intrinsic
  atomic physics behind x-ray absorption line shifts in warm dense silicon
  plasmas},\ }\href {https://doi.org/10.1103/PhysRevE.103.033202} {\bibfield
  {journal} {\bibinfo  {journal} {Phys. Rev. E}\ }\textbf {\bibinfo {volume}
  {103}},\ \bibinfo {pages} {033202} (\bibinfo {year} {2021})}\BibitemShut
  {NoStop}%
\bibitem [{\citenamefont {Glenzer}\ and\ \citenamefont
  {Redmer}(2009)}]{09RMP-Glenzer}%
  \BibitemOpen
  \bibfield  {author} {\bibinfo {author} {\bibfnamefont {S.~H.}\ \bibnamefont
  {Glenzer}}\ and\ \bibinfo {author} {\bibfnamefont {R.}~\bibnamefont
  {Redmer}},\ }\bibfield  {title} {\bibinfo {title} {X-ray thomson scattering
  in high energy density plasmas},\ }\href
  {https://doi.org/10.1103/RevModPhys.81.1625} {\bibfield  {journal} {\bibinfo
  {journal} {Rev. Mod. Phys.}\ }\textbf {\bibinfo {volume} {81}},\ \bibinfo
  {pages} {1625} (\bibinfo {year} {2009})}\BibitemShut {NoStop}%
\bibitem [{\citenamefont {Vorberger}\ and\ \citenamefont
  {Gericke}(2015)}]{15E-Vorberger}%
  \BibitemOpen
  \bibfield  {author} {\bibinfo {author} {\bibfnamefont {J.}~\bibnamefont
  {Vorberger}}\ and\ \bibinfo {author} {\bibfnamefont {D.~O.}\ \bibnamefont
  {Gericke}},\ }\bibfield  {title} {\bibinfo {title} {Ab initio approach to
  model x-ray diffraction in warm dense matter},\ }\href
  {https://doi.org/10.1103/PhysRevE.91.033112} {\bibfield  {journal} {\bibinfo
  {journal} {Phys. Rev. E}\ }\textbf {\bibinfo {volume} {91}},\ \bibinfo
  {pages} {033112} (\bibinfo {year} {2015})}\BibitemShut {NoStop}%
\bibitem [{\citenamefont {Mo}\ \emph {et~al.}(2018)\citenamefont {Mo},
  \citenamefont {Fu}, \citenamefont {Kang}, \citenamefont {Zhang},\ and\
  \citenamefont {He}}]{18L-Mo}%
  \BibitemOpen
  \bibfield  {author} {\bibinfo {author} {\bibfnamefont {C.}~\bibnamefont
  {Mo}}, \bibinfo {author} {\bibfnamefont {Z.}~\bibnamefont {Fu}}, \bibinfo
  {author} {\bibfnamefont {W.}~\bibnamefont {Kang}}, \bibinfo {author}
  {\bibfnamefont {P.}~\bibnamefont {Zhang}},\ and\ \bibinfo {author}
  {\bibfnamefont {X.~T.}\ \bibnamefont {He}},\ }\bibfield  {title} {\bibinfo
  {title} {First-principles estimation of electronic temperature from x-ray
  thomson scattering spectrum of isochorically heated warm dense matter},\
  }\href {https://doi.org/10.1103/PhysRevLett.120.205002} {\bibfield  {journal}
  {\bibinfo  {journal} {Phys. Rev. Lett.}\ }\textbf {\bibinfo {volume} {120}},\
  \bibinfo {pages} {205002} (\bibinfo {year} {2018})}\BibitemShut {NoStop}%
\bibitem [{\citenamefont {Desjarlais}\ \emph {et~al.}(2002)\citenamefont
  {Desjarlais}, \citenamefont {Kress},\ and\ \citenamefont
  {Collins}}]{02E-Desjarlais}%
  \BibitemOpen
  \bibfield  {author} {\bibinfo {author} {\bibfnamefont {M.~P.}\ \bibnamefont
  {Desjarlais}}, \bibinfo {author} {\bibfnamefont {J.~D.}\ \bibnamefont
  {Kress}},\ and\ \bibinfo {author} {\bibfnamefont {L.~A.}\ \bibnamefont
  {Collins}},\ }\bibfield  {title} {\bibinfo {title} {Electrical conductivity
  for warm, dense aluminum plasmas and liquids},\ }\href
  {https://doi.org/10.1103/PhysRevE.66.025401} {\bibfield  {journal} {\bibinfo
  {journal} {Phys. Rev. E}\ }\textbf {\bibinfo {volume} {66}},\ \bibinfo
  {pages} {025401} (\bibinfo {year} {2002})}\BibitemShut {NoStop}%
\bibitem [{\citenamefont {Witte}\ \emph {et~al.}(2018)\citenamefont {Witte},
  \citenamefont {Sperling}, \citenamefont {French}, \citenamefont {Recoules},
  \citenamefont {Glenzer},\ and\ \citenamefont {Redmer}}]{18PP-Witte}%
  \BibitemOpen
  \bibfield  {author} {\bibinfo {author} {\bibfnamefont {B.~B.~L.}\
  \bibnamefont {Witte}}, \bibinfo {author} {\bibfnamefont {P.}~\bibnamefont
  {Sperling}}, \bibinfo {author} {\bibfnamefont {M.}~\bibnamefont {French}},
  \bibinfo {author} {\bibfnamefont {V.}~\bibnamefont {Recoules}}, \bibinfo
  {author} {\bibfnamefont {S.~H.}\ \bibnamefont {Glenzer}},\ and\ \bibinfo
  {author} {\bibfnamefont {R.}~\bibnamefont {Redmer}},\ }\bibfield  {title}
  {\bibinfo {title} {Observations of non-linear plasmon damping in dense
  plasmas},\ }\href {https://doi.org/10.1063/1.5017889} {\bibfield  {journal}
  {\bibinfo  {journal} {Phys. Plasmas}\ }\textbf {\bibinfo {volume} {25}},\
  \bibinfo {pages} {056901} (\bibinfo {year} {2018})}\BibitemShut {NoStop}%
\bibitem [{\citenamefont {Karasiev}\ \emph {et~al.}(2014)\citenamefont
  {Karasiev}, \citenamefont {Sjostrom}, \citenamefont {Dufty},\ and\
  \citenamefont {Trickey}}]{14L-Karasiev}%
  \BibitemOpen
  \bibfield  {author} {\bibinfo {author} {\bibfnamefont {V.~V.}\ \bibnamefont
  {Karasiev}}, \bibinfo {author} {\bibfnamefont {T.}~\bibnamefont {Sjostrom}},
  \bibinfo {author} {\bibfnamefont {J.}~\bibnamefont {Dufty}},\ and\ \bibinfo
  {author} {\bibfnamefont {S.~B.}\ \bibnamefont {Trickey}},\ }\bibfield
  {title} {\bibinfo {title} {Accurate homogeneous electron gas
  exchange-correlation free energy for local spin-density calculations},\
  }\href {https://doi.org/10.1103/PhysRevLett.112.076403} {\bibfield  {journal}
  {\bibinfo  {journal} {Phys. Rev. Lett.}\ }\textbf {\bibinfo {volume} {112}},\
  \bibinfo {pages} {076403} (\bibinfo {year} {2014})}\BibitemShut {NoStop}%
\bibitem [{\citenamefont {Karasiev}\ \emph {et~al.}(2016)\citenamefont
  {Karasiev}, \citenamefont {Calder\'{\i}n},\ and\ \citenamefont
  {Trickey}}]{16E-Karasiev}%
  \BibitemOpen
  \bibfield  {author} {\bibinfo {author} {\bibfnamefont {V.~V.}\ \bibnamefont
  {Karasiev}}, \bibinfo {author} {\bibfnamefont {L.}~\bibnamefont
  {Calder\'{\i}n}},\ and\ \bibinfo {author} {\bibfnamefont {S.~B.}\
  \bibnamefont {Trickey}},\ }\bibfield  {title} {\bibinfo {title} {Importance
  of finite-temperature exchange correlation for warm dense matter
  calculations},\ }\href {https://doi.org/10.1103/PhysRevE.93.063207}
  {\bibfield  {journal} {\bibinfo  {journal} {Phys. Rev. E}\ }\textbf {\bibinfo
  {volume} {93}},\ \bibinfo {pages} {063207} (\bibinfo {year}
  {2016})}\BibitemShut {NoStop}%
\bibitem [{\citenamefont {Karasiev}\ \emph {et~al.}(2019)\citenamefont
  {Karasiev}, \citenamefont {Hu}, \citenamefont {Zaghoo},\ and\ \citenamefont
  {Boehly}}]{19B-Karasiev}%
  \BibitemOpen
  \bibfield  {author} {\bibinfo {author} {\bibfnamefont {V.~V.}\ \bibnamefont
  {Karasiev}}, \bibinfo {author} {\bibfnamefont {S.~X.}\ \bibnamefont {Hu}},
  \bibinfo {author} {\bibfnamefont {M.}~\bibnamefont {Zaghoo}},\ and\ \bibinfo
  {author} {\bibfnamefont {T.~R.}\ \bibnamefont {Boehly}},\ }\bibfield  {title}
  {\bibinfo {title} {Exchange-correlation thermal effects in shocked deuterium:
  Softening the principal hugoniot and thermophysical properties},\ }\href
  {https://doi.org/10.1103/PhysRevB.99.214110} {\bibfield  {journal} {\bibinfo
  {journal} {Phys. Rev. B}\ }\textbf {\bibinfo {volume} {99}},\ \bibinfo
  {pages} {214110} (\bibinfo {year} {2019})}\BibitemShut {NoStop}%
\bibitem [{\citenamefont {Surh}\ \emph {et~al.}(2001)\citenamefont {Surh},
  \citenamefont {Barbee},\ and\ \citenamefont {Yang}}]{01L-Surh}%
  \BibitemOpen
  \bibfield  {author} {\bibinfo {author} {\bibfnamefont {M.~P.}\ \bibnamefont
  {Surh}}, \bibinfo {author} {\bibfnamefont {T.~W.}\ \bibnamefont {Barbee}},\
  and\ \bibinfo {author} {\bibfnamefont {L.~H.}\ \bibnamefont {Yang}},\
  }\bibfield  {title} {\bibinfo {title} {First principles molecular dynamics of
  dense plasmas},\ }\href {https://doi.org/10.1103/PhysRevLett.86.5958}
  {\bibfield  {journal} {\bibinfo  {journal} {Phys. Rev. Lett.}\ }\textbf
  {\bibinfo {volume} {86}},\ \bibinfo {pages} {5958} (\bibinfo {year}
  {2001})}\BibitemShut {NoStop}%
\bibitem [{\citenamefont {Hu}\ \emph {et~al.}(2010)\citenamefont {Hu},
  \citenamefont {Militzer}, \citenamefont {Goncharov},\ and\ \citenamefont
  {Skupsky}}]{10L-Hu}%
  \BibitemOpen
  \bibfield  {author} {\bibinfo {author} {\bibfnamefont {S.~X.}\ \bibnamefont
  {Hu}}, \bibinfo {author} {\bibfnamefont {B.}~\bibnamefont {Militzer}},
  \bibinfo {author} {\bibfnamefont {V.~N.}\ \bibnamefont {Goncharov}},\ and\
  \bibinfo {author} {\bibfnamefont {S.}~\bibnamefont {Skupsky}},\ }\bibfield
  {title} {\bibinfo {title} {Strong coupling and degeneracy effects in inertial
  confinement fusion implosions},\ }\href
  {https://doi.org/10.1103/PhysRevLett.104.235003} {\bibfield  {journal}
  {\bibinfo  {journal} {Phys. Rev. Lett.}\ }\textbf {\bibinfo {volume} {104}},\
  \bibinfo {pages} {235003} (\bibinfo {year} {2010})}\BibitemShut {NoStop}%
\bibitem [{\citenamefont {Cytter}\ \emph {et~al.}(2018)\citenamefont {Cytter},
  \citenamefont {Rabani}, \citenamefont {Neuhauser},\ and\ \citenamefont
  {Baer}}]{18B-Cytter}%
  \BibitemOpen
  \bibfield  {author} {\bibinfo {author} {\bibfnamefont {Y.}~\bibnamefont
  {Cytter}}, \bibinfo {author} {\bibfnamefont {E.}~\bibnamefont {Rabani}},
  \bibinfo {author} {\bibfnamefont {D.}~\bibnamefont {Neuhauser}},\ and\
  \bibinfo {author} {\bibfnamefont {R.}~\bibnamefont {Baer}},\ }\bibfield
  {title} {\bibinfo {title} {Stochastic density functional theory at finite
  temperatures},\ }\href {https://doi.org/10.1103/PhysRevB.97.115207}
  {\bibfield  {journal} {\bibinfo  {journal} {Phys. Rev. B}\ }\textbf {\bibinfo
  {volume} {97}},\ \bibinfo {pages} {115207} (\bibinfo {year}
  {2018})}\BibitemShut {NoStop}%
\bibitem [{\citenamefont {Pollock}\ and\ \citenamefont
  {Ceperley}(1984)}]{84B-Pollock}%
  \BibitemOpen
  \bibfield  {author} {\bibinfo {author} {\bibfnamefont {E.~L.}\ \bibnamefont
  {Pollock}}\ and\ \bibinfo {author} {\bibfnamefont {D.~M.}\ \bibnamefont
  {Ceperley}},\ }\bibfield  {title} {\bibinfo {title} {Simulation of quantum
  many-body systems by path-integral methods},\ }\href
  {https://doi.org/10.1103/PhysRevB.30.2555} {\bibfield  {journal} {\bibinfo
  {journal} {Phys. Rev. B}\ }\textbf {\bibinfo {volume} {30}},\ \bibinfo
  {pages} {2555} (\bibinfo {year} {1984})}\BibitemShut {NoStop}%
\bibitem [{\citenamefont {Takahashi}\ and\ \citenamefont
  {Imada}(1984)}]{84JPSJ-Takahashi}%
  \BibitemOpen
  \bibfield  {author} {\bibinfo {author} {\bibfnamefont {M.}~\bibnamefont
  {Takahashi}}\ and\ \bibinfo {author} {\bibfnamefont {M.}~\bibnamefont
  {Imada}},\ }\bibfield  {title} {\bibinfo {title} {Monte carlo calculation of
  quantum systems},\ }\href
  {https://doi.org/https://doi.org/10.1143/JPSJ.53.963} {\bibfield  {journal}
  {\bibinfo  {journal} {J. Phys. Soc. Jpn.}\ }\textbf {\bibinfo {volume}
  {53}},\ \bibinfo {pages} {963} (\bibinfo {year} {1984})}\BibitemShut
  {NoStop}%
\bibitem [{\citenamefont {Ceperley}\ and\ \citenamefont
  {Pollock}(1986)}]{86L-Ceperley}%
  \BibitemOpen
  \bibfield  {author} {\bibinfo {author} {\bibfnamefont {D.~M.}\ \bibnamefont
  {Ceperley}}\ and\ \bibinfo {author} {\bibfnamefont {E.~L.}\ \bibnamefont
  {Pollock}},\ }\bibfield  {title} {\bibinfo {title} {Path-integral computation
  of the low-temperature properties of liquid $^{4}\mathrm{He}$},\ }\href
  {https://doi.org/10.1103/PhysRevLett.56.351} {\bibfield  {journal} {\bibinfo
  {journal} {Phys. Rev. Lett.}\ }\textbf {\bibinfo {volume} {56}},\ \bibinfo
  {pages} {351} (\bibinfo {year} {1986})}\BibitemShut {NoStop}%
\bibitem [{\citenamefont {Ceperley}(1995)}]{95RMP-Ceperley}%
  \BibitemOpen
  \bibfield  {author} {\bibinfo {author} {\bibfnamefont {D.~M.}\ \bibnamefont
  {Ceperley}},\ }\bibfield  {title} {\bibinfo {title} {Path integrals in the
  theory of condensed helium},\ }\href
  {https://doi.org/10.1103/RevModPhys.67.279} {\bibfield  {journal} {\bibinfo
  {journal} {Rev. Mod. Phys.}\ }\textbf {\bibinfo {volume} {67}},\ \bibinfo
  {pages} {279} (\bibinfo {year} {1995})}\BibitemShut {NoStop}%
\bibitem [{\citenamefont {Militzer}\ and\ \citenamefont
  {Driver}(2015)}]{15L-Militzer}%
  \BibitemOpen
  \bibfield  {author} {\bibinfo {author} {\bibfnamefont {B.}~\bibnamefont
  {Militzer}}\ and\ \bibinfo {author} {\bibfnamefont {K.~P.}\ \bibnamefont
  {Driver}},\ }\bibfield  {title} {\bibinfo {title} {Development of path
  integral monte carlo simulations with localized nodal surfaces for second-row
  elements},\ }\href {https://doi.org/10.1103/PhysRevLett.115.176403}
  {\bibfield  {journal} {\bibinfo  {journal} {Phys. Rev. Lett.}\ }\textbf
  {\bibinfo {volume} {115}},\ \bibinfo {pages} {176403} (\bibinfo {year}
  {2015})}\BibitemShut {NoStop}%
\bibitem [{\citenamefont {Zhang}\ \emph {et~al.}(2017)\citenamefont {Zhang},
  \citenamefont {Driver}, \citenamefont {Soubiran},\ and\ \citenamefont
  {Militzer}}]{17E-Zhang}%
  \BibitemOpen
  \bibfield  {author} {\bibinfo {author} {\bibfnamefont {S.}~\bibnamefont
  {Zhang}}, \bibinfo {author} {\bibfnamefont {K.~P.}\ \bibnamefont {Driver}},
  \bibinfo {author} {\bibfnamefont {F.~m.~c.}\ \bibnamefont {Soubiran}},\ and\
  \bibinfo {author} {\bibfnamefont {B.}~\bibnamefont {Militzer}},\ }\bibfield
  {title} {\bibinfo {title} {First-principles equation of state and shock
  compression predictions of warm dense hydrocarbons},\ }\href
  {https://doi.org/10.1103/PhysRevE.96.013204} {\bibfield  {journal} {\bibinfo
  {journal} {Phys. Rev. E}\ }\textbf {\bibinfo {volume} {96}},\ \bibinfo
  {pages} {013204} (\bibinfo {year} {2017})}\BibitemShut {NoStop}%
\bibitem [{\citenamefont {Militzer}\ \emph {et~al.}(2021)\citenamefont
  {Militzer}, \citenamefont {Gonz\'alez-Cataldo}, \citenamefont {Zhang},
  \citenamefont {Driver},\ and\ \citenamefont {Soubiran}}]{21E-Militzer}%
  \BibitemOpen
  \bibfield  {author} {\bibinfo {author} {\bibfnamefont {B.}~\bibnamefont
  {Militzer}}, \bibinfo {author} {\bibfnamefont {F.}~\bibnamefont
  {Gonz\'alez-Cataldo}}, \bibinfo {author} {\bibfnamefont {S.}~\bibnamefont
  {Zhang}}, \bibinfo {author} {\bibfnamefont {K.~P.}\ \bibnamefont {Driver}},\
  and\ \bibinfo {author} {\bibfnamefont {F.~m.~c.}\ \bibnamefont {Soubiran}},\
  }\bibfield  {title} {\bibinfo {title} {First-principles equation of state
  database for warm dense matter computation},\ }\href
  {https://doi.org/10.1103/PhysRevE.103.013203} {\bibfield  {journal} {\bibinfo
   {journal} {Phys. Rev. E}\ }\textbf {\bibinfo {volume} {103}},\ \bibinfo
  {pages} {013203} (\bibinfo {year} {2021})}\BibitemShut {NoStop}%
\bibitem [{\citenamefont {Wang}\ and\ \citenamefont {Carter}(2002)}]{02-Wang}%
  \BibitemOpen
  \bibfield  {author} {\bibinfo {author} {\bibfnamefont {Y.~A.}\ \bibnamefont
  {Wang}}\ and\ \bibinfo {author} {\bibfnamefont {E.~A.}\ \bibnamefont
  {Carter}},\ }\bibinfo {title} {Orbital-free kinetic-energy density functional
  theory},\ in\ \href {https://doi.org/10.1007/0-306-46949-9_5} {\emph
  {\bibinfo {booktitle} {Theoretical Methods in Condensed Phase Chemistry}}},\
  \bibinfo {editor} {edited by\ \bibinfo {editor} {\bibfnamefont {S.~D.}\
  \bibnamefont {Schwartz}}}\ (\bibinfo  {publisher} {Springer Netherlands},\
  \bibinfo {address} {Dordrecht},\ \bibinfo {year} {2002})\ pp.\ \bibinfo
  {pages} {117--184}\BibitemShut {NoStop}%
\bibitem [{\citenamefont {Karasiev}\ \emph {et~al.}(2012)\citenamefont
  {Karasiev}, \citenamefont {Sjostrom},\ and\ \citenamefont
  {Trickey}}]{12B-Karasiev}%
  \BibitemOpen
  \bibfield  {author} {\bibinfo {author} {\bibfnamefont {V.~V.}\ \bibnamefont
  {Karasiev}}, \bibinfo {author} {\bibfnamefont {T.}~\bibnamefont {Sjostrom}},\
  and\ \bibinfo {author} {\bibfnamefont {S.~B.}\ \bibnamefont {Trickey}},\
  }\bibfield  {title} {\bibinfo {title} {Generalized-gradient-approximation
  noninteracting free-energy functionals for orbital-free density functional
  calculations},\ }\href {https://doi.org/10.1103/PhysRevB.86.115101}
  {\bibfield  {journal} {\bibinfo  {journal} {Phys. Rev. B}\ }\textbf {\bibinfo
  {volume} {86}},\ \bibinfo {pages} {115101} (\bibinfo {year}
  {2012})}\BibitemShut {NoStop}%
\bibitem [{\citenamefont {Sjostrom}\ and\ \citenamefont
  {Daligault}(2014)}]{14L-Sjostrom}%
  \BibitemOpen
  \bibfield  {author} {\bibinfo {author} {\bibfnamefont {T.}~\bibnamefont
  {Sjostrom}}\ and\ \bibinfo {author} {\bibfnamefont {J.}~\bibnamefont
  {Daligault}},\ }\bibfield  {title} {\bibinfo {title} {Fast and accurate
  quantum molecular dynamics of dense plasmas across temperature regimes},\
  }\href {https://doi.org/10.1103/PhysRevLett.113.155006} {\bibfield  {journal}
  {\bibinfo  {journal} {Phys. Rev. Lett.}\ }\textbf {\bibinfo {volume} {113}},\
  \bibinfo {pages} {155006} (\bibinfo {year} {2014})}\BibitemShut {NoStop}%
\bibitem [{\citenamefont {Luo}\ \emph {et~al.}(2020)\citenamefont {Luo},
  \citenamefont {Karasiev},\ and\ \citenamefont {Trickey}}]{20B-Luo}%
  \BibitemOpen
  \bibfield  {author} {\bibinfo {author} {\bibfnamefont {K.}~\bibnamefont
  {Luo}}, \bibinfo {author} {\bibfnamefont {V.~V.}\ \bibnamefont {Karasiev}},\
  and\ \bibinfo {author} {\bibfnamefont {S.~B.}\ \bibnamefont {Trickey}},\
  }\bibfield  {title} {\bibinfo {title} {Towards accurate orbital-free
  simulations: A generalized gradient approximation for the noninteracting free
  energy density functional},\ }\href
  {https://doi.org/10.1103/PhysRevB.101.075116} {\bibfield  {journal} {\bibinfo
   {journal} {Phys. Rev. B}\ }\textbf {\bibinfo {volume} {101}},\ \bibinfo
  {pages} {075116} (\bibinfo {year} {2020})}\BibitemShut {NoStop}%
\bibitem [{\citenamefont {Thomas}(1927)}]{27MPCPS-Thomas}%
  \BibitemOpen
  \bibfield  {author} {\bibinfo {author} {\bibfnamefont {L.~H.}\ \bibnamefont
  {Thomas}},\ }\bibfield  {title} {\bibinfo {title} {The calculation of atomic
  fields},\ }\href {https://doi.org/10.1017/S0305004100011683} {\bibfield
  {journal} {\bibinfo  {journal} {Math. Proc. Cambridge Philos. Soc.}\ }\textbf
  {\bibinfo {volume} {23}},\ \bibinfo {pages} {542–548} (\bibinfo {year}
  {1927})}\BibitemShut {NoStop}%
\bibitem [{\citenamefont {Wang}\ and\ \citenamefont {Teter}(1992)}]{92B-Wang}%
  \BibitemOpen
  \bibfield  {author} {\bibinfo {author} {\bibfnamefont {L.-W.}\ \bibnamefont
  {Wang}}\ and\ \bibinfo {author} {\bibfnamefont {M.~P.}\ \bibnamefont
  {Teter}},\ }\bibfield  {title} {\bibinfo {title} {Kinetic-energy functional
  of the electron density},\ }\href {https://doi.org/10.1103/PhysRevB.45.13196}
  {\bibfield  {journal} {\bibinfo  {journal} {Phys. Rev. B}\ }\textbf {\bibinfo
  {volume} {45}},\ \bibinfo {pages} {13196} (\bibinfo {year}
  {1992})}\BibitemShut {NoStop}%
\bibitem [{\citenamefont {Huang}\ and\ \citenamefont
  {Carter}(2010)}]{10B-Huang}%
  \BibitemOpen
  \bibfield  {author} {\bibinfo {author} {\bibfnamefont {C.}~\bibnamefont
  {Huang}}\ and\ \bibinfo {author} {\bibfnamefont {E.~A.}\ \bibnamefont
  {Carter}},\ }\bibfield  {title} {\bibinfo {title} {Nonlocal orbital-free
  kinetic energy density functional for semiconductors},\ }\href
  {https://doi.org/10.1103/PhysRevB.81.045206} {\bibfield  {journal} {\bibinfo
  {journal} {Phys. Rev. B}\ }\textbf {\bibinfo {volume} {81}},\ \bibinfo
  {pages} {045206} (\bibinfo {year} {2010})}\BibitemShut {NoStop}%
\bibitem [{\citenamefont {Sjostrom}\ and\ \citenamefont
  {Daligault}(2013)}]{13B-Sjostrom}%
  \BibitemOpen
  \bibfield  {author} {\bibinfo {author} {\bibfnamefont {T.}~\bibnamefont
  {Sjostrom}}\ and\ \bibinfo {author} {\bibfnamefont {J.}~\bibnamefont
  {Daligault}},\ }\bibfield  {title} {\bibinfo {title} {Nonlocal orbital-free
  noninteracting free-energy functional for warm dense matter},\ }\href
  {https://doi.org/10.1103/PhysRevB.88.195103} {\bibfield  {journal} {\bibinfo
  {journal} {Phys. Rev. B}\ }\textbf {\bibinfo {volume} {88}},\ \bibinfo
  {pages} {195103} (\bibinfo {year} {2013})}\BibitemShut {NoStop}%
\bibitem [{\citenamefont {Liu}\ \emph {et~al.}(2020)\citenamefont {Liu},
  \citenamefont {Lu},\ and\ \citenamefont {Chen}}]{20JPCM-Liu}%
  \BibitemOpen
  \bibfield  {author} {\bibinfo {author} {\bibfnamefont {Q.}~\bibnamefont
  {Liu}}, \bibinfo {author} {\bibfnamefont {D.}~\bibnamefont {Lu}},\ and\
  \bibinfo {author} {\bibfnamefont {M.}~\bibnamefont {Chen}},\ }\bibfield
  {title} {\bibinfo {title} {Structure and dynamics of warm dense aluminum: a
  molecular dynamics study with density functional theory and deep potential},\
  }\href {https://doi.org/10.1088/1361-648x/ab5890} {\bibfield  {journal}
  {\bibinfo  {journal} {J. Phys.: Condens. Matter}\ }\textbf {\bibinfo {volume}
  {32}},\ \bibinfo {pages} {144002} (\bibinfo {year} {2020})}\BibitemShut
  {NoStop}%
\bibitem [{\citenamefont {Zhang}\ \emph
  {et~al.}(2016{\natexlab{b}})\citenamefont {Zhang}, \citenamefont {Wang},
  \citenamefont {Kang}, \citenamefont {Zhang},\ and\ \citenamefont
  {He}}]{16PP-Zhang}%
  \BibitemOpen
  \bibfield  {author} {\bibinfo {author} {\bibfnamefont {S.}~\bibnamefont
  {Zhang}}, \bibinfo {author} {\bibfnamefont {H.}~\bibnamefont {Wang}},
  \bibinfo {author} {\bibfnamefont {W.}~\bibnamefont {Kang}}, \bibinfo {author}
  {\bibfnamefont {P.}~\bibnamefont {Zhang}},\ and\ \bibinfo {author}
  {\bibfnamefont {X.~T.}\ \bibnamefont {He}},\ }\bibfield  {title} {\bibinfo
  {title} {Extended application of kohn-sham first-principles molecular
  dynamics method with plane wave approximation at high energy—from cold
  materials to hot dense plasmas},\ }\href {https://doi.org/10.1063/1.4947212}
  {\bibfield  {journal} {\bibinfo  {journal} {Phys. Plasmas}\ }\textbf
  {\bibinfo {volume} {23}},\ \bibinfo {pages} {042707} (\bibinfo {year}
  {2016}{\natexlab{b}})}\BibitemShut {NoStop}%
\bibitem [{\citenamefont {Blanchet}\ \emph {et~al.}(2020)\citenamefont
  {Blanchet}, \citenamefont {Torrent},\ and\ \citenamefont
  {Clérouin}}]{20PP-Blanchet}%
  \BibitemOpen
  \bibfield  {author} {\bibinfo {author} {\bibfnamefont {A.}~\bibnamefont
  {Blanchet}}, \bibinfo {author} {\bibfnamefont {M.}~\bibnamefont {Torrent}},\
  and\ \bibinfo {author} {\bibfnamefont {J.}~\bibnamefont {Clérouin}},\
  }\bibfield  {title} {\bibinfo {title} {Requirements for very high temperature
  kohn–sham dft simulations and how to bypass them},\ }\href
  {https://doi.org/10.1063/5.0016538} {\bibfield  {journal} {\bibinfo
  {journal} {Phys. Plasmas}\ }\textbf {\bibinfo {volume} {27}},\ \bibinfo
  {pages} {122706} (\bibinfo {year} {2020})}\BibitemShut {NoStop}%
\bibitem [{\citenamefont {Liu}\ \emph {et~al.}(2021{\natexlab{a}})\citenamefont
  {Liu}, \citenamefont {Zhang}, \citenamefont {Gao}, \citenamefont {Zhang},
  \citenamefont {Wang}, \citenamefont {Li}, \citenamefont {Zhang},
  \citenamefont {Kang}, \citenamefont {Zhang},\ and\ \citenamefont
  {He}}]{21B-Liu}%
  \BibitemOpen
  \bibfield  {author} {\bibinfo {author} {\bibfnamefont {X.}~\bibnamefont
  {Liu}}, \bibinfo {author} {\bibfnamefont {X.}~\bibnamefont {Zhang}}, \bibinfo
  {author} {\bibfnamefont {C.}~\bibnamefont {Gao}}, \bibinfo {author}
  {\bibfnamefont {S.}~\bibnamefont {Zhang}}, \bibinfo {author} {\bibfnamefont
  {C.}~\bibnamefont {Wang}}, \bibinfo {author} {\bibfnamefont {D.}~\bibnamefont
  {Li}}, \bibinfo {author} {\bibfnamefont {P.}~\bibnamefont {Zhang}}, \bibinfo
  {author} {\bibfnamefont {W.}~\bibnamefont {Kang}}, \bibinfo {author}
  {\bibfnamefont {W.}~\bibnamefont {Zhang}},\ and\ \bibinfo {author}
  {\bibfnamefont {X.~T.}\ \bibnamefont {He}},\ }\bibfield  {title} {\bibinfo
  {title} {Equations of state of poly-$\ensuremath{\alpha}$-methylstyrene and
  polystyrene: First-principles calculations versus precision measurements},\
  }\href {https://doi.org/10.1103/PhysRevB.103.174111} {\bibfield  {journal}
  {\bibinfo  {journal} {Phys. Rev. B}\ }\textbf {\bibinfo {volume} {103}},\
  \bibinfo {pages} {174111} (\bibinfo {year} {2021}{\natexlab{a}})}\BibitemShut
  {NoStop}%
\bibitem [{\citenamefont {Gao}\ \emph {et~al.}(2016)\citenamefont {Gao},
  \citenamefont {Zhang}, \citenamefont {Kang}, \citenamefont {Wang},
  \citenamefont {Zhang},\ and\ \citenamefont {He}}]{16B-Gao}%
  \BibitemOpen
  \bibfield  {author} {\bibinfo {author} {\bibfnamefont {C.}~\bibnamefont
  {Gao}}, \bibinfo {author} {\bibfnamefont {S.}~\bibnamefont {Zhang}}, \bibinfo
  {author} {\bibfnamefont {W.}~\bibnamefont {Kang}}, \bibinfo {author}
  {\bibfnamefont {C.}~\bibnamefont {Wang}}, \bibinfo {author} {\bibfnamefont
  {P.}~\bibnamefont {Zhang}},\ and\ \bibinfo {author} {\bibfnamefont {X.~T.}\
  \bibnamefont {He}},\ }\bibfield  {title} {\bibinfo {title} {Validity boundary
  of orbital-free molecular dynamics method corresponding to thermal ionization
  of shell structure},\ }\href {https://doi.org/10.1103/PhysRevB.94.205115}
  {\bibfield  {journal} {\bibinfo  {journal} {Phys. Rev. B}\ }\textbf {\bibinfo
  {volume} {94}},\ \bibinfo {pages} {205115} (\bibinfo {year}
  {2016})}\BibitemShut {NoStop}%
\bibitem [{\citenamefont {Mo}\ \emph {et~al.}(2020)\citenamefont {Mo},
  \citenamefont {Fu}, \citenamefont {Zhang}, \citenamefont {Kang},
  \citenamefont {Zhang},\ and\ \citenamefont {He}}]{20B-Mo}%
  \BibitemOpen
  \bibfield  {author} {\bibinfo {author} {\bibfnamefont {C.}~\bibnamefont
  {Mo}}, \bibinfo {author} {\bibfnamefont {Z.-G.}\ \bibnamefont {Fu}}, \bibinfo
  {author} {\bibfnamefont {P.}~\bibnamefont {Zhang}}, \bibinfo {author}
  {\bibfnamefont {W.}~\bibnamefont {Kang}}, \bibinfo {author} {\bibfnamefont
  {W.}~\bibnamefont {Zhang}},\ and\ \bibinfo {author} {\bibfnamefont {X.~T.}\
  \bibnamefont {He}},\ }\bibfield  {title} {\bibinfo {title} {First-principles
  method for x-ray thomson scattering including both elastic and inelastic
  features in warm dense matter},\ }\href
  {https://doi.org/10.1103/PhysRevB.102.195127} {\bibfield  {journal} {\bibinfo
   {journal} {Phys. Rev. B}\ }\textbf {\bibinfo {volume} {102}},\ \bibinfo
  {pages} {195127} (\bibinfo {year} {2020})}\BibitemShut {NoStop}%
\bibitem [{\citenamefont {Baer}\ \emph {et~al.}(2013)\citenamefont {Baer},
  \citenamefont {Neuhauser},\ and\ \citenamefont {Rabani}}]{13L-Baer}%
  \BibitemOpen
  \bibfield  {author} {\bibinfo {author} {\bibfnamefont {R.}~\bibnamefont
  {Baer}}, \bibinfo {author} {\bibfnamefont {D.}~\bibnamefont {Neuhauser}},\
  and\ \bibinfo {author} {\bibfnamefont {E.}~\bibnamefont {Rabani}},\
  }\bibfield  {title} {\bibinfo {title} {Self-averaging stochastic kohn-sham
  density-functional theory},\ }\href
  {https://doi.org/10.1103/PhysRevLett.111.106402} {\bibfield  {journal}
  {\bibinfo  {journal} {Phys. Rev. Lett.}\ }\textbf {\bibinfo {volume} {111}},\
  \bibinfo {pages} {106402} (\bibinfo {year} {2013})}\BibitemShut {NoStop}%
\bibitem [{\citenamefont {Neuhauser}\ \emph
  {et~al.}(2014{\natexlab{a}})\citenamefont {Neuhauser}, \citenamefont {Baer},\
  and\ \citenamefont {Rabani}}]{14JCP-Daniel}%
  \BibitemOpen
  \bibfield  {author} {\bibinfo {author} {\bibfnamefont {D.}~\bibnamefont
  {Neuhauser}}, \bibinfo {author} {\bibfnamefont {R.}~\bibnamefont {Baer}},\
  and\ \bibinfo {author} {\bibfnamefont {E.}~\bibnamefont {Rabani}},\
  }\bibfield  {title} {\bibinfo {title} {Communication: Embedded fragment
  stochastic density functional theory},\ }\href
  {https://doi.org/10.1063/1.4890651} {\bibfield  {journal} {\bibinfo
  {journal} {J. Chem. Phys.}\ }\textbf {\bibinfo {volume} {141}},\ \bibinfo
  {pages} {041102} (\bibinfo {year} {2014}{\natexlab{a}})}\BibitemShut
  {NoStop}%
\bibitem [{\citenamefont {Arnon}\ \emph {et~al.}(2017)\citenamefont {Arnon},
  \citenamefont {Rabani}, \citenamefont {Neuhauser},\ and\ \citenamefont
  {Baer}}]{17JCP-Aronon}%
  \BibitemOpen
  \bibfield  {author} {\bibinfo {author} {\bibfnamefont {E.}~\bibnamefont
  {Arnon}}, \bibinfo {author} {\bibfnamefont {E.}~\bibnamefont {Rabani}},
  \bibinfo {author} {\bibfnamefont {D.}~\bibnamefont {Neuhauser}},\ and\
  \bibinfo {author} {\bibfnamefont {R.}~\bibnamefont {Baer}},\ }\bibfield
  {title} {\bibinfo {title} {Equilibrium configurations of large nanostructures
  using the embedded saturated-fragments stochastic density functional
  theory},\ }\href {https://doi.org/10.1063/1.4984931} {\bibfield  {journal}
  {\bibinfo  {journal} {J. Chem. Phys.}\ }\textbf {\bibinfo {volume} {146}},\
  \bibinfo {pages} {224111} (\bibinfo {year} {2017})}\BibitemShut {NoStop}%
\bibitem [{\citenamefont {Cytter}\ \emph {et~al.}(2019)\citenamefont {Cytter},
  \citenamefont {Rabani}, \citenamefont {Neuhauser}, \citenamefont {Preising},
  \citenamefont {Redmer},\ and\ \citenamefont {Baer}}]{19B-Cytter}%
  \BibitemOpen
  \bibfield  {author} {\bibinfo {author} {\bibfnamefont {Y.}~\bibnamefont
  {Cytter}}, \bibinfo {author} {\bibfnamefont {E.}~\bibnamefont {Rabani}},
  \bibinfo {author} {\bibfnamefont {D.}~\bibnamefont {Neuhauser}}, \bibinfo
  {author} {\bibfnamefont {M.}~\bibnamefont {Preising}}, \bibinfo {author}
  {\bibfnamefont {R.}~\bibnamefont {Redmer}},\ and\ \bibinfo {author}
  {\bibfnamefont {R.}~\bibnamefont {Baer}},\ }\bibfield  {title} {\bibinfo
  {title} {Transition to metallization in warm dense helium-hydrogen mixtures
  using stochastic density functional theory within the kubo-greenwood
  formalism},\ }\href {https://doi.org/10.1103/PhysRevB.100.195101} {\bibfield
  {journal} {\bibinfo  {journal} {Phys. Rev. B}\ }\textbf {\bibinfo {volume}
  {100}},\ \bibinfo {pages} {195101} (\bibinfo {year} {2019})}\BibitemShut
  {NoStop}%
\bibitem [{\citenamefont {Chen}\ \emph
  {et~al.}(2019{\natexlab{a}})\citenamefont {Chen}, \citenamefont {Baer},
  \citenamefont {Neuhauser},\ and\ \citenamefont {Rabani}}]{19JCP-Ming}%
  \BibitemOpen
  \bibfield  {author} {\bibinfo {author} {\bibfnamefont {M.}~\bibnamefont
  {Chen}}, \bibinfo {author} {\bibfnamefont {R.}~\bibnamefont {Baer}}, \bibinfo
  {author} {\bibfnamefont {D.}~\bibnamefont {Neuhauser}},\ and\ \bibinfo
  {author} {\bibfnamefont {E.}~\bibnamefont {Rabani}},\ }\bibfield  {title}
  {\bibinfo {title} {Overlapped embedded fragment stochastic density functional
  theory for covalently-bonded materials},\ }\href
  {https://doi.org/10.1063/1.5064472} {\bibfield  {journal} {\bibinfo
  {journal} {J. Chem. Phys.}\ }\textbf {\bibinfo {volume} {150}},\ \bibinfo
  {pages} {034106} (\bibinfo {year} {2019}{\natexlab{a}})}\BibitemShut
  {NoStop}%
\bibitem [{\citenamefont {Chen}\ \emph
  {et~al.}(2019{\natexlab{b}})\citenamefont {Chen}, \citenamefont {Baer},
  \citenamefont {Neuhauser},\ and\ \citenamefont {Rabani}}]{19JCP-Ming2}%
  \BibitemOpen
  \bibfield  {author} {\bibinfo {author} {\bibfnamefont {M.}~\bibnamefont
  {Chen}}, \bibinfo {author} {\bibfnamefont {R.}~\bibnamefont {Baer}}, \bibinfo
  {author} {\bibfnamefont {D.}~\bibnamefont {Neuhauser}},\ and\ \bibinfo
  {author} {\bibfnamefont {E.}~\bibnamefont {Rabani}},\ }\bibfield  {title}
  {\bibinfo {title} {Energy window stochastic density functional theory},\
  }\href {https://doi.org/10.1063/1.5114984} {\bibfield  {journal} {\bibinfo
  {journal} {J. Chem. Phys.}\ }\textbf {\bibinfo {volume} {151}},\ \bibinfo
  {pages} {114116} (\bibinfo {year} {2019}{\natexlab{b}})}\BibitemShut
  {NoStop}%
\bibitem [{\citenamefont {Li}\ \emph {et~al.}(2019)\citenamefont {Li},
  \citenamefont {Chen}, \citenamefont {Rabani}, \citenamefont {Baer},\ and\
  \citenamefont {Neuhauser}}]{19JCP-Li}%
  \BibitemOpen
  \bibfield  {author} {\bibinfo {author} {\bibfnamefont {W.}~\bibnamefont
  {Li}}, \bibinfo {author} {\bibfnamefont {M.}~\bibnamefont {Chen}}, \bibinfo
  {author} {\bibfnamefont {E.}~\bibnamefont {Rabani}}, \bibinfo {author}
  {\bibfnamefont {R.}~\bibnamefont {Baer}},\ and\ \bibinfo {author}
  {\bibfnamefont {D.}~\bibnamefont {Neuhauser}},\ }\bibfield  {title} {\bibinfo
  {title} {Stochastic embedding dft: Theory and application to p-nitroaniline
  in water},\ }\href {https://doi.org/10.1063/1.5110226} {\bibfield  {journal}
  {\bibinfo  {journal} {J. Chem. Phys.}\ }\textbf {\bibinfo {volume} {151}},\
  \bibinfo {pages} {174115} (\bibinfo {year} {2019})}\BibitemShut {NoStop}%
\bibitem [{\citenamefont {Fabian}\ \emph {et~al.}(2019)\citenamefont {Fabian},
  \citenamefont {Shpiro}, \citenamefont {Rabani}, \citenamefont {Neuhauser},\
  and\ \citenamefont {Baer}}]{19WCMS-Fabian}%
  \BibitemOpen
  \bibfield  {author} {\bibinfo {author} {\bibfnamefont {M.~D.}\ \bibnamefont
  {Fabian}}, \bibinfo {author} {\bibfnamefont {B.}~\bibnamefont {Shpiro}},
  \bibinfo {author} {\bibfnamefont {E.}~\bibnamefont {Rabani}}, \bibinfo
  {author} {\bibfnamefont {D.}~\bibnamefont {Neuhauser}},\ and\ \bibinfo
  {author} {\bibfnamefont {R.}~\bibnamefont {Baer}},\ }\bibfield  {title}
  {\bibinfo {title} {Stochastic density functional theory},\ }\href
  {https://doi.org/https://doi.org/10.1002/wcms.1412} {\bibfield  {journal}
  {\bibinfo  {journal} {WIREs Comput. Mol. Sci.}\ }\textbf {\bibinfo {volume}
  {9}},\ \bibinfo {pages} {e1412} (\bibinfo {year} {2019})}\BibitemShut
  {NoStop}%
\bibitem [{\citenamefont {White}\ and\ \citenamefont
  {Collins}(2020)}]{20L-White}%
  \BibitemOpen
  \bibfield  {author} {\bibinfo {author} {\bibfnamefont {A.~J.}\ \bibnamefont
  {White}}\ and\ \bibinfo {author} {\bibfnamefont {L.~A.}\ \bibnamefont
  {Collins}},\ }\bibfield  {title} {\bibinfo {title} {Fast and universal
  kohn-sham density functional theory algorithm for warm dense matter to hot
  dense plasma},\ }\href {https://doi.org/10.1103/PhysRevLett.125.055002}
  {\bibfield  {journal} {\bibinfo  {journal} {Phys. Rev. Lett.}\ }\textbf
  {\bibinfo {volume} {125}},\ \bibinfo {pages} {055002} (\bibinfo {year}
  {2020})}\BibitemShut {NoStop}%
\bibitem [{\citenamefont {Martin}(2004)}]{04-Martin}%
  \BibitemOpen
  \bibfield  {author} {\bibinfo {author} {\bibfnamefont {R.~M.}\ \bibnamefont
  {Martin}},\ }\href@noop {} {\emph {\bibinfo {title} {Electronic structure:
  basic theory and practical methods}}}\ (\bibinfo  {publisher} {Cambridge
  university press},\ \bibinfo {year} {2004})\ p.\ \bibinfo {pages}
  {204}\BibitemShut {NoStop}%
\bibitem [{\citenamefont {Hamann}\ \emph {et~al.}(1979)\citenamefont {Hamann},
  \citenamefont {Schl\"uter},\ and\ \citenamefont {Chiang}}]{79L-Hamann}%
  \BibitemOpen
  \bibfield  {author} {\bibinfo {author} {\bibfnamefont {D.~R.}\ \bibnamefont
  {Hamann}}, \bibinfo {author} {\bibfnamefont {M.}~\bibnamefont {Schl\"uter}},\
  and\ \bibinfo {author} {\bibfnamefont {C.}~\bibnamefont {Chiang}},\
  }\bibfield  {title} {\bibinfo {title} {Norm-conserving pseudopotentials},\
  }\href {https://doi.org/10.1103/PhysRevLett.43.1494} {\bibfield  {journal}
  {\bibinfo  {journal} {Phys. Rev. Lett.}\ }\textbf {\bibinfo {volume} {43}},\
  \bibinfo {pages} {1494} (\bibinfo {year} {1979})}\BibitemShut {NoStop}%
\bibitem [{\citenamefont {Payne}\ \emph {et~al.}(1992)\citenamefont {Payne},
  \citenamefont {Teter}, \citenamefont {Allan}, \citenamefont {Arias},\ and\
  \citenamefont {Joannopoulos}}]{92RMP-Payne}%
  \BibitemOpen
  \bibfield  {author} {\bibinfo {author} {\bibfnamefont {M.~C.}\ \bibnamefont
  {Payne}}, \bibinfo {author} {\bibfnamefont {M.~P.}\ \bibnamefont {Teter}},
  \bibinfo {author} {\bibfnamefont {D.~C.}\ \bibnamefont {Allan}}, \bibinfo
  {author} {\bibfnamefont {T.~A.}\ \bibnamefont {Arias}},\ and\ \bibinfo
  {author} {\bibfnamefont {J.~D.}\ \bibnamefont {Joannopoulos}},\ }\bibfield
  {title} {\bibinfo {title} {Iterative minimization techniques for ab initio
  total-energy calculations: molecular dynamics and conjugate gradients},\
  }\href {https://doi.org/10.1103/RevModPhys.64.1045} {\bibfield  {journal}
  {\bibinfo  {journal} {Rev. Mod. Phys.}\ }\textbf {\bibinfo {volume} {64}},\
  \bibinfo {pages} {1045} (\bibinfo {year} {1992})}\BibitemShut {NoStop}%
\bibitem [{\citenamefont {Hellmann}(1935)}]{35AP-Hellmann}%
  \BibitemOpen
  \bibfield  {author} {\bibinfo {author} {\bibfnamefont {H.}~\bibnamefont
  {Hellmann}},\ }\bibfield  {title} {\bibinfo {title} {A combined approximation
  method for the energy calculation in the many-electron problem},\ }\href@noop
  {} {\bibfield  {journal} {\bibinfo  {journal} {Acta Physicochim. URSS}\
  }\textbf {\bibinfo {volume} {1}},\ \bibinfo {pages} {913} (\bibinfo {year}
  {1935})}\BibitemShut {NoStop}%
\bibitem [{\citenamefont {Feynman}(1939)}]{39-Feynman}%
  \BibitemOpen
  \bibfield  {author} {\bibinfo {author} {\bibfnamefont {R.~P.}\ \bibnamefont
  {Feynman}},\ }\bibfield  {title} {\bibinfo {title} {Forces in molecules},\
  }\href {https://doi.org/10.1103/PhysRev.56.340} {\bibfield  {journal}
  {\bibinfo  {journal} {Phys. Rev.}\ }\textbf {\bibinfo {volume} {56}},\
  \bibinfo {pages} {340} (\bibinfo {year} {1939})}\BibitemShut {NoStop}%
\bibitem [{\citenamefont {Chadi}\ and\ \citenamefont
  {Cohen}(1973)}]{73B-Chadi}%
  \BibitemOpen
  \bibfield  {author} {\bibinfo {author} {\bibfnamefont {D.~J.}\ \bibnamefont
  {Chadi}}\ and\ \bibinfo {author} {\bibfnamefont {M.~L.}\ \bibnamefont
  {Cohen}},\ }\bibfield  {title} {\bibinfo {title} {Special points in the
  brillouin zone},\ }\href {https://doi.org/10.1103/PhysRevB.8.5747} {\bibfield
   {journal} {\bibinfo  {journal} {Phys. Rev. B}\ }\textbf {\bibinfo {volume}
  {8}},\ \bibinfo {pages} {5747} (\bibinfo {year} {1973})}\BibitemShut
  {NoStop}%
\bibitem [{\citenamefont {Monkhorst}\ and\ \citenamefont
  {Pack}(1976)}]{76B-Monkhorst}%
  \BibitemOpen
  \bibfield  {author} {\bibinfo {author} {\bibfnamefont {H.~J.}\ \bibnamefont
  {Monkhorst}}\ and\ \bibinfo {author} {\bibfnamefont {J.~D.}\ \bibnamefont
  {Pack}},\ }\bibfield  {title} {\bibinfo {title} {Special points for
  brillouin-zone integrations},\ }\href
  {https://doi.org/10.1103/PhysRevB.13.5188} {\bibfield  {journal} {\bibinfo
  {journal} {Phys. Rev. B}\ }\textbf {\bibinfo {volume} {13}},\ \bibinfo
  {pages} {5188} (\bibinfo {year} {1976})}\BibitemShut {NoStop}%
\bibitem [{\citenamefont {Li}\ \emph {et~al.}(2016)\citenamefont {Li},
  \citenamefont {Liu}, \citenamefont {Chen}, \citenamefont {Lin}, \citenamefont
  {Ren}, \citenamefont {Lin}, \citenamefont {Yang},\ and\ \citenamefont
  {He}}]{16CMS-Li}%
  \BibitemOpen
  \bibfield  {author} {\bibinfo {author} {\bibfnamefont {P.}~\bibnamefont
  {Li}}, \bibinfo {author} {\bibfnamefont {X.}~\bibnamefont {Liu}}, \bibinfo
  {author} {\bibfnamefont {M.}~\bibnamefont {Chen}}, \bibinfo {author}
  {\bibfnamefont {P.}~\bibnamefont {Lin}}, \bibinfo {author} {\bibfnamefont
  {X.}~\bibnamefont {Ren}}, \bibinfo {author} {\bibfnamefont {L.}~\bibnamefont
  {Lin}}, \bibinfo {author} {\bibfnamefont {C.}~\bibnamefont {Yang}},\ and\
  \bibinfo {author} {\bibfnamefont {L.}~\bibnamefont {He}},\ }\bibfield
  {title} {\bibinfo {title} {Large-scale ab initio simulations based on
  systematically improvable atomic basis},\ }\href
  {https://doi.org/10.1016/j.commatsci.2015.07.004} {\bibfield  {journal}
  {\bibinfo  {journal} {Comput. Mater. Sci.}\ }\textbf {\bibinfo {volume}
  {112}},\ \bibinfo {pages} {503} (\bibinfo {year} {2016})}\BibitemShut
  {NoStop}%
\bibitem [{\citenamefont {Chen}\ \emph {et~al.}(2010)\citenamefont {Chen},
  \citenamefont {Guo},\ and\ \citenamefont {He}}]{10JPCM-Mohan}%
  \BibitemOpen
  \bibfield  {author} {\bibinfo {author} {\bibfnamefont {M.}~\bibnamefont
  {Chen}}, \bibinfo {author} {\bibfnamefont {G.-C.}\ \bibnamefont {Guo}},\ and\
  \bibinfo {author} {\bibfnamefont {L.}~\bibnamefont {He}},\ }\bibfield
  {title} {\bibinfo {title} {Systematically improvable optimized atomic basis
  sets forab initiocalculations},\ }\href
  {https://doi.org/10.1088/0953-8984/22/44/445501} {\bibfield  {journal}
  {\bibinfo  {journal} {J. Phys.: Condens. Matter}\ }\textbf {\bibinfo {volume}
  {22}},\ \bibinfo {pages} {445501} (\bibinfo {year} {2010})}\BibitemShut
  {NoStop}%
\bibitem [{\citenamefont {Perdew}\ and\ \citenamefont
  {Zunger}(1981)}]{81B-Perdew}%
  \BibitemOpen
  \bibfield  {author} {\bibinfo {author} {\bibfnamefont {J.~P.}\ \bibnamefont
  {Perdew}}\ and\ \bibinfo {author} {\bibfnamefont {A.}~\bibnamefont
  {Zunger}},\ }\bibfield  {title} {\bibinfo {title} {Self-interaction
  correction to density-functional approximations for many-electron systems},\
  }\href {https://doi.org/10.1103/PhysRevB.23.5048} {\bibfield  {journal}
  {\bibinfo  {journal} {Phys. Rev. B}\ }\textbf {\bibinfo {volume} {23}},\
  \bibinfo {pages} {5048} (\bibinfo {year} {1981})}\BibitemShut {NoStop}%
\bibitem [{\citenamefont {Perdew}\ \emph {et~al.}(1996)\citenamefont {Perdew},
  \citenamefont {Burke},\ and\ \citenamefont {Ernzerhof}}]{96L-PBE}%
  \BibitemOpen
  \bibfield  {author} {\bibinfo {author} {\bibfnamefont {J.~P.}\ \bibnamefont
  {Perdew}}, \bibinfo {author} {\bibfnamefont {K.}~\bibnamefont {Burke}},\ and\
  \bibinfo {author} {\bibfnamefont {M.}~\bibnamefont {Ernzerhof}},\ }\bibfield
  {title} {\bibinfo {title} {Generalized gradient approximation made simple},\
  }\href@noop {} {\bibfield  {journal} {\bibinfo  {journal} {Phys. Rev. L}\
  }\textbf {\bibinfo {volume} {77}},\ \bibinfo {pages} {3865} (\bibinfo {year}
  {1996})}\BibitemShut {NoStop}%
\bibitem [{\citenamefont {Pulay}(1980)}]{80CPL-Pulay}%
  \BibitemOpen
  \bibfield  {author} {\bibinfo {author} {\bibfnamefont {P.}~\bibnamefont
  {Pulay}},\ }\bibfield  {title} {\bibinfo {title} {Convergence acceleration of
  iterative sequences. the case of scf iteration},\ }\href
  {https://doi.org/https://doi.org/10.1016/0009-2614(80)80396-4} {\bibfield
  {journal} {\bibinfo  {journal} {Chem. Phys. Lett.}\ }\textbf {\bibinfo
  {volume} {73}},\ \bibinfo {pages} {393} (\bibinfo {year} {1980})}\BibitemShut
  {NoStop}%
\bibitem [{\citenamefont {Johnson}(1988)}]{88B-Johnson}%
  \BibitemOpen
  \bibfield  {author} {\bibinfo {author} {\bibfnamefont {D.~D.}\ \bibnamefont
  {Johnson}},\ }\bibfield  {title} {\bibinfo {title} {Modified broyden's method
  for accelerating convergence in self-consistent calculations},\ }\href
  {https://doi.org/10.1103/PhysRevB.38.12807} {\bibfield  {journal} {\bibinfo
  {journal} {Phys. Rev. B}\ }\textbf {\bibinfo {volume} {38}},\ \bibinfo
  {pages} {12807} (\bibinfo {year} {1988})}\BibitemShut {NoStop}%
\bibitem [{\citenamefont {Huang}\ \emph {et~al.}(1995)\citenamefont {Huang},
  \citenamefont {Kouri},\ and\ \citenamefont {Hoffman}}]{95CPL-Huang}%
  \BibitemOpen
  \bibfield  {author} {\bibinfo {author} {\bibfnamefont {Y.}~\bibnamefont
  {Huang}}, \bibinfo {author} {\bibfnamefont {D.~J.}\ \bibnamefont {Kouri}},\
  and\ \bibinfo {author} {\bibfnamefont {D.~K.}\ \bibnamefont {Hoffman}},\
  }\bibfield  {title} {\bibinfo {title} {Direct approach to density functional
  theory: iterative treatment using a polynomial representation of the
  heaviside step function operator},\ }\href
  {https://doi.org/https://doi.org/10.1016/0009-2614(95)00900-O} {\bibfield
  {journal} {\bibinfo  {journal} {Chem. Phys. Lett.}\ }\textbf {\bibinfo
  {volume} {243}},\ \bibinfo {pages} {367} (\bibinfo {year}
  {1995})}\BibitemShut {NoStop}%
\bibitem [{\citenamefont {Baer}\ and\ \citenamefont
  {Head-Gordon}(1997{\natexlab{a}})}]{97JCP-Baer}%
  \BibitemOpen
  \bibfield  {author} {\bibinfo {author} {\bibfnamefont {R.}~\bibnamefont
  {Baer}}\ and\ \bibinfo {author} {\bibfnamefont {M.}~\bibnamefont
  {Head-Gordon}},\ }\bibfield  {title} {\bibinfo {title} {Chebyshev expansion
  methods for electronic structure calculations on large molecular systems},\
  }\href {https://doi.org/10.1063/1.474158} {\bibfield  {journal} {\bibinfo
  {journal} {J. Chem. Phys.}\ }\textbf {\bibinfo {volume} {107}},\ \bibinfo
  {pages} {10003} (\bibinfo {year} {1997}{\natexlab{a}})}\BibitemShut {NoStop}%
\bibitem [{\citenamefont {Baer}\ and\ \citenamefont
  {Head-Gordon}(1997{\natexlab{b}})}]{97L-Baer}%
  \BibitemOpen
  \bibfield  {author} {\bibinfo {author} {\bibfnamefont {R.}~\bibnamefont
  {Baer}}\ and\ \bibinfo {author} {\bibfnamefont {M.}~\bibnamefont
  {Head-Gordon}},\ }\bibfield  {title} {\bibinfo {title} {Sparsity of the
  density matrix in kohn-sham density functional theory and an assessment of
  linear system-size scaling methods},\ }\href
  {https://doi.org/10.1103/PhysRevLett.79.3962} {\bibfield  {journal} {\bibinfo
   {journal} {Phys. Rev. Lett.}\ }\textbf {\bibinfo {volume} {79}},\ \bibinfo
  {pages} {3962} (\bibinfo {year} {1997}{\natexlab{b}})}\BibitemShut {NoStop}%
\bibitem [{\citenamefont {Kosloff}\ and\ \citenamefont
  {Tal-Ezer}(1986)}]{86CPL-Kosloff}%
  \BibitemOpen
  \bibfield  {author} {\bibinfo {author} {\bibfnamefont {R.}~\bibnamefont
  {Kosloff}}\ and\ \bibinfo {author} {\bibfnamefont {H.}~\bibnamefont
  {Tal-Ezer}},\ }\bibfield  {title} {\bibinfo {title} {A direct relaxation
  method for calculating eigenfunctions and eigenvalues of the schrödinger
  equation on a grid},\ }\href
  {https://doi.org/https://doi.org/10.1016/0009-2614(86)80262-7} {\bibfield
  {journal} {\bibinfo  {journal} {Chem. Phys. Lett.}\ }\textbf {\bibinfo
  {volume} {127}},\ \bibinfo {pages} {223} (\bibinfo {year}
  {1986})}\BibitemShut {NoStop}%
\bibitem [{\citenamefont {Wang}(1994)}]{94B-Wang}%
  \BibitemOpen
  \bibfield  {author} {\bibinfo {author} {\bibfnamefont {L.-W.}\ \bibnamefont
  {Wang}},\ }\bibfield  {title} {\bibinfo {title} {Calculating the density of
  states and optical-absorption spectra of large quantum systems by the
  plane-wave moments method},\ }\href
  {https://doi.org/10.1103/PhysRevB.49.10154} {\bibfield  {journal} {\bibinfo
  {journal} {Phys. Rev. B}\ }\textbf {\bibinfo {volume} {49}},\ \bibinfo
  {pages} {10154} (\bibinfo {year} {1994})}\BibitemShut {NoStop}%
\bibitem [{\citenamefont {Atkinson}(1989)}]{89-Atkinson}%
  \BibitemOpen
  \bibfield  {author} {\bibinfo {author} {\bibfnamefont {K.~E.}\ \bibnamefont
  {Atkinson}},\ }\href@noop {} {\emph {\bibinfo {title} {An introduction to
  numerical analysis}}}\ (\bibinfo  {publisher} {John wiley \& sons},\ \bibinfo
  {year} {1989})\BibitemShut {NoStop}%
\bibitem [{\citenamefont {Hutchinson}(1989)}]{89CSSC-Hutchinson}%
  \BibitemOpen
  \bibfield  {author} {\bibinfo {author} {\bibfnamefont {M.}~\bibnamefont
  {Hutchinson}},\ }\bibfield  {title} {\bibinfo {title} {A stochastic estimator
  of the trace of the influence matrix for laplacian smoothing splines},\
  }\href {https://doi.org/10.1080/03610918908812806} {\bibfield  {journal}
  {\bibinfo  {journal} {Comm. Stat. Sim. Comp.}\ }\textbf {\bibinfo {volume}
  {18}},\ \bibinfo {pages} {1059} (\bibinfo {year} {1989})}\BibitemShut
  {NoStop}%
\bibitem [{\citenamefont {Baer}\ \emph {et~al.}(2022)\citenamefont {Baer},
  \citenamefont {Neuhauser},\ and\ \citenamefont {Rabani}}]{22ARPC-Baer}%
  \BibitemOpen
  \bibfield  {author} {\bibinfo {author} {\bibfnamefont {R.}~\bibnamefont
  {Baer}}, \bibinfo {author} {\bibfnamefont {D.}~\bibnamefont {Neuhauser}},\
  and\ \bibinfo {author} {\bibfnamefont {E.}~\bibnamefont {Rabani}},\
  }\bibfield  {title} {\bibinfo {title} {Stochastic vector techniques in
  ground-state electronic structure},\ }\href
  {https://doi.org/10.1146/annurev-physchem-090519-045916} {\bibfield
  {journal} {\bibinfo  {journal} {Annu. Rev. Phys. Chem.}\ }\textbf {\bibinfo
  {volume} {73}},\ \bibinfo {pages} {null} (\bibinfo {year}
  {2022})}\BibitemShut {NoStop}%
\bibitem [{\citenamefont {Neuhauser}\ \emph
  {et~al.}(2014{\natexlab{b}})\citenamefont {Neuhauser}, \citenamefont {Gao},
  \citenamefont {Arntsen}, \citenamefont {Karshenas}, \citenamefont {Rabani},\
  and\ \citenamefont {Baer}}]{14L-Neuhauser}%
  \BibitemOpen
  \bibfield  {author} {\bibinfo {author} {\bibfnamefont {D.}~\bibnamefont
  {Neuhauser}}, \bibinfo {author} {\bibfnamefont {Y.}~\bibnamefont {Gao}},
  \bibinfo {author} {\bibfnamefont {C.}~\bibnamefont {Arntsen}}, \bibinfo
  {author} {\bibfnamefont {C.}~\bibnamefont {Karshenas}}, \bibinfo {author}
  {\bibfnamefont {E.}~\bibnamefont {Rabani}},\ and\ \bibinfo {author}
  {\bibfnamefont {R.}~\bibnamefont {Baer}},\ }\bibfield  {title} {\bibinfo
  {title} {Breaking the theoretical scaling limit for predicting quasiparticle
  energies: The stochastic $gw$ approach},\ }\href
  {https://doi.org/10.1103/PhysRevLett.113.076402} {\bibfield  {journal}
  {\bibinfo  {journal} {Phys. Rev. Lett.}\ }\textbf {\bibinfo {volume} {113}},\
  \bibinfo {pages} {076402} (\bibinfo {year} {2014}{\natexlab{b}})}\BibitemShut
  {NoStop}%
\bibitem [{\citenamefont {Vlček}\ \emph {et~al.}(2017)\citenamefont {Vlček},
  \citenamefont {Rabani}, \citenamefont {Neuhauser},\ and\ \citenamefont
  {Baer}}]{17JCTC-Vlcek}%
  \BibitemOpen
  \bibfield  {author} {\bibinfo {author} {\bibfnamefont {V.}~\bibnamefont
  {Vlček}}, \bibinfo {author} {\bibfnamefont {E.}~\bibnamefont {Rabani}},
  \bibinfo {author} {\bibfnamefont {D.}~\bibnamefont {Neuhauser}},\ and\
  \bibinfo {author} {\bibfnamefont {R.}~\bibnamefont {Baer}},\ }\bibfield
  {title} {\bibinfo {title} {Stochastic gw calculations for molecules},\ }\href
  {https://doi.org/10.1021/acs.jctc.7b00770} {\bibfield  {journal} {\bibinfo
  {journal} {J. Chem. Theory Comput.}\ }\textbf {\bibinfo {volume} {13}},\
  \bibinfo {pages} {4997} (\bibinfo {year} {2017})}\BibitemShut {NoStop}%
\bibitem [{\citenamefont {Vl\v{c}ek}\ \emph {et~al.}(2018)\citenamefont
  {Vl\v{c}ek}, \citenamefont {Li}, \citenamefont {Baer}, \citenamefont
  {Rabani},\ and\ \citenamefont {Neuhauser}}]{18B-Vlcek}%
  \BibitemOpen
  \bibfield  {author} {\bibinfo {author} {\bibfnamefont {V.}~\bibnamefont
  {Vl\v{c}ek}}, \bibinfo {author} {\bibfnamefont {W.}~\bibnamefont {Li}},
  \bibinfo {author} {\bibfnamefont {R.}~\bibnamefont {Baer}}, \bibinfo {author}
  {\bibfnamefont {E.}~\bibnamefont {Rabani}},\ and\ \bibinfo {author}
  {\bibfnamefont {D.}~\bibnamefont {Neuhauser}},\ }\bibfield  {title} {\bibinfo
  {title} {Swift $gw$ beyond 10,000 electrons using sparse stochastic
  compression},\ }\href {https://doi.org/10.1103/PhysRevB.98.075107} {\bibfield
   {journal} {\bibinfo  {journal} {Phys. Rev. B}\ }\textbf {\bibinfo {volume}
  {98}},\ \bibinfo {pages} {075107} (\bibinfo {year} {2018})}\BibitemShut
  {NoStop}%
\bibitem [{\citenamefont {Baer}\ and\ \citenamefont
  {Rabani}(2012)}]{12NL-Baer}%
  \BibitemOpen
  \bibfield  {author} {\bibinfo {author} {\bibfnamefont {R.}~\bibnamefont
  {Baer}}\ and\ \bibinfo {author} {\bibfnamefont {E.}~\bibnamefont {Rabani}},\
  }\bibfield  {title} {\bibinfo {title} {Expeditious stochastic calculation of
  multiexciton generation rates in semiconductor nanocrystals},\ }\href
  {https://doi.org/10.1021/nl300452c} {\bibfield  {journal} {\bibinfo
  {journal} {Nano Lett.}\ }\textbf {\bibinfo {volume} {12}},\ \bibinfo {pages}
  {2123} (\bibinfo {year} {2012})}\BibitemShut {NoStop}%
\bibitem [{\citenamefont {Ge}\ \emph {et~al.}(2014)\citenamefont {Ge},
  \citenamefont {Gao}, \citenamefont {Baer}, \citenamefont {Rabani},\ and\
  \citenamefont {Neuhauser}}]{14JPCL-Ge}%
  \BibitemOpen
  \bibfield  {author} {\bibinfo {author} {\bibfnamefont {Q.}~\bibnamefont
  {Ge}}, \bibinfo {author} {\bibfnamefont {Y.}~\bibnamefont {Gao}}, \bibinfo
  {author} {\bibfnamefont {R.}~\bibnamefont {Baer}}, \bibinfo {author}
  {\bibfnamefont {E.}~\bibnamefont {Rabani}},\ and\ \bibinfo {author}
  {\bibfnamefont {D.}~\bibnamefont {Neuhauser}},\ }\bibfield  {title} {\bibinfo
  {title} {A guided stochastic energy-domain formulation of the second order
  møller–plesset perturbation theory},\ }\href
  {https://doi.org/10.1021/jz402206m} {\bibfield  {journal} {\bibinfo
  {journal} {J. Phys. Chem. Lett.}\ }\textbf {\bibinfo {volume} {5}},\ \bibinfo
  {pages} {185} (\bibinfo {year} {2014})}\BibitemShut {NoStop}%
\bibitem [{\citenamefont {Neuhauser}\ \emph {et~al.}(2016)\citenamefont
  {Neuhauser}, \citenamefont {Rabani}, \citenamefont {Cytter},\ and\
  \citenamefont {Baer}}]{16JPCA-Neuhauser}%
  \BibitemOpen
  \bibfield  {author} {\bibinfo {author} {\bibfnamefont {D.}~\bibnamefont
  {Neuhauser}}, \bibinfo {author} {\bibfnamefont {E.}~\bibnamefont {Rabani}},
  \bibinfo {author} {\bibfnamefont {Y.}~\bibnamefont {Cytter}},\ and\ \bibinfo
  {author} {\bibfnamefont {R.}~\bibnamefont {Baer}},\ }\bibfield  {title}
  {\bibinfo {title} {Stochastic optimally tuned range-separated hybrid density
  functional theory},\ }\href {https://doi.org/10.1021/acs.jpca.5b10573}
  {\bibfield  {journal} {\bibinfo  {journal} {J. Phys. Chem. A}\ }\textbf
  {\bibinfo {volume} {120}},\ \bibinfo {pages} {3071} (\bibinfo {year}
  {2016})}\BibitemShut {NoStop}%
\bibitem [{\citenamefont {Pozzo}\ \emph {et~al.}(2011)\citenamefont {Pozzo},
  \citenamefont {Desjarlais},\ and\ \citenamefont {Alf\`e}}]{11B-Pozzo}%
  \BibitemOpen
  \bibfield  {author} {\bibinfo {author} {\bibfnamefont {M.}~\bibnamefont
  {Pozzo}}, \bibinfo {author} {\bibfnamefont {M.~P.}\ \bibnamefont
  {Desjarlais}},\ and\ \bibinfo {author} {\bibfnamefont {D.}~\bibnamefont
  {Alf\`e}},\ }\bibfield  {title} {\bibinfo {title} {Electrical and thermal
  conductivity of liquid sodium from first-principles calculations},\ }\href
  {https://doi.org/10.1103/PhysRevB.84.054203} {\bibfield  {journal} {\bibinfo
  {journal} {Phys. Rev. B}\ }\textbf {\bibinfo {volume} {84}},\ \bibinfo
  {pages} {054203} (\bibinfo {year} {2011})}\BibitemShut {NoStop}%
\bibitem [{\citenamefont {Liu}\ \emph {et~al.}(2021{\natexlab{b}})\citenamefont
  {Liu}, \citenamefont {Li},\ and\ \citenamefont {Chen}}]{21MRE-Liu}%
  \BibitemOpen
  \bibfield  {author} {\bibinfo {author} {\bibfnamefont {Q.}~\bibnamefont
  {Liu}}, \bibinfo {author} {\bibfnamefont {J.}~\bibnamefont {Li}},\ and\
  \bibinfo {author} {\bibfnamefont {M.}~\bibnamefont {Chen}},\ }\bibfield
  {title} {\bibinfo {title} {Thermal transport by electrons and ions in warm
  dense aluminum: A combined density functional theory and deep potential
  study},\ }\href {https://doi.org/10.1063/5.0030123} {\bibfield  {journal}
  {\bibinfo  {journal} {Matter Radiat. Extremes}\ }\textbf {\bibinfo {volume}
  {6}},\ \bibinfo {pages} {026902} (\bibinfo {year}
  {2021}{\natexlab{b}})}\BibitemShut {NoStop}%
\bibitem [{\citenamefont {Nielsen}\ and\ \citenamefont
  {Martin}(1985)}]{85B-Nielsen}%
  \BibitemOpen
  \bibfield  {author} {\bibinfo {author} {\bibfnamefont {O.~H.}\ \bibnamefont
  {Nielsen}}\ and\ \bibinfo {author} {\bibfnamefont {R.~M.}\ \bibnamefont
  {Martin}},\ }\bibfield  {title} {\bibinfo {title} {Stresses in
  semiconductors: Ab initio calculations on si, ge, and gaas},\ }\href
  {https://doi.org/10.1103/PhysRevB.32.3792} {\bibfield  {journal} {\bibinfo
  {journal} {Phys. Rev. B}\ }\textbf {\bibinfo {volume} {32}},\ \bibinfo
  {pages} {3792} (\bibinfo {year} {1985})}\BibitemShut {NoStop}%
\end{thebibliography}%

\end{document}